%% file: DIPS.tex
\documentclass[12pt]{article}
\usepackage[utf8]{inputenc}
\usepackage{amsmath, amssymb, amsthm, bbm, mathrsfs, mathtools}
\usepackage{hyperref}
\usepackage{geometry}
\usepackage{natbib}
\usepackage[linesnumbered,ruled,vlined]{algorithm2e}
\usepackage{verbatim}
\usepackage{multirow}
\usepackage{booktabs, array, threeparttable}
\usepackage{rotating}
\usepackage{graphicx}
\usepackage{subfigure}
\usepackage{float}
\usepackage{caption}
\usepackage{amsfonts}
\usepackage{xcolor}
\usepackage{xr}
\allowdisplaybreaks

\mathtoolsset{showonlyrefs=true}

\usepackage{geometry}
\input{CustomizedCommands.tex}

\SetKwInput{KwInput}{Input}                
\SetKwInput{KwOutput}{Output}              
\renewcommand{\theequation} {\arabic{section}.\arabic{equation}}
\newtheorem{theorem}{Theorem}[section]
\newtheorem{lemma}{Lemma}[section]

\newtheorem{proposition}{Proposition}[section]
\newtheorem{definition}{Definition}[section]

\newtheorem{assumption}{Assumption}[section]
\newtheorem{example}{Example}[section]

\newcommand{\Varpi}[1]{{\operatorname{Var}^\pi\left\{#1\right\}}}

\usepackage{xargs}
\renewcommandx{\E}[3][1= , 2= ]{\bbE_{#1}^{#2}\lt\{#3\rt\}}

\makeatletter
\newcommand*{\addFileDependency}[1]{
  \typeout{(#1)}
  \@addtofilelist{#1}
  \IfFileExists{#1}{}{\typeout{No file #1.}}
}

\title{Asymptotic theory of the quadratic assignment procedure for dyadic data analysis}
\author{Lei Shi and Peng Ding
\footnote{Lei Shi, Division of Biostatistics, University of California, Berkeley, CA 94720  (E-mail: leishi@berkeley.edu). Peng Ding, Department of Statistics, University of California, Berkeley, CA 94720 (E-mail: pengdingpku@berkeley.edu). Peng Ding is partially supported by the National Science Foundation grant \# 1945136.
}
}
\date{}

\begin{document}

\maketitle

\begin{abstract}
\renewcommand{\baselinestretch}{1.5}

The quadratic assignment procedure (QAP) is a popular tool for analyzing dyadic data in medical and social sciences. To test the association between two dyadic measurements represented by two symmetric matrices, QAP calculates the $p$-value by permuting the units, or equivalently, by permuting the rows and columns of one matrix in the same way. Its extension to the regression setting, known as the multiple regression QAP, has also gained popularity, especially in psychometrics. However, the statistics theory for QAP has not been fully established in the literature. We fill the gap in this paper. We formulate the network models underlying various QAPs. We derive (a) the asymptotic sampling distributions of some canonical test statistics and (b) the corresponding asymptotic permutation distributions induced by QAP under strong and weak null hypotheses. Task (a) relies on applying the theory of U-statistics, and task (b) relies on applying the theory of double-indexed permutation statistics. The combination of tasks (a) and (b) provides a relatively complete picture of QAP. Overall, our asymptotic theory suggests that using properly studentized statistics in QAP is a robust choice in that it is finite-sample exact under the strong null hypothesis and preserves the asymptotic type one error rate under the weak null hypothesis.
\vspace{3ex}

    \end{abstract}

\renewcommand{\baselinestretch}{1.5}

\section{Introduction to QAP for dyadic data}


The Quadratic Assignment Procedure (QAP) is widely used in dyadic data analysis, with applications in biology, psychology, economics and other fields \citep{mantel1967detection, winkler2014permutation, silk2017application, broekel2014modeling, elmer2020depressive, lewis2008tastes}. QAP aims to test the relationship between two dyadic measurements, typically represented by symmetric matrices that capture pairwise measurements for the relationship between units. Computationally, as a permutation test, QAP has a double permutation scheme. It permutes the rows and columns of one dyadic matrix in the same way for calculating the $p$-value, preserving the dyadic structure across permutations.


There have been extensive methodological developments and empirical applications of QAP and its extensions. The double permutation idea behind QAP has its roots in the Mantel test \citep{mantel1967detection}, a classical method for testing time-space correlations which was further developed by others \citep{hubert1976quadratic, hubert1981heuristic}. \citet{krackhardt1988predicting} further proposed the multiple regression quadratic assignment procedure (MRQAP) by extending QAP to a multiple regression setting where the parameter of interests are the regression coefficients. Moreover, \citet{krackhardt1988predicting} used simulation studies to demonstrate the importance of the double permutation with respect to the dyadic data structure, as using the standard F-test with linear models on dyadic data can inflate the type one error rate. \citet{dekker2007sensitivity} studied the problem of testing partial regression coefficients in MRQAP. They discussed various double permutation strategies and proposed a new permutation method to address the shortcomings of previously used versions of MRQAP.
Besides, they conducted extensive simulation studies to evaluate the performance of different permutation strategies.



Nevertheless, there are still gaps in the literature on (MR)QAP. First, there is potential for further improvement in the statistical formulation underlying (MR)QAP for analyzing dyadic data. The statistical hypothesis in the literature is usually formulated as testing the independence between an outcome network and covariate networks. In a dyadic linear model, this hypothesis is usually characterized by zero regression coefficients. However, linear regression coefficients are measures of partial correlations instead of independence \citep{chung2013exact, diciccio2017robust}. Therefore, without a rigorous formulation of (MR)QAP, what researchers want to test and what they truly test can differ. Second, although (MR)QAP has attractive empirical performance in many simulation studies \citep{krackhardt1988predicting, dekker2007sensitivity}, its asymptotic theory has not been fully established in the literature. \citet{toulis2019life} established some properties of a residual permutation version of (MR)QAP, yet many fundamental questions remain. For example, how do different test statistics affect the test results? In the multiple regression setting, how do we compare different permutation strategies, such as permuting the outcome networks and permuting the residual networks? Such a theoretical characterization is crucial for understanding the advantages and limitations of different versions of (MR)QAP and can inspire the development of new methodologies. 

We fill the aforementioned gaps. First, we formulate a dyadic model and the statistical hypotheses that motivate (MR)QAP. The model has probabilistic connections to weakly exchangeable arrays \citep{hoover1979relations, aldous1985exchangeability, 
silverman1976limit, eagleson1978limit}, as well as statistical connections to some earlier formulations of dyadic data \citep{barbour1986tests}. Second, we establish the asymptotic properties of several statistical tests and make comparisons among these methods. Our results highlight that permutation tests with proper studentization of the test statistics can have appealing finite-sample and asymptotic guarantees. It serves as a dyadic network generalization of the theoretical discussion on robust permutation tests in the classic two-sample test and least-squares regression settings  \citep{janssen1997studentized, chung2013exact, diciccio2017robust}. We show that the variance estimator used for the studentized permutation test can be conveniently implemented by utilizing the classic cluster-robust variance estimator \citep{liang1986longitudinal} multiplied by a factor of four. Third, we made some technical contributions. Our technical analysis bridges the sampling and permutation distributions of (MR)QAP with a delicate combination of the theory of U-statistics and the theory of double-indexed permutation statistics. Additionally, we developed some new technical results on permutations to complete the technical arguments. These results are of independent interest for other permutation-related problems.







The rest of the paper is organized as follows. Section \ref{sec:background} introduces the background of QAP and formulates the statistical model and hypotheses. Section \ref{sec:method} focuses on the scenario with two dyadic networks and characterizes the properties of QAP, also known as the Mantel test. Section \ref{sec:QAP} studies a dyadic linear regression and presents the asymptotic theory for QAP. Section \ref{sec:MRQAP} extends the discussion to MRQAP. Section \ref{sec:simulation} complements the asymptotic theory with finite-sample simulation studies. Section \ref{sec:case} presents a case study. Section \ref{sec:discussion} concludes. The supplementary material contains all the proofs. 

\textbf{Notation.} 
We use $\cL(t; W) = \bbP\{W \le t\}$ to denote the cumulative distribution function of the random variable $W$. 
Let $[n]$ denote the set of integers $\{1,\dots, n\}$. Let $\bbS_n = \{ \pi \mid \pi: [n] \rightarrow [n] \text{ is a one-to-one mapping} \}$ denote the set of all permutations over $[n]$. We use $\pi$ to denote a particular permutation $\pi \in \mathbb{S}_n$ or a random permutation uniformly over $\mathbb{S}_n$, with meanings clarified in the context. For a matrix $A = (a_{ij}) \in \mathbb{R}^{n\times n}$ with $0$ diagonal elements, we use $\bar{a} = \{ n(n-1) \}^{-1} \sum_{i\neq j} a_{ij}  $ to denote the average of the $n(n-1)$ off-diagonal elements, with $\sum_{i\neq j}$ for the summation over $i, j = 1, \ldots, n$ with $i\neq j$.

\section{Basic setting and model for QAP}\label{sec:background}

Many research problems involve dyadic data, which encode pairwise relationships between units as a symmetric matrix. This includes examples such as trading conditions between countries \citep{broekel2014modeling}, closeness in friendship \citep{felmlee2016toxic}, and similarity in environmental conditions \citep{mbizah2020effect}. In this section, we focus on two symmetric matrices with zero diagonals, $A = (a_{ij}) \in \mathbb{R}^{n\times n}$ and $B = (b_{ij}) \in \mathbb{R}^{n\times n}$, which represent dyadic data between pairs of $n$ units. A classic question in dyadic data analysis is whether the measures in $A$ and $B$ are related. 

Many statistical methods have been proposed to test the relationship between matrices $A$ and $B$. A classic procedure is QAP, which is a permutation test that compares the observed value of the Pearson correlation coefficient between two dyadic matrices against its permutation distribution. 
In particular, QAP adopts a double permutation strategy, which permutes the columns and rows of one matrix in the same way and fully respects the dependence between the pairwise relations. It proceeds as follows. 
\begin{definition}[QAP]\label{def:mantel} Let $W = W(A,B)$ be the test statistic. 
\begin{enumerate}
    \item Compute $W = W(A,B)$, e.g., $W$ can be the Pearson correlation coefficient \begin{align}
        \hrho = \frac{\sum_{i\neq j}(a_{ij} - \oa)(b_{ij} - \ob)}{\{\sum_{i\neq j}(a_{ij} - \oa)^2\}^{1/2} \{\sum_{i\neq j}(b_{ij} - \ob)^2\}^{1/2}};
    \end{align}

    \item Permute the columns and rows of $A$ in the same way to obtain the permutation distribution with $A_\pi = (a_{\pi(i)\pi(j)})_{i,j\in[n]}$ and $B$'s:
    \begin{align*}
        {\cL}(t; W^\pi) = \frac{1}{n!}\sum_{\pi\in\bbS_n} \ind{W^\pi \le  t}, \text{ where }W^\pi = W(A_\pi,B).
    \end{align*}
    When $|\bbS_n| = n!$ is large, we can approximate ${\cL}(t; W^\pi)$ by Monte Carlo. 
    
    \item Compute the two-sided p-value based on the observed value of $W$ and its permutation distribution:
    \begin{align}
        p_{\textup{QAP}} = 1 - {\cL}(|W|; W^\pi) + {\cL}(-|W|; W^\pi).
    \end{align}
\end{enumerate}    
\end{definition}

The notation in Definition \ref{def:mantel} follows \cite{freedman1983nonstochastic}, where the subscript $\pi$ denotes the actual permutation on the indices $[n]$, while the superscript $\pi$ means the variable or distribution is derived from random permutation.

Below we give two concrete examples. 

\begin{example}[Time-space association]  
    A widely-used statistical technique for testing time-space association is the Knox test \citep{knox1964epidemiology, kulldorff1999knox, barbour1986tests}. For the studied cases, the distances between them are calculated in terms of time and space. The time-space association arises if many of the cases that are ``close'' in time are also ``close'' in space or vice versa. Formally, with unit-level time-space data $\{R_i, S_i\}_{i\in[n]}$, we take $a_{ij} = \ind{|R_i - R_j| < \tau}$ and $b_{ij} = \ind{|S_i - S_j| < \delta}$ for some pre-specified constants $\tau$ and $\delta$. The statistic $\sum_{i\neq j} a_{ij}b_{ij}$ 
    measures time-space association. 
\end{example}

\begin{example}[Network analysis] 
    In network analysis, researchers want to study the relationship between two sets of dyadic observations $(a_{ij}) \in \mathbb{R}^{n\times n}$ and $(b_{ij}) \in \mathbb{R}^{n\times n}$ \citep{hubert1976quadratic, hubert1981heuristic, krackhardt1988predicting}. 
    For example, $(a_{ij}, b_{ij})$ can represent the trading relationship and political similarity between countries, friend closeness, and age similarity, the similarity in species composition between different sites, and the similarity in environmental conditions, etc. The statistic $\sum_{i\neq j} (a_{ij} - \overline{a})(b_{ij} - \ob)$ measures the association between the dyadic observations.

\end{example}

Due to the dyadic structure of the observed data, it is natural to formulate a statistical model for the observed dyadic measurements $A$ and $B$ based on \textit{unit-specific features}. For example, in many ecological applications, the observations are distance measures between locations and thus can be characterized by the coordinates of the locations together with a distance metric. As another example, in many international trade applications, the observations are total trading volumes between countries, and the unit-specific features are the country's characteristics such as demographics, economic conditions, technological advancements, etc. Motivated by these examples, we propose the following model for dyadic data:
\begin{assumption}\label{asp:qap}
    Let $\{R_i, S_i\}_{i=1}^n$ be a sequence of independently identically distributed (i.i.d.) copies of $(R,S)$, which are some random variables or vectors. Let $\alpha(r,r')$ and $\beta(s,s')$ be possibly unknown functions that are symmetric in the sense that
    $\alpha(r,r') = \alpha(r',r)$ and $\beta(s,s') = \beta(s',s)$. 
    The elements of the dyadic data matrices $A =(a_{ij}) \in \bbR^{n\times n}$ and $B = (b_{ij}) \in \bbR^{n\times n}$ are characterized by:
    \begin{align}
         a_{ii} = b_{ii} = 0, \text{ for } i\in[n]; \quad  a_{ij} = \alpha(R_i, R_j),~ b_{ij} = \beta(S_i,S_j), \text{ for } i\neq j. 
    \end{align}  
\end{assumption}
A similar model appeared in \cite{barbour1986tests}. The model in Assumption \ref{asp:qap} is motivated by the representation of jointly exchangeable arrays \citep{hoover1979relations}, also termed weak exchangeable arrays \citep{aldous1985exchangeability}. 
We could also extend Assumption \ref{asp:qap} to include idiosyncratic errors in $\alpha$ and $\beta$, that is, $a_{ij} = \alpha(R_i, R_j, u_{ij})$ and $b_{ij} = \beta(S_i, S_j, v_{ij})$, where $u_{ij}$'s and $v_{ij}$'s are i.i.d. errors. We conjecture that the main messages of the paper will remain the same, although the general formulation will further complicates the technical details. We leave it to future research. 



For the aforementioned examples of distance and trading volume, the features $R$, $S$, and the functions $\alpha$, and $\beta$ are known. In other applications, these elements might be unknown. For example, in some psychological studies, the distances are provided by some score that measures the closeness or similarity of units, which usually depends on unknown unit-level features in an unknown way. In data, we still observe $A$ and $B$ in those cases, but not $\alpha,\beta$ and the random features $R, S$. 

Despite the common presence of dyadic data and the popularity of QAP, there are still gaps in the literature. First, the statistical hypothesis has not been formulated rigorously. Testing the independence and zero correlation between two dyadic data matrices are two different problems that need different statistical theories and methods. Second, while extensive simulation studies demonstrated attractive performance of QAP, a rigorous theoretical discussion is missing. 

We start with the first question. Given the two arrays as in Assumption \ref{asp:qap}, the high-level statistical question is to test whether they are related. Let $(R',S')$ be an independent copy of $R,S$. We can formulate two testing problems under Assumption \ref{asp:qap}: 
\begin{align}\label{eqn:null-qap}
    \mathrm{H}_{0\textsc{w}}: \Corr{\alpha(R, R')}{\beta(S, S')} = 0
    \quad \text{and} \quad 
    \mathrm{H}_{0\textsc{s}}: R \indep S. 
\end{align}
We will call $\mathrm{H}_{0\textsc{s}}$ the strong null hypothesis of independence, and call $\mathrm{H}_{0\textsc{w}}$ the weak null hypothesis of zero correlation.
$\mathrm{H}_{0\textsc{w}}$ assumes zero correlation between two distance measures, which generally does not imply the independence between $R$ and $S$. Moreover, $\mathrm{H}_{0\textsc{w}}$ is motivated by the following population linear projection that minimizes the mean square error:
\begin{align*}
(\vartheta_0^\star, \vartheta_1^\star) = \arg\min_{\vartheta_0, \vartheta_1} \E{(\alpha(R,R') - \vartheta_0 - \vartheta_1 \cdot \beta(S,S'))^2}.
\end{align*}
$\mathrm{H}_{0\textsc{s}}$ assumes independence between the two unit-specific features. $\mathrm{H}_{0\textsc{s}}$ is motivated by problems such as testing whether two ecological spatial data are independent, or whether one social dyadic measure is independent of some social-economic similarity metric. $\mathrm{H}_{0\textsc{s}}$ implies $\mathrm{H}_{0\textsc{w}}$. 
By the standard least-squares theory, we have $\vartheta_1^\star = \rho(A,B) \sqrt{\Var{\alpha(R,R')}/\Var{\beta(S,S')}} $, where
\begin{align}\label{eqn:corrAB}
    \rho(A, B) = \frac{\Cov{\alpha(R,R')}{ \beta(S,S')}}{\sqrt{\Var{\alpha(R,R')}}\sqrt{\Var{\beta(S,S')}}}
\end{align}
measures the correlation between $A$ and $B$.
$\mathrm{H}_{0\textsc{w}}$ is equivalent to $\rho(A, B) = 0$. 

\section{Theory for QAP}\label{sec:method}



\subsection{Test statistic and its asymptotic distribution}
We start with the problem of estimation and inference of $\rho(A, B)$ in \eqref{eqn:corrAB}. For the ease of presentation, let $\Phi(x,x')$ denote a general kernel of degree $2$, and $X, X'$ denote two i.i.d. copies from a general distribution $\bbP$. Table \ref{tab:notation-phi} introduces general quantities that are related to the kernel $\Phi(x,x')$ and will be used throughout the paper.
\begin{table}[!ht]
    \centering
    \caption{Notation based on kernel $\Phi(x,x')$ and $X,X'\overset{i.i.d.}{\sim} \bbP$}
    \label{tab:notation-phi}
    \begin{tabular}{cc}
    \toprule
        Full kernel & $\Phi(x,x')$ \\
        Expectation & $\Phi_0 = \E{\Phi(X,X')}$ \\
        First-order projection & $\Phi_1(x) = \E{\Phi(x,X')}$ \\
        Centered kernel & $\tPhi(x,x') = \Phi(x,x') - \Phi_0$\\
        Centered first-order projection & $\tPhi_1(x) = \Phi_1(x) - \Phi_0$\\
        First-order variance & $\eta_{1,\Phi} = \E{\tPhi_1^2(X)} $ \\
        Second-order variance & $\eta_{2,\Phi} = \E{\tPhi^2(X,X')}$ \\
    \bottomrule
    \end{tabular}
\end{table}

Applying the notation in Table \ref{tab:notation-phi} to the kernels $\alpha$ and $\beta$, we have
\begin{gather}
    \alpha_0 = \E{\alpha(R,R')}, \quad \beta_0 = \E{\beta(S,S')}, \\
    \tilde{\alpha}(r,r') = \alpha(r,r') - \alpha_0, \quad \tilde{\beta}(s,s') = \beta(s,s') - \beta_0, \\
    \quad  \eta_{2,\alpha} = \Var{\alpha(R,R')} = 
 \E{\talpha^2(R,R')}, \quad  \eta_{2,\beta} = \Var{\beta(S,S')} = \E{\tbeta^2(S,S')}.
\end{gather} 
Define another kernel of degree $2$:
\begin{align}\label{eqn:phi}
    \phi(r,s;r',s') = \tilde{\alpha}(r,r')\tilde{\beta}(s,s'),
\end{align}
with expectation identical to the covariance between $ \alpha(R,R')$ and $\beta(S,S')$:
\begin{align*}
     \phi_0 = 
 \E{\phi(R,S;R',S')}  = \Cov{\alpha(R,R')}{ \beta(S,S')}.
\end{align*}
Therefore, we call $\phi$ \textit{the covariance kernel} of $\alpha$ and $\beta$. 


A plug-in estimator for $\rho(A,B)$ in  \eqref{eqn:corrAB} is:
\begin{align}\label{eqn:hrho}
\hat{\rho} =  \frac{\hphi_0}{ \sqrt{\heta_{1,\alpha}} \cdot \sqrt{\heta_{1,\beta}} },
\end{align}
where
\begin{gather}
 \hphi_0 = {\frac{1}{n(n-1) - 1} \sum_{i\neq j} (a_{ij} - \overline{a} )( b_{ij} - \overline{b} ) }, \\
 \heta_{2,\alpha} = \frac{1}{n(n-1) - 1} \sum_{i\neq j} (a_{ij} - \overline{a})^2, \label{eqn:heta-alpha}\\
 \heta_{2,\beta} = \frac{1}{n(n-1) - 1} \sum_{i\neq j} (b_{ij} - \overline{b})^2 \label{eqn:heta-beta}
\end{gather}
are the moment estimators of $\phi_0$, $\eta_{1,\alpha}$ and $\eta_{1,\beta}$, respectively. Based on the theory of U-statistics, the above estimators are consistent for their population counterparts. Theorem \ref{thm:asp-super} below presents the asymptotic distribution of the estimator $\hrho$ in  \eqref{eqn:hrho}.

\begin{theorem}[Asymptotic sampling distribution of $\hrho$]\label{thm:asp-super} 
Assume that the covariance kernel in \eqref{eqn:phi} is not degenerate in the sense that 
$\eta_{1,\phi} > 0$,
and the following moments are finite:
\begin{align}
    \E{\alpha^2(R,R')} <\infty, \quad 
    \E{\beta^2(S,S')} <\infty, \quad 
    \E{\phi^2(R,S;R',S')} < \infty.
\end{align}
\begin{enumerate}
    \item Under $\mathrm{H}_{0\textsc{w}}$ in \eqref{eqn:null-qap}, we have: 
\begin{align}\label{eqn:L1star}
    {\sqrt{n}} \hrho\rightsquigarrow \cN\lt(0, v_\textsc{w}\rt), \text{ where } v_\textsc{w} = \frac{4\eta_{1, \phi}}{\eta_{2,\alpha} \eta_{2,\beta}}. 
\end{align}
 
    \item Under $\mathrm{H}_{0\textsc{s}}$ in \eqref{eqn:null-qap}, we further have
\begin{align}\label{eqn:L2star}
    \eta_{1, \phi} = \eta_{1,\alpha}\eta_{1,\beta}, \quad 
    {\sqrt{n} \hrho} \rightsquigarrow \cN\lt(0, v_\textsc{s}\rt), \text{ where } v_\textsc{s} = \frac{4\eta_{1,\alpha}\eta_{1,\beta}}{\eta_{2,\alpha}\eta_{2,\beta}}. 
\end{align}
\end{enumerate}

\end{theorem}

Theorem \ref{thm:asp-super} establishes the asymptotic sampling distribution of $\hrho$ under both the weak null hypothesis and the strong null hypothesis. The non-degeneracy assumption, $\eta_{1,\phi} > 0$, ensures a positive asymptotic variance for $\hrho$. It is motivated by the asymptotic theory of U-statistics, which requires the first-order projection of the kernel to be non-degenerate. It holds for many examples. For example, if $(R,S)$ are generated from a multivariate normal distribution, and $\alpha(\cdot,\cdot)$, $\beta(\cdot,\cdot)$ are any two metrics in $\bbR^p$ and $\bbR^q$ that are derived from norms \citep{kelley2017general}, then $\eta_{1,\phi} > 0$ holds by Proposition \ref{prop:non-degeneracy-normal} in the supplementary material.

Below we use an example to illustrate Theorem \ref{thm:asp-super}. We design it to give closed-form expressions, which can provide insights into the Theorem \ref{thm:asp-super} and will be helpful for the simulation studies in Section \ref{sec:simulation}.


\begin{example}[Dyadic data as pairwise averages]\label{exp:walsh-average}
Consider a random vector $(R, S)$ with mean zero, unit variance, and correlation coefficient $\rho_{R,S}$. Define pairwise sums of the features
    \begin{align*}
       \alpha(r,r') = \frac{r+r'}{\sqrt{2}}, \quad  
       \beta(s,s') = \frac{s+s'}{\sqrt{2}},
    \end{align*}
    up to a scaling of $1/\sqrt{2}$, which ensures that $\Var{\alpha(R, R')} = \Var{\beta(S, S')} = 1$. The covariance kernel is 
    \begin{gather}
      \phi(r,s;r',s') = \alpha(r,r')\beta(s,s') = \frac{(r+r')(s+s')}{2}.
    \end{gather}
    The correlation between these two dyadic data is $\rho^\star = 
    \E{\phi(R,S;R',S')} = \rho_{R,S}$,
which is identical to the correlation between $R$ and $S$. We can compute $\eta_{1,\alpha} =  \eta_{1,\beta} = {1}/{2}$ and $\eta_{2,\alpha} =  \eta_{2,\beta} = 1$. Besides, $\tphi_1(r,s) = (rs + \rho_{R,S})/\sqrt{2}$. Under $\mathrm{H}_{0\textsc{w}}$ in \eqref{eqn:null-qap}, $\rho_{R,S} = 0$, and $\eta_{1,\phi} = \E{(RS)^2}$. 
Applying Theorem \ref{thm:asp-super}, we obtain
\begin{align*}
    \sqrt{n}\hrho \rightsquigarrow
    \left\{
    \begin{array}{cc}
        \cN(0,\E{(RS)^2}), & \text{under } \mathrm{H}_{0\textsc{w}}: \rho_{R,S} = 0; \\
        \cN(0,1), & \text{under } \mathrm{H}_{0\textsc{s}}: R\indep S. 
    \end{array}
    \right.
\end{align*}

\end{example}

We then construct a variance estimator for $\hrho$ in \eqref{eqn:hrho}. 
Based on the standard variance estimation for U-statistics (see Lemma \ref{lem:var-est-1-dim} in the supplementary material), we have the variance estimator:
\begin{align}\label{eqn:hvp}
\hv = \frac{4\heta_{1,\phi}}{\heta_{2,\alpha} \heta_{2,\beta}},
\end{align}
where
\begin{align*}
    \heta_{1,\phi} = \frac{1}{n}\sum_{i=1}^n \lt(\frac{1}{n-1}\sum_{j\neq i}^n ({a}_{ij} - \overline{a})({b}_{ij} - \overline{b})\rt)^2,
\end{align*}
and $\heta_{2,\alpha}$, $\heta_{2,\beta}$ are given by \eqref{eqn:heta-alpha} and \eqref{eqn:heta-beta} respectively. The key insight in constructing $\widehat{\eta}_{1,\phi}$ is to approximate the centered first-order projection of $\phi(r,s; r', s') = \widetilde{\alpha}(r, r') \widetilde{\beta}(s,s')$ by ${(n-1)^{-1}}\sum_{j:j\neq i}^n ({a}_{ij} - \overline{a})({b}_{ij} - \overline{b})$, because
\begin{align}
    \frac{1}{n-1}\sum_{j:j\neq i}^n ({a}_{ij} - \overline{a})({b}_{ij} - \overline{b})
    =&
    \frac{1}{n-1}\sum_{j:j\neq i}^n (\alpha(R_i,R_j) - \overline{a})(\beta(S_i, S_j) - \overline{b})\\
    \approx&
    \E{\talpha(R,R')\tbeta(S,S')\mid R=R_i, S=S_i}\\
    = & \phi_1(R_i, S_i),
\end{align}
The following Theorem \ref{thm:asp-super-s} establishes the consistency of the variance estimator $\hv$, which implies the asymptotic sampling distribution of the studentized $\widehat{\rho}$.

\begin{theorem}[Asymptotic sampling distribution of the studentized $\hat\rho$]\label{thm:asp-super-s}
    Under the same conditions as Theorem \ref{thm:asp-super}, we have 
    \begin{align*}
        \heta_{1,\phi} \rightarrow \eta_{1,\phi}, \quad \heta_{1,\alpha} \rightarrow \eta_{1,\alpha}, \quad 
        \heta_{1,\beta} \rightarrow \eta_{1,\beta},\quad
        \hv \rightarrow \frac{4\eta_{1,\phi}}{\eta_{2,\alpha}\eta_{2,\beta}} \quad \bbP\text{-a.s.}
    \end{align*}
Therefore,
under $\mathrm{H}_{0\textsc{w}}$ (and thus under $\mathrm{H}_{0\textsc{s}}$) in \eqref{eqn:null-qap}, we have
\begin{align*}
    \frac{\sqrt{n}\hrho}{{\hv^{1/2}}} \rightsquigarrow \cN(0,1).
\end{align*}
    
\end{theorem}


\subsection{QAP based on $\hrho$}

Recall QAP in Definition \ref{def:mantel}. This permutation test is motivated by the fact that, under $\mathrm{H}_{0\textsc{s}}$ in \eqref{eqn:null-qap}, $(a_{ij}, b_{ij})_{i,j\in[n]}$ and $(a_{ij}, b_{\pi(i)\pi(j)})_{i,j\in[n]}$ have the same distribution for any permutation $\pi$. Therefore, QAP yields a finite-sample exact $p$-value under $\mathrm{H}_{0\textsc{s}}$ in \eqref{eqn:null-qap}. The following theorem characterizes the limit of ${\cL}(t; \sqrt{n}\hrho^\pi)$ introduced in Definition \ref{def:permute-b}: 
\begin{theorem}[Permutation distribution of $\hrho$]\label{thm:asp-mantel-ns} 
Assume the non-degeneracy of the kernels $\alpha$ and $\beta$:  $\eta_{1,\alpha} > 0$ and $\eta_{1,\beta} > 0$. 
Assume there are univariate functions $\kappa_\alpha(\cdot)$ and $\kappa_\beta(\cdot)$ such that 
\begin{align}\label{eqn:kernel-bd}
    |\alpha(r,r')| \le \kappa_\alpha(r) + \kappa_\alpha(r'), \quad |\beta(s,s')| \le \kappa_\beta(s) + \kappa_\beta(s'),
\end{align}
with $\E{\kappa_\alpha^4(R)} < \infty$ and $\E{\kappa_\beta^4(S)} < \infty$. 
Then we have
    \begin{align*}
        \lim_{n\to\infty} \sup_{t\in\bbR} \lt|{{\cL}(t; \sqrt{n}\hrho^\pi)} - \cL(t;\cN(0,v_\textsc{s}))\rt| = 0, \quad \bbP\text{-a.s.}
    \end{align*}
where $\cN(0,v_\textsc{s})$ is the asymptotic sampling distribution of $\sqrt{n}\hrho$ under $\mathrm{H}_{0\textsc{s}}$ from \eqref{eqn:L2star}.     
\end{theorem}

The proof of Theorem \ref{thm:asp-mantel-ns} relies on the probabilistic results on double-indexed permutation statistics. Compare Theorem \ref{thm:asp-mantel-ns} with Theorem \ref{thm:asp-super}. The permutation distribution of $\sqrt{n}\hrho$ is valid for testing $\mathrm{H}_{0\textsc{s}}$ but is not valid for testing $\mathrm{H}_{0\textsc{w}}$, because $\cN(0,v_\textsc{w})$ generally differs from $\cN(0,v_\textsc{s})$ when $R$ and $S$ are not independent. 

In terms of the regularity conditions, Theorem \ref{thm:asp-mantel-ns} does not require the non-degeneracy of the kernel $\phi$ in \eqref{eqn:phi}, but the non-degeneracy of $\alpha$ and $\beta$ separately. Also, Theorem \ref{thm:asp-mantel-ns} imposes an additional condition \eqref{eqn:kernel-bd}, which states that the kernels $\alpha$ and $\beta$ are bounded by some functions of their marginal variables. This is a mild condition.  We give two examples. First, if $\alpha(r, r')$ is a metric for points in $\bbR^p$ \citep{kelley2017general}, the triangle inequality implies that $\alpha(r,r') \le \alpha(r_0, r) + \alpha(r_0, r')$, 
where $r_0$ is some fixed point in the domain of $R$ (for example, $r_0 = \E{R}$). Then we can take $\kappa_\alpha(r) = \alpha(r_0, r)$. Second, for bounded kernels $|\alpha(r,r')| \le M$, we can choose $\kappa_\alpha(r) = M$ to ensure \eqref{eqn:kernel-bd}. 


\subsection{QAP based on the studentized Pearson correlation}

From Theorem \ref{thm:asp-super} and Theorem \ref{thm:asp-mantel-ns}, the usual QAP targets $\mathrm{H}_{0\textsc{s}}$ but is not robust for testing $\mathrm{H}_{0\textsc{w}}$ with $\sqrt{n}\hrho$ as the test statistic. 
This section studies QAP with the studentized test statistic $\sqrt{n}\hrho/\hv^{1/2}$ and shows its robustness for testing $\mathrm{H}_{0\textsc{w}}$.

\begin{theorem}[Permutation distribution of the studentized $\widehat{\rho}$]\label{thm:asp-mantel-s}  Under the same condition as Theorem \ref{thm:asp-mantel-ns}, we have
    \begin{align*}
        \lim_{n\to\infty} \sup_{t\in\bbR} \lt|\cL(t;\sqrt{n}(\hrho/\hv^{1/2})^\pi) - \cL(t;\cN(0,1))\rt| = 0, \quad \bbP\text{-a.s.}
    \end{align*}
\end{theorem}

Theorem \ref{thm:asp-mantel-s} shows that, for the studentized statistic $\sqrt{n}\hrho/\hv^{1/2}$, the asymptotic permutation distribution matches its asymptotic sampling distribution under $\mathrm{H}_{0\textsc{w}}$ in Theorem \ref{thm:asp-super-s}. Therefore, QAP with the studentized $\widehat{\rho}$ is asymptotically valid for testing zero correlation between $\alpha(R,R')$ and $\beta(S,S')$, in addition to its finite-sample exactness for testing the independence between $R$ and $S$.

To conclude this section, we use Table \ref{tab:compare} to summarize the results. Column 3 and Column 4 have two ``YES'', whereas Column 5 has three ``YES''. Overall, QAP with the studentized statistic is our recommendation, given its superior properties specified in Column 5 of Table \ref{tab:compare}. 

\begin{table}[ht!]
\caption{Comparison of asymptotic and permutation tests with or without studentization}
\label{tab:compare}
\centering
{\small
\begin{tabular}{P{1.3cm}cP{2.4cm}P{2.2cm}P{3.2cm}}
\toprule
Hypothesis              & Criteria & Asymptotic Test & QAP with $\sqrt{n}\hrho$ & QAP with $\sqrt{n}\hrho/\hv^{1/2}$ \\ \midrule
\multirow{2}{*}{$\mathrm{H}_{0\textsc{s}}$} & Finite-Sample Exact?      & NO                               & YES                    & \textbf{YES}                 \\
                                               & Asymptotically Valid?     & YES                              & YES                    & \textbf{YES}                 \\
\multirow{2}{*}{$\mathrm{H}_{0\textsc{w}}$} & Finite-Sample Exact?      & NO                               & NO                     & NO                  \\
                                               & Asymptotically Valid?     & YES                              & NO                     & \textbf{YES}                 \\ \bottomrule
\end{tabular}
}
\end{table}


\section{Linear regression with dyadic data}\label{sec:QAP}

\subsection{A dyadic linear model}\label{sec:model-inference-lr}
In this section, we will generalize the theory of QAP in Section \ref{sec:method} to MRQAP, which targets a multiple linear regression setting. Formally, suppose we have an outcome network $A$ and $p$ covariate networks $B_{1},\dots, B_{p}$. The goal is to test whether $A$ is related to the $B_k$'s. Then MRQAP proceeds as follows.  
\begin{definition}[MRQAP]\label{def:QAP-perm}
    Let $W = W(A, B_{1},\dots, B_{p})$ be the test statistic. 
    \begin{enumerate}
        \item Compute the test statistic $W$ based on the observe data, for example, $W = {n}\cdot\hat{\vartheta}^\top \hV^{-1}\hat{\vartheta}$ where $\hvartheta$ is the coefficient of the $B_k$'s from regressing $A$ on $(1, B_1, \dots, B_p)$ and $\hV$ is some covariance estimator (see \eqref{eqn:hV} below for details); 
        
        \item Permute the columns and rows of $A$ in the same way and obtain the permutation distribution with $A_\pi = (a_{\pi(i)\pi(j)})_{i,j\in[n]}$ and the $B_k$'s:
        \begin{align}
            {\cL}(t; W^\pi) = \frac{1}{n!}\sum_{\pi\in\bbS_n} \ind{W^\pi \le  t}, \text{ where }W^\pi = W(A_\pi, B_1, \dots, B_p).
        \end{align}

        \item Compute the p-value based on the observed $W$ and its permutation distribution. 
    \end{enumerate}
\end{definition}

To understand the properties of MRQAP in Definition \ref{def:QAP-perm}, we start with a mathematical formulation of a dyadic linear model. 
\begin{assumption}\label{asp:noise-lm}
Consider i.i.d. unit-specific features $(R_1, S_1),\cdots, (R_n, S_n) \sim (R, S)$, where $(R, S)$ are possibly unknown multi-dimensional features. 
The dyadic data $\{a_{ij}, (b_{kij})_{k\in[p]}\}_{i,j\in[n]}$ follow the linear model below:
\begin{align}
    a_{ij} = \vartheta_0 + \sum_{k=1}^p \vartheta_k b_{kij} + e_{ij}, 
\end{align}
where the regressors $b_{kij}$'s are generated from kernels $\beta^{(k)}$'s: 
\begin{align}\label{eqn:bkij}
b_{kij} = \beta^{(k)}(S_i, S_j), 
\end{align}
and the noise term $e_{ij}$'s are defined based on the kernel $\epsilon$:
\begin{gather}
    e_{ij} = \epsilon(R_i, R_j) + \zeta_{ij}, \quad 
    \E{\epsilon(R_i, R_j) \mid \{b_{kij}\}_{k=1}^p} = 0, \\
    \{\zeta_{ij}\}_{i<j\in[n]} \overset{i.i.d.}{\sim} \zeta, \quad \{\zeta_{ij}\}_{i<j\in[n]}\indep \{(R_i, S_i)\}_{i\in[n]}.
\end{gather}    
\end{assumption}
Assumption \ref{asp:noise-lm} introduces a set of kernels $\beta^{(k)}$, but the same set of unit-specific features $S_i$ across the $p$ kernels. This general representation covers the special case where each $\beta^{(k)}$ takes a different $S^{(k)}$, as one can always concatenate the $S^{(k)}$'s into a single vector and modify the definition of $\beta^{(k)}$ to obtain a form of \eqref{eqn:bkij}.






The above model on $e_{ij}$ is again motivated by the general representation of the weakly exchangeable arrays \citep{aldous1985exchangeability}, which is given by
$
    e_{ij} = f(R_i, R_j, \zeta_{ij})
$,
where $\{R_i\}_{i\in[n]}\overset{i.i.d.}{\sim} R$, $\{\zeta_{ij}\}_{i,j\in[n]} \overset{i.i.d.}{\sim} \zeta$, $R_i$'s and $\zeta_{ij}$'s are jointly independent. Assumption \ref{asp:noise-lm} takes the partially additive form for simplicity.
The $ \epsilon(R_i, R_j) $ can be interpreted as pairwise noise, which is impacted simultaneously by $R_i$ and $R_j$. The $\zeta_{ij}$ can be interpreted as entrywise noise, which is mostly due to idiosyncratic error. 

Let $\vartheta = (\vartheta_1, \dots, \vartheta_p)^\top$. We are interested in testing the following hypotheses based on the dyadic linear model in Assumption \ref{asp:noise-lm}:
\begin{gather}\label{eqn:null-mrqap-full} 
    \mathrm{H}_{0\textsc{w}}: \vartheta = 0
    \quad \text{and} \quad 
    \mathrm{H}_{0\textsc{s}}: \vartheta = 0 \text{ with }R\indep S,
\end{gather}
which generalizes the weak and strong null hypotheses \eqref{eqn:null-qap} in Section \ref{sec:method} to the multiple regression setting. $\mathrm{H}_{0\textsc{w}}$ tests for a zero coefficient, which implies the outcome and covariate networks are uncorrelated under the model in Assumption \ref{asp:noise-lm}. $\mathrm{H}_{0\textsc{s}}$ tests in addition for the independence between the features $R$ and $S$, which further implies the independence between the outcome and covariate networks.  

\subsection{Asymptotic inference}

In this subsection, we study asymptotic inference based on the dyadic linear model. Under Assumption \ref{asp:noise-lm}, the population-level ordinary least squares gives $\vartheta = \Sigma_{bb}^{-1}\Sigma_{ba}$, 
where $\Sigma_{bb} = (\Cov{b_{kij}}{b_{lij}})_{k,l\in[p]}$ and  $\Sigma_{ba} = (\Cov{a_{ij}}{b_{lij}})_{l\in[p]}$.
This motivates the estimator by regressing $A$ on the $B_k$'s with intercept: $\hvartheta = \hSigma_{bb}^{-1}\hSigma_{ba}$, 
where
\begin{gather}
    \hSigma_{bb} = \lt(\frac{1}{n(n-1)}\sum_{i\neq j} (b_{kij} - \ob_k)(b_{lij} - \ob_l)\rt)_{k,l\in[p]}, \label{eqn:hSigma-bb}\\
     \hSigma_{ba} = \lt( \frac{1}{n(n-1)}\sum_{i\neq j} (a_{ij} - \oa)(b_{kij} - \ob_k) \rt)_{k\in[p]}. \label{eqn:hSigma-ab}
\end{gather} 
Introduce the kernels:
\begin{align*}
    \phi^{(k)}(r,s;r',s') = \epsilon(r,r')\tbeta^{(k)}(s,s'),\quad k=1, \ldots, p. 
\end{align*}
Based on Table \ref{tab:notation-phi}, recall the definition $\tphi^{(k)}(r,s;r',s') = \phi^{(k)}(r,s; r',s') - \phi^{(k)}_{0}$, where $\phi_{0}^{(k)} = \E{\phi^{(k)}(R,S; R',S')}$. 
Also, recall the first-order projection kernels
    \begin{align}
        \tphi^{(k)}_{1}(r,s) = \E{\tphi^{(k)}(r, s; R', S')},\quad k = 1, \ldots, p.
    \end{align}
Define the covariance matrix of the first-order projection kernels
    \begin{align}\label{eqn:H1-phi}
        H_{1,\phi} = \lt( \eta^{(kl)}_{1,\phi} \rt)_{k,l\in[p]},\ \text{with }\eta^{(kl)}_{1,\phi} = \E{\tphi^{(k)}_{1}(R, S) \tphi^{(l)}_{1}(R, S)}.
    \end{align}
Recall $\eta_{1,\epsilon}$ is the covariance of the first-order projection kernel $\tilde{\epsilon}(\cdot,\cdot)$ and $H_{1,\beta}$ is the covariance matrix of the first-order projection kernel $\tilde{\beta}(\cdot,\cdot)$, as introduced in Table \ref{tab:notation-phi}. Using the theory for multi-dimensional U-statistics, we have the following results for the sampling distribution of $\hvartheta$:
\begin{theorem}[Sampling distribution of $\hvartheta$ under the null hypotheses]\label{thm:asp-super-ns-p}
Assume Assumption \ref{asp:noise-lm}. Assume
\begin{align}
    \E{\zeta^2}<\infty, \quad \E{\beta^{(k)}(S, S')^2}<\infty, \quad \E{\epsilon(R, R')^2}<\infty.
\end{align}
Assume each kernel $\phi^{(k)}$ satisfies: $\eta_{1,\phi}^{(kk)} > 0$ and $ \E{\phi^{(k)}(R,S;R',S')^2} < \infty$. 
\begin{enumerate}
    \item Under $\mathrm{H}_{0\textsc{w}}$ in \eqref{eqn:null-mrqap-full}, we have 
    \begin{align}
        \sqrt{n}\hvartheta \rightsquigarrow \cN(0, V_\textsc{w}), \text{ where } V_\textsc{w} = 4\Sigma_{bb}^{-1} H_{1,\phi} \Sigma_{bb}^{-1}.
    \end{align} 
    \item Under $\mathrm{H}_{0\textsc{s}}$ in \eqref{eqn:null-mrqap-full}, we further have
    \begin{align}
        \sqrt{n}\hvartheta \rightsquigarrow \cN(0, V_\textsc{s}), \text{ where } V_\textsc{s} = 4\eta_{1,\epsilon}\Sigma_{bb}^{-1} H_{1,\beta} \Sigma_{bb}^{-1}.
    \end{align}
    
\end{enumerate}

\end{theorem}
We construct a plug-in variance estimator for $V_\textsc{w}$:
\begin{align}\label{eqn:hV}
    \hV = 4\hSigma_{bb}^{-1} \hH_{1,\phi} \hSigma_{bb}^{-1}, 
\end{align}
where $\hSigma_{bb}$ is defined in \eqref{eqn:hSigma-bb} and $\hH_{1,\phi}$ is defined below. 
First, similar to our intuition with a single regressor following Theorem \ref{thm:asp-super-s}, we construct an estimator for the first-order projected kernels $\tphi_1=(\widetilde{\phi}_1^{(1)}, \ldots, \widetilde{\phi}_1^{(p)})^\top$ using the row-wise means:
\begin{align}
    \hat{\phi}_{1,i}
    = \frac{1}{n-1}\sum_{j: j\neq i} \hat{e}_{ij} (\bsb_{ij} - \overline{\bsb}),
\end{align}
where $\hat{e}_{ij} = y_{ij} - \oy - (\bsb_{ij} - \overline{\bsb})^\top \hvartheta$ is the residual from the least-squares fit.
Then, we estimate $H_{1,\phi}$ by the sample covariance of the estimated projected kernels:
\begin{align}
\hH_{1,\phi} = \frac{1}{n} \sum_{i=1}^n \hat{\phi}_{1,i}\hat{\phi}_{1,i}^\top.
\end{align}

In Theorem \ref{thm:asp-super-s-p} below, we verify that $\hV$ converges in probability to the asymptotic variance of $\hvartheta$ under $\mathrm{H}_{0\textsc{w}}$ in \eqref{eqn:null-mrqap-full}. Under $\mathrm{H}_{0\textsc{s}}$ in \eqref{eqn:null-mrqap-full}, $\hV$ works automatically for estimating $V_\textsc{s}$.




\begin{theorem}[Asymptotic sampling distribution for the studentized $\hvartheta$]\label{thm:asp-super-s-p}
Under the same conditions as Theorem \ref{asp:noise-lm}, we have
\begin{align*}
    \hV \rightarrow V_\textsc{w} \quad \bbP\text{-a.s.}
\end{align*}
Therefore, under $ \mathrm{H}_{0\textsc{w}}$ (and thus under $\mathrm{H}_{0\textsc{s}}$) in \eqref{eqn:null-mrqap-full}, we have 
\begin{align*}
    {n}\cdot\hvartheta^\top\hV^{-1}\hvartheta \rightsquigarrow \chi^2_p.
\end{align*}
\end{theorem}

Theorem \ref{thm:asp-super-s-p} ensures an asymptotic test for $\mathrm{H}_{0\textsc{w}}$: we reject $\mathrm{H}_{0\textsc{w}}$ if $
n\cdot \hvartheta^\top \hV^{-1} \hvartheta > q_\alpha$, 
where $q_\alpha$ is the upper $\alpha$ quantile of the $\chi^2_p$ distribution. 








\subsection{MRQAP and its asymptotic properties}



Below we give the permutation distribution of the Wald statistic, $W = {n}\cdot(\hvartheta^\top\hV^{-1}\hvartheta)$.

\begin{theorem}[Permutation distribution of $W$]\label{thm:asp-mantel-s-p}
    Assume Assumption \ref{asp:noise-lm} with $\E{\zeta^4} <\infty$. Assume there are univariate functions $\kappa_\epsilon,\kappa_\beta$ such that 
\begin{align}\label{eqn:kernel-bd-lm-1}
    |\epsilon(r,r')| \le \kappa_\epsilon(r) + \kappa_\epsilon(r'), \quad |\beta^{(k)}(s,s')| \le \kappa_\beta^{(k)}(s) + \kappa_\beta^{(k)}(s'),
\end{align}
where $\E{\kappa_\epsilon(R)^4} < \infty$ and $\bbE\{\kappa_\beta^{(k)}(S)^4\} < \infty$.
Also, assume non-degeneracy of the kernels $\epsilon$ and $\beta^{(k)}$'s in the sense that 
$\eta_{1,\epsilon} > 0$ and $H_{1,\beta} = (\eta_{1,\beta}^{(kl)})_{k,l\in[p]}$ is positive definite.
Under $\mathrm{H}_{0\textsc{w}}$ in \eqref{eqn:null-mrqap-full}, we have
\begin{align*}
    \lim_{n\to\infty} \sup_{t\in\bbR} \lt|\cL(t; W^\pi) - \cL(t;\chi^2_p)\rt| = 0 \quad \bbP\text{-a.s.}
\end{align*}    
\end{theorem}

Theorem \ref{thm:asp-mantel-s-p} shows that the permutation distribution of the Wald statistic matches its sampling distribution under $\mathrm{H}_{0\textsc{w}}$ in \eqref{eqn:null-mrqap-full} asymptotically, in addition to its finite-sample exactness under $\mathrm{H}_{0\textsc{s}}$ in \eqref{eqn:null-mrqap-full}. However, MRQAP with non-Wald test statistics may not be valid for $\mathrm{H}_{0\textsc{w}}$. In the proof of Theorem \ref{thm:asp-mantel-s-p} in the supplementary material, we have shown that, under $\mathrm{H}_{0\textsc{w}}$, the permutation distribution of the regression coefficients $\sqrt{n}\hvartheta$ has an asymptotic normal distribution with variance $V_{\textsc{s}}$, which fails to match $V_{\textsc{w}}$ of the sampling distribution and may lead to Type I error. Therefore, we recommend MRQAP with the Wald statistic $W$ to obtain robust test results.


\subsection{Implementing $\hV$ with the cluster-robust variance estimator}\label{sec:cluster-var-est}
Although $\hV$ in \eqref{eqn:hV} appears to have a complex form, it is easy to implement in practice. We will show that asymptotically, the variance estimator $\hV$ in \eqref{eqn:hV} is equivalent to the cluster-robust variance estimator \citep{liang1986longitudinal}, multiplied by a factor of $4$. The factor $4$ is due to the data matrices being symmetric and half of the components being duplicated.  Intuitively, in the dyadic setting, the observations for each unit should be treated as a ``cluster'' and the variance estimator needs to account for the correlation within the clusters. To see this, recall that for the following linear regression model with $G$ clusters:
    \begin{align}
        y_g = X_g\vartheta + e_g, \quad g = 1,\dots, G,
    \end{align}
    where $y_g \in \bbR^{n_g}$, $e_g \in \bbR^{n_g}$, $X_g \in \bbR^{n_g \times p}$ and $\vartheta \in \bbR^p$. 
    The least-squares estimator and cluster-robust variance estimator are given by 
    \begin{gather}
        \hvartheta 
        = \lt(\sum_{g=1}^G X_g^\top X_g\rt)^{-1} \sum_{g=1}^G X_g^\top y_g,\\
        \hV_{\textup{LZ}} 
        = 
        \lt(\sum_{g=1}^G X_g^\top X_g\rt)^{-1}
        \lt(\sum_{g=1}^G X_g^\top\hat{e}_g \hat{e}_g^\top X_g \rt)
        \lt(\sum_{g=1}^G X_g^\top X_g\rt)^{-1},
    \end{gather}
    where $\hat e_g = y_g - X_g \hvartheta \in \bbR^{n_g}$ for $g = 1,\dots,G$.

    Fill the diagonal elements of $A$ and the $B_k$'s with the mean of the off-diagonal values $\oa$ and $\ob_k$. 
    If we take $G = n$, $y_g$ as the $g$-th column of the outcome matrix, and $X_g$ as the stacked columns of the covariate matrices that involve the $g$-th unit plus an intercept :
    \begin{align}
        y_g = 
        \begin{pmatrix}
            a_{1g}\\
            \vdots\\
            a_{ig}\\
            \vdots\\
            a_{ng}
        \end{pmatrix} \in \bbR^n,
        \quad 
        X_g = 
        \begin{pmatrix}
            1 & b_{11g} & \cdots & b_{k1g} & \cdots & b_{p1g}\\
            \vdots & \vdots & \vdots & \vdots & \vdots & \vdots\\
            1 & b_{1ig} & \cdots & b_{kig} & \cdots & b_{pig}\\
            \vdots & \vdots & \vdots & \vdots & \vdots & \vdots\\
            1 & b_{1ng} & \cdots & b_{kng} & \cdots & b_{png}
        \end{pmatrix} \in \bbR^{n\times (p+1)},
    \end{align}
    then we can recover $\hV$ in \eqref{eqn:hV} by $4 \cdot \widehat{V}_{\textup{LZ}}$ asymptotically. See Section \ref{sec:LZ} in the supplementary material for algebraic details. 
    The equivalence of $\hV$ and cluster-robust variance estimator facilitates the implementation of $\hV$ and the QAP via standard software packages.

We briefly discuss some connections between the variance estimator and the literature.  $\hV$ in \eqref{eqn:hV} is motivated from a U-statistics perspective, and it has a deep connection to cluster-robust variance estimation, which aligns with the findings in the literature of dyadic robust variance estimation \citep{aronow2017cluster, tabord2019inference}. More generally, it is connected to network robust variance estimation \citep{conley1999gmm, fafchamps2007formation}, which is also revealed in previous works. For example, \citet{graham2020dyadic} analyzed a dyadic regression model and proposed a variance estimator that is numerically equivalent to the proposal of \cite{fafchamps2007formation} up to a finite sample correction. Our results verify that the QAP setting allows for simpler implementation via the cluster robust standard error, which is a new message given the literature.

\section{MRQAP for testing partial regression coefficients}\label{sec:MRQAP}

In this section, we generalize the discussion in Section \ref{sec:QAP} to perform inference on partial regression coefficients. In Section \ref{sec:MRQAP-model}, the asymptotic inference of the partial correlation coefficient is a straightforward application of the result in Section \ref{sec:QAP}. However, the corresponding permutation tests have much richer results. In particular, we have multiple permutation strategies under MRQAP. In Section \ref{sec:MRQAP-permute-b}, we focus on the strategy of permuting the covariate network. In Section \ref{sec:MRQAP-other}, we discuss other permutation strategies.

\subsection{A dyadic linear model and asymptotic inference}\label{sec:MRQAP-model}
We still assume the dyadic linear model in Assumption \ref{asp:noise-lm}, but test only a subset of the coefficients. For easier presentation, we distinguish the covariates of interest from the rest by using the notation $b$ for the former and $c$ for the latter, in contrast to Definition \ref{def:QAP-perm}. The following assumption further clarifies this idea:
\begin{assumption}\label{asp:plm}
Consider i.i.d. unit-specific features $\{(R_i, S_i, T_i)\}_{i=1}^n\sim (R,S,T)$. 
The dyadic data $\{a_{ij}, (b_{kij})_{k\in [p]}, (c_{lij})_{l\in [q]}\}_{i,j\in[n]}$ follows the linear model:
\begin{align}
    a_{ij} = \vartheta_0 + \sum_{k=1}^p \vartheta_k b_{kij} + \sum_{l=1}^q \varrho_l c_{lij} + e_{ij}, 
\end{align}
where the regressors $b_{ij}$'s and $c_{ij}$'s are generated from kernels $\beta^{(k)}$'s and $\gamma^{(l)}$'s, respectively:
\begin{align}
    b_{kij} = \beta^{(k)}(S_{i}, S_{j}), \quad c_{lij} = \gamma^{(l)}(T_{i}, T_{j}),
\end{align}
and the noise term $e_{ij}$'s are formulated with another kernel, $\epsilon$: 
\begin{gather}
    e_{ij} = \epsilon(R_i, R_j) + \zeta_{ij}, \quad 
    \E{\epsilon(R_i, R_j) \mid \{b_{kij}\}_{k=1}^p, \{c_{lij}\}_{l=1}^q} = 0, \\
    \{\zeta_{ij}\}_{i<j\in[n]} \overset{i.i.d.}{\sim} \zeta, \quad \{\zeta_{ij}\}_{i<j\in[n]}\indep \{(R_i, S_i, T_i)\}_{i\in[n]}.
\end{gather}    
\end{assumption}
Let $\vartheta = (\vartheta_1,\dots,\vartheta_p)^\top$ and $\varrho = (\varrho_1,\dots,\varrho_q)^\top$. We are interested in testing the following two hypotheses:
\begin{gather}\label{eqn:null-mrqap-partial}
    \mathrm{H}_{0\textsc{w}}: \vartheta = 0  
    \quad \text{and} \quad 
    \mathrm{H}_{0\textsc{s}}: \vartheta = 0 \text{ with } (R,T)\indep S.
\end{gather}
$\mathrm{H}_{0\textsc{w}}$ is the weak null hypothesis, which tests for a zero partial regression coefficient. $\mathrm{H}_{0\textsc{s}}$ is the strong null hypothesis, which in addition tests for independence between features: $(R, T)\indep S$. The formulation of the strong null hypothesis is related to the permutation strategy in Section \ref{sec:MRQAP-permute-b}. See Sections \ref{sec:MRQAP-permute-b} and \ref{sec:MRQAP-other} for more discussions.

Under Assumption \ref{asp:plm}, the population level least squares gives
\begin{align*}
    w = (\vartheta^\top, \varrho^\top)^\top = \Sigma^{-1}\Sigma_{a},
\end{align*}
where
\begin{align}\label{eqn:mrqap-Sigma}
    \Sigma =
    \begin{pmatrix}
        \Sigma_{bb} & \Sigma_{bc} \\
        \Sigma_{cb} & \Sigma_{cc} 
    \end{pmatrix}, 
    \quad
    \Sigma_a = 
    \begin{pmatrix}
        \Sigma_{ba}\\
        \Sigma_{ca}
    \end{pmatrix},
\end{align}
with $\Sigma_{bb} = (\Cov{b_{k}}{b_{k'}})_{k,k'\in[p]}$, 
    $\Sigma_{cc} = (\Cov{c_{l}}{c_{l'}})_{l,l'\in[p]}$,
    $\Sigma_{bc} = (\Cov{b_{k}}{c_{l}})_{k\in[p],l\in[p]}$, 
and $\Sigma_{ba} = (\Cov{a_{ij}}{b_{kij}})_{k\in[p]}$, 
$\Sigma_{ca} = (\Cov{a_{ij}}{c_{lij}})_{l\in[q]}$. 
The least-squares estimator then equals: 
\begin{align}\label{eqn:hw}
    \hw = 
    \begin{pmatrix}
        \hvartheta \\
        \hvarrho
    \end{pmatrix} =
    \begin{pmatrix}
        \hSigma_{bb} ~ \hSigma_{bc} \\
        \hSigma_{bc}^\top ~ \hSigma_{cc}
    \end{pmatrix}^{-1}
    \begin{pmatrix}
        \hSigma_{ba}\\
        \hSigma_{ca}
    \end{pmatrix} 
    \triangleq 
    \hSigma^{-1} \hSigma_{a}.
\end{align}
For convenience, we introduce the matrix $F = (I_p, 0_{p\times q})^\top\in\bbR^{(p+q)\times p} $ to write $\vartheta = F^\top w$, $\hvartheta = F^\top \hw$. We can use the least-squares coefficient $\hat{\vartheta}$ with the variance estimator $F^\top \hV F $ for inference. We use the Wald statistic
\begin{align}\label{eqn:plm-W}
    W = n 
    \cdot 
    (F^\top \hw)^\top
    (F^\top \hV  F)^{-1}
    (F^\top \hw),
\end{align}
and reject the null hypotheses if $W > q_\alpha$ where $q_\alpha$ is the upper $\alpha$ quantile of the $\chi^2_p$ distribution. The asymptotic distributions of $\hat{\vartheta}$ and $W$ under both hypotheses are fully characterized by the Theorem \ref{thm:asp-super-ns-p} in Section \ref{sec:model-inference-lr}. For conciseness, we relegate the technical details to Section \ref{sec:hatvartheta} of the Supplementary Material.

\subsection{MRQAP with the permute-$b$ strategy}\label{sec:MRQAP-permute-b}




In this section, we discuss the properties of permutation tests for partial regression coefficients. In particular, we will study the ``permute-$b$'' strategy, where we only permute the covariates of interest while keeping the remaining covariates $c$ and outcomes $a$ unchanged. Formally, the permutation test proceeds as follows:
\begin{definition}[MRQAP with the permute-$b$ strategy]\label{def:permute-b}
    Let $W = W(A, \{B_k\}_{k\in[p]}, \{C_k\}_{k\in[q]})$ be the test statistic. 
    \begin{enumerate}
        \item Compute the test statistic based on the observed data, for example, the Wald statistic in \eqref{eqn:plm-W}.

        \item Permute the columns and rows of $\{B_k\}_{k\in[p]}$ in the same way and obtain the permutation distribution with $A$ and the $B_{k,\pi}$'s and $C_{k}$'s:
        \begin{align}
            {\cL}(t; W^\pi) = \frac{1}{n!}\sum_{\pi\in\bbS_n} \ind{W^\pi \le  t}, \text{ where }W^\pi = W(A, \{B_{k, \pi}\}_{k\in[p]}, \{C_k\}_{k\in[q]}).
        \end{align}

        \item Compute the p-value based on the observed $W$ and its permutation distribution. 
    \end{enumerate}
\end{definition}

Theorem \ref{thm:permute-b-W} below presents the asymptotic permutation distribution for the Wald statistic:
\begin{theorem}\label{thm:permute-b-W}
    Assume Assumption \ref{asp:plm} with $\E{\zeta^4} <\infty$. Assume there are univariate functions $\kappa_\epsilon, \kappa_\beta^{(k)}, \kappa_\gamma^{(l)}$ such that 
\begin{align}\label{eqn:kernel-bd-lm}
    |\epsilon(r,r')| \le \kappa_\epsilon(r) + \kappa_\epsilon(r'), 
    ~ 
    |\beta^{(k)}(s,s')| \le \kappa_\beta^{(k)}(s) + \kappa_\beta^{(k)}(s'),
    ~ 
    |\gamma^{(l)}(t,t')| \le \kappa_\gamma^{(l)}(t) + \kappa_\gamma^{(l)}(t'),
\end{align}
where $\E{\kappa_\epsilon(R)^4} < \infty$, $\bbE\{\kappa_\beta^{(k)}(S)^4\} < \infty$ and $\bbE\{\kappa_\gamma^{(l)}(T)^4\} < \infty$.
Also, assume non-degeneracy of the kernels $\epsilon$ and $\beta^{(k)}$'s in the sense that 
$\eta_{1,\epsilon} > 0$ and $H_{1,(\beta,\gamma)} = (\eta_{1, (\beta, \gamma)}^{(kl)})_{k,l\in[p]}$ is positive definite. The Wald statistic in \eqref{eqn:plm-W} has the following asymptotic distribution under the permute-$b$ strategy in Definition \ref{def:permute-b}:
    \begin{align}
        \lim_{n\to\infty}\sup_{t\in\bbR}|\cL(t; W^\pi) - \cL(t;\chi^2_p)| = 0 \quad \bbP\text{-a.s.}
    \end{align}
\end{theorem}

Theorem \ref{thm:permute-b-W} shows that asymptotically, the permutation distribution matches the sampling distribution with the Wald test statistic under both the weak and strong null hypotheses in \eqref{eqn:null-mrqap-partial}. It ensures that MRQAP is asymptotically valid for testing the weak null hypothesis, in addition to its finite-sample exactness under the strong null hypothesis. However, non-Wald test statistics may not be valid, as confirmed by the simulation results in Section \ref{sec:simulation-mrqap} as well as our theoretical discussion in Section \ref{sec:more-mrqap} in the supplementary material. Moreover, the permute-$b$ strategy guarantees the finite-sample exactness of the studentized permutation test under $\mathrm{H}_{0\textsc{s}}$ in \eqref{eqn:null-mrqap-partial}. This property coincides with the conclusion in linear regression models \citep{diciccio2017robust} and Fisher randomization test in completely randomized experiments \citep{wu2021randomization, zhao2021covariate}. 

\subsection{MRQAP with other permutation strategies}\label{sec:MRQAP-other}
In Section \ref{sec:compare-permutation} of the Supplementary Material, we also discuss other permutation strategies including permuting outcomes, permuting residuals as a generalization of the Freedman--Lane approach \citep{freedman1983nonstochastic, dekker2007sensitivity}, or permuting residual of covariates as proposed by \citet{dekker2007sensitivity}. More concretely, we compare four permutation strategies: (i) Permuting $a$: permute the $(a_{ij})$ matrix; (ii) Permuting $b$: permute the $(b_{kij})$ matrices, as introduced in Definition \ref{def:permute-b}; (iii) Permuting $\hepsilon_b$: permute the residuals from regressing each of the $(b_{kij})$'s on $1$ and the $(c_{lij})$'s, as proposed by \cite{dekker2007sensitivity}; (iv) Permuting $\hepsilon$: permute the residuals from regressing the outcomes $(a_{ij})$ on $1$ and the $(c_{lij})$'s, which extends the Freedman--Lane approach under the classic least-squares setting \citep{freedman1983nonstochastic, dekker2007sensitivity}. 

Our theoretical analysis characterizes the performance of different permutation strategies. Table \ref{tab:mrqap-partial} summarizes the results and delivers some important messages. First, studentization ensures asymptotic validity for all permutation strategies under both the weak and strong null hypotheses in \eqref{eqn:null-mrqap-partial}. In contrast, without studentization, none of the strategies achieve asymptotic type I error control under $\mathrm{H}_{0\textsc{w}}$. This echoes the empirical results in previous literature such as \cite{dekker2007sensitivity} as well as the theoretical discussions in \cite{diciccio2017robust} and \cite{zhao2021covariate} for simpler settings without network structure.  Second, the permute-$b$ strategy is the only one that achieves finite-sample exactness under $\mathrm{H}_{0\textsc{s}}:(R, T)\indep S$, which is a stronger guarantee than the asymptotic error rate control. It is particularly natural when the $B_k$'s are randomized with $S\indep T$ so that $\mathrm{H}_{0\textsc{s}}: (R, T)\indep S$ is equivalent to $R\indep S \mid T$ \citep{zhao2021covariate}. Therefore, the null hypothesis $\mathrm{H}_{0\textsc{s}}: (R, T)\indep S$ is reminiscence of the conditional independence hypothesis $A \indep B \mid C$ in classic linear regression \citep{diciccio2017robust}. In a nutshell, in practice, we recommend permuting $b$ with the Wald statistic, which guarantees asymptotic validity under $\mathrm{H}_{0\textsc{w}}$ and finite sample exactness under $\mathrm{H}_{0\textsc{s}}$.


\begin{table}[ht!]
\caption{Comparison of permutation strategies for testing partial coefficients with MRQAP }
\label{tab:mrqap-partial}
\begin{tabular}{cccccccccc}
\toprule
                                                  &                       & \multicolumn{4}{c}{Without Studentization} & \multicolumn{4}{c}{With Studentization} \\ \hline
\multicolumn{2}{c}{Permuting Strategy}                                    & $a$  & $b$  & $\hepsilon_b$  & $\hepsilon$ & $a$ & $b$ & $\hepsilon_b$ & $\hepsilon$ \\ \hline
\multirow{2}{*}{Under $\mathrm{H}_{0\textsc{w}}$} & Finite Sample Exact?  & NO   & NO   & NO             & NO          & NO  & NO  & NO            & NO          \\ \cline{2-10} 
                                                  & Asymptotically Valid? & NO   & NO   & NO             & NO          & YES & \textbf{YES} & YES           & YES         \\ \hline
\multirow{2}{*}{Under $\mathrm{H}_{0\textsc{s}}$} & Finite Sample Exact?  & NO   & YES  & NO             & NO          & NO$^\star$  & \textbf{YES} & NO            & NO          \\ \cline{2-10} 
                                                  & Asymptotically Valid? & NO   & YES  & YES            & YES         & YES & \textbf{YES} & YES           & YES         \\ \bottomrule
\end{tabular}

\begin{minipage}{\textwidth}

\vspace{0.1cm}

\footnotesize  $\star$ Although permuting outcomes is not finite-sample exact for $\mathrm{H}_{0\textsc{s}}$ in \eqref{eqn:null-mrqap-partial}, it is finite-sample exact for another strong null hypothesis, $\mathrm{H}_{0\textsc{s}}: \vartheta = 0, ~\varrho = 0, ~R\indep (S, T)$. 

\end{minipage}

\end{table}


\section{Simulation}\label{sec:simulation}
\subsection{QAP for the weak null hypothesis}
In this subsection, we investigate the validity of using QAP to test the weak null hypothesis in \eqref{eqn:null-qap} based on Example \ref{exp:walsh-average}. We consider two ways for generating $(R,S)$: with non-studentized $\widehat{\rho}$, the first one leads to a conservative QAP, while the second one leads to an anti-conservative QAP. We will show that using a studentized $\widehat{\rho}$ fixes the problems in both scenarios.

\begin{figure}[ht!]
\centering
\subfigure[]{\includegraphics[width=0.42\textwidth]{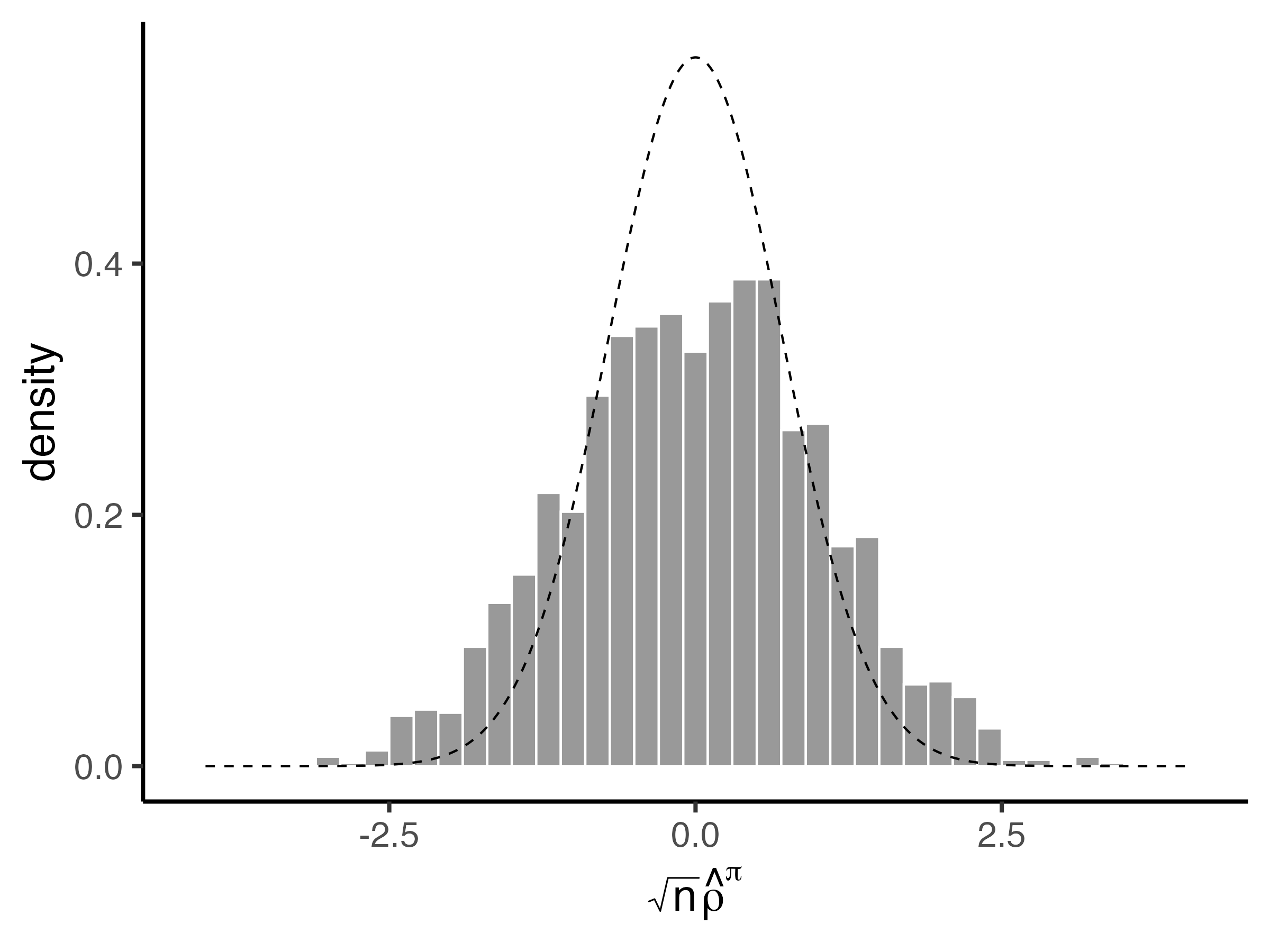}}\quad
\subfigure[]{\includegraphics[width=0.42\textwidth]{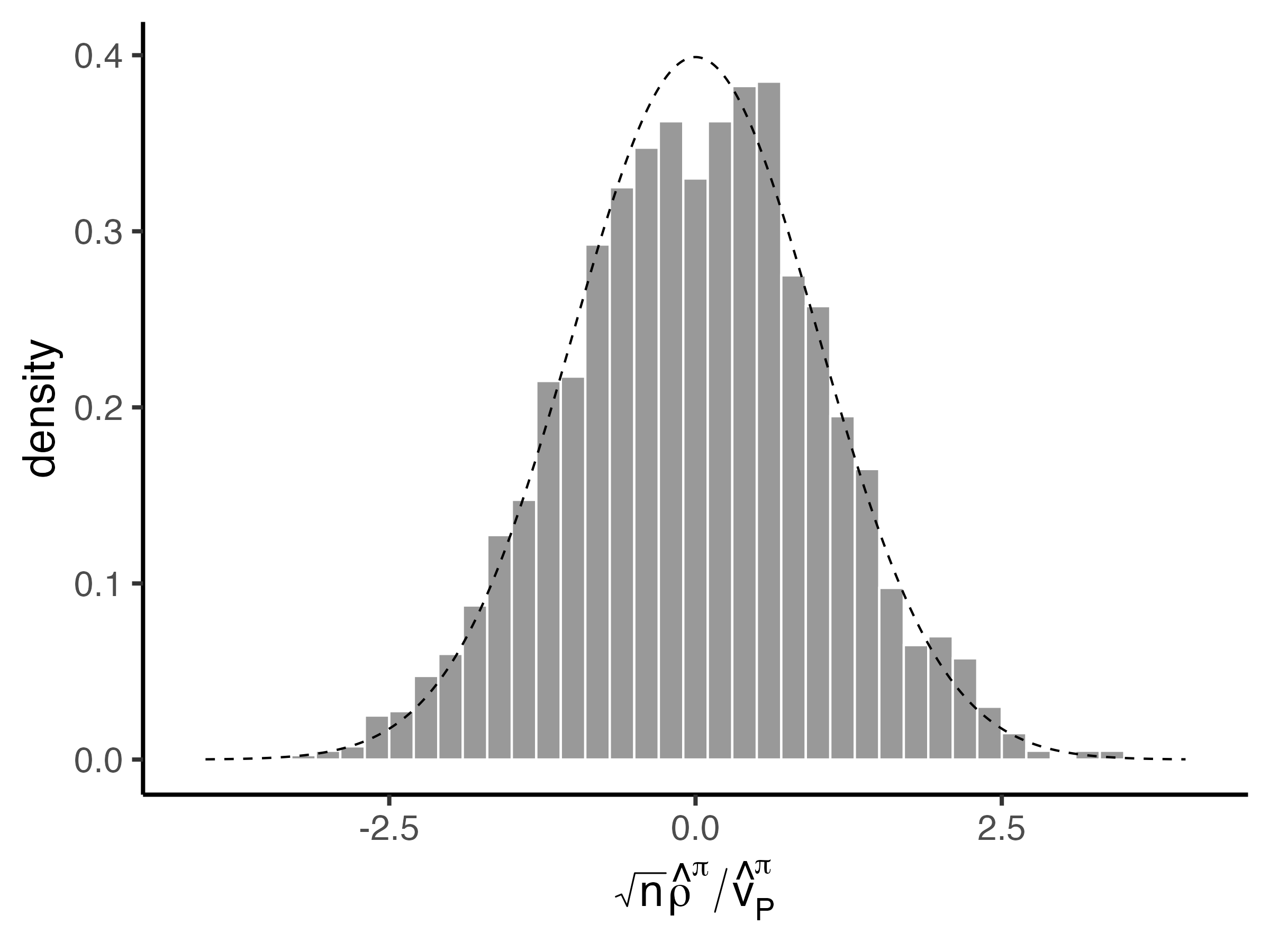}}
\subfigure[]{\includegraphics[width=0.42\textwidth]{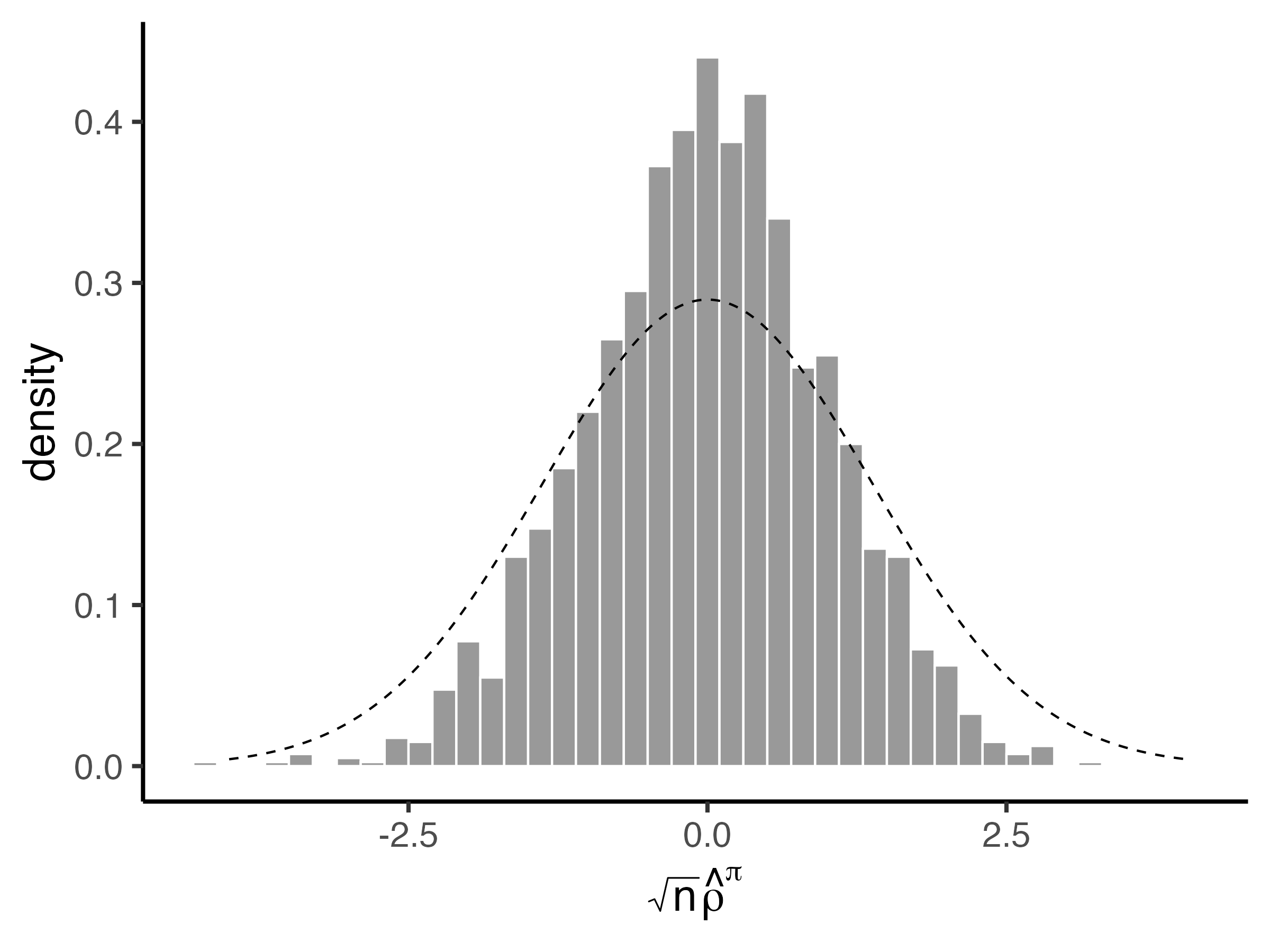}}\quad
\subfigure[]{\includegraphics[width=0.42\textwidth]{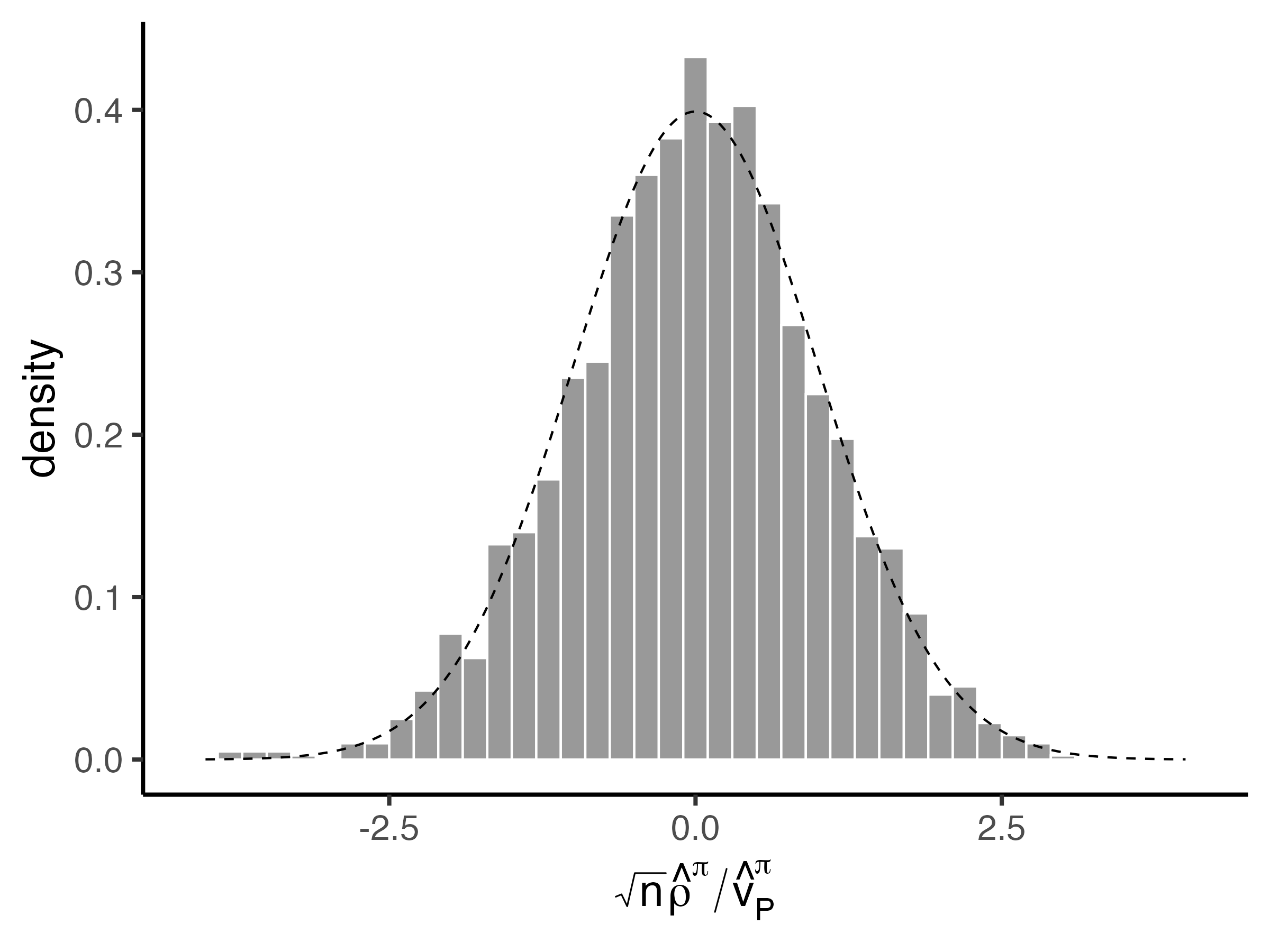}}
\caption{Studentization can fix permutation tests for testing $\mathrm{H}_{0\textsc{w}}$ in \eqref{eqn:null-qap}. The histograms are permutation distributions and the dashed lines are asymptotic sampling distributions. Panel (a) and (b) compare permutation tests with non-studentized and studentized $\widehat{\rho}$ under Setting 1, while Panel (c) and (d) make the comparison under Setting 2.  }
\label{fig:test-weak-null}
\end{figure}

\noindent\textbf{Setting 1. Fix conservative QAP by studentization.}
First generate $U\sim\mathrm{Unif}([0,2\pi])$, and then set $(R,S) = (\sqrt{2}\sin(U), \sqrt{2}\cos(U))$. $R$ and $S$ are dependent because $R^2 + S^2 = 2$, but they have zero correlation $\rho_{R,S} = 0$, which gives $\rho^\star = \rho_{R,S} = 0$. 
By Example \ref{exp:walsh-average}, we can compute $4\sigma_1^2 = \frac{1}{2}$. 
Hence we have $\sqrt{n}\hrho \rightsquigarrow
\cN(0, \frac{1}{2})$. Figure \ref{fig:test-weak-null}(a) and \ref{fig:test-weak-null}(b) show the results.  

\noindent\textbf{Setting 2. Fix anti-conservative QAP by studentization.} 
First generate $U\sim\mathrm{Unif}([-2\pi,2\pi])$, and then set 
$$
(R,S) = \lt(\frac{\sinh(3U)}{\sqrt{\sinh(12\pi)/(24\pi)-1/2}}, \sqrt{2}\cos(U)\rt).
$$
We can check that $R$ and $S$ have zero mean, unit variance, and zero correlation, which gives $\rho^\star = \rho_{R,S} = 0$. 
By Example \ref{exp:walsh-average}, we can compute
$4\sigma_1^2 \approx 1.90$. 
Hence we have $\sqrt{n}\hrho \rightsquigarrow \cN(0, 4\sigma_1^2)$. 
Figure \ref{fig:test-weak-null}(c) and \ref{fig:test-weak-null}(d) show the results. 

From Figure \ref{fig:test-weak-null}, we can see that when using the non-studentized $\widehat{\rho}$, the permutation distributions fail to match those under the weak null hypothesis. On the contrary, when using the studentized $\widehat{\rho}$, the permutation distributions match the sampling distributions well under the weak null hypothesis.  

\subsection{MRQAP for the weak null hypothesis}\label{sec:simulation-mrqap}
In this section, we conduct the simulation for using MRQAP to test the partial regression coefficients. The goal is to validate the theoretical results in Section \ref{sec:MRQAP} by comparing the asymptotic sampling distributions and permutation distributions with or without studentization. 

\begin{figure}[ht!]
\centering
\begin{subfigure}
  \centering
\includegraphics[width=0.48\textwidth]{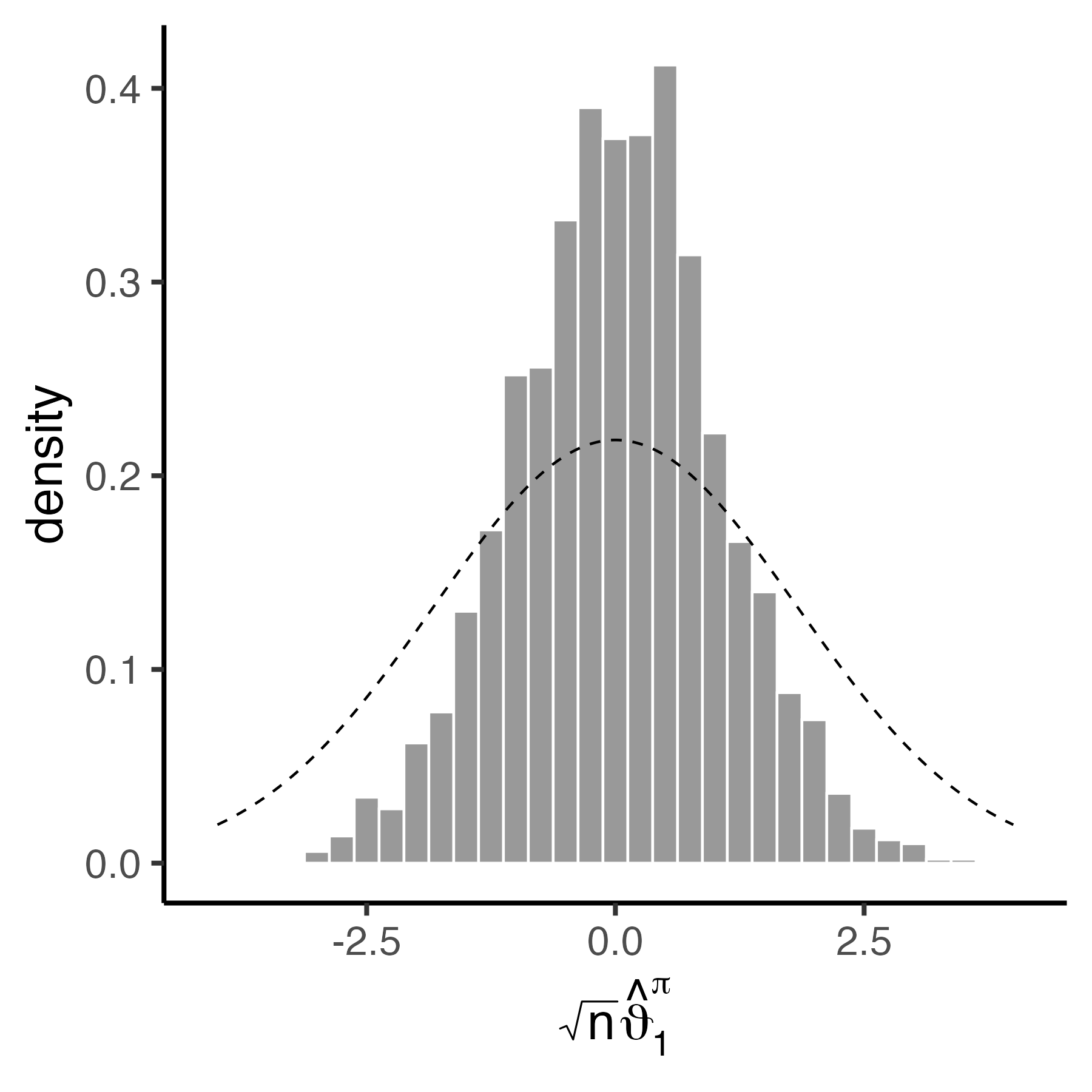}
\end{subfigure}%
\begin{subfigure}
  \centering \includegraphics[width=0.49\textwidth]{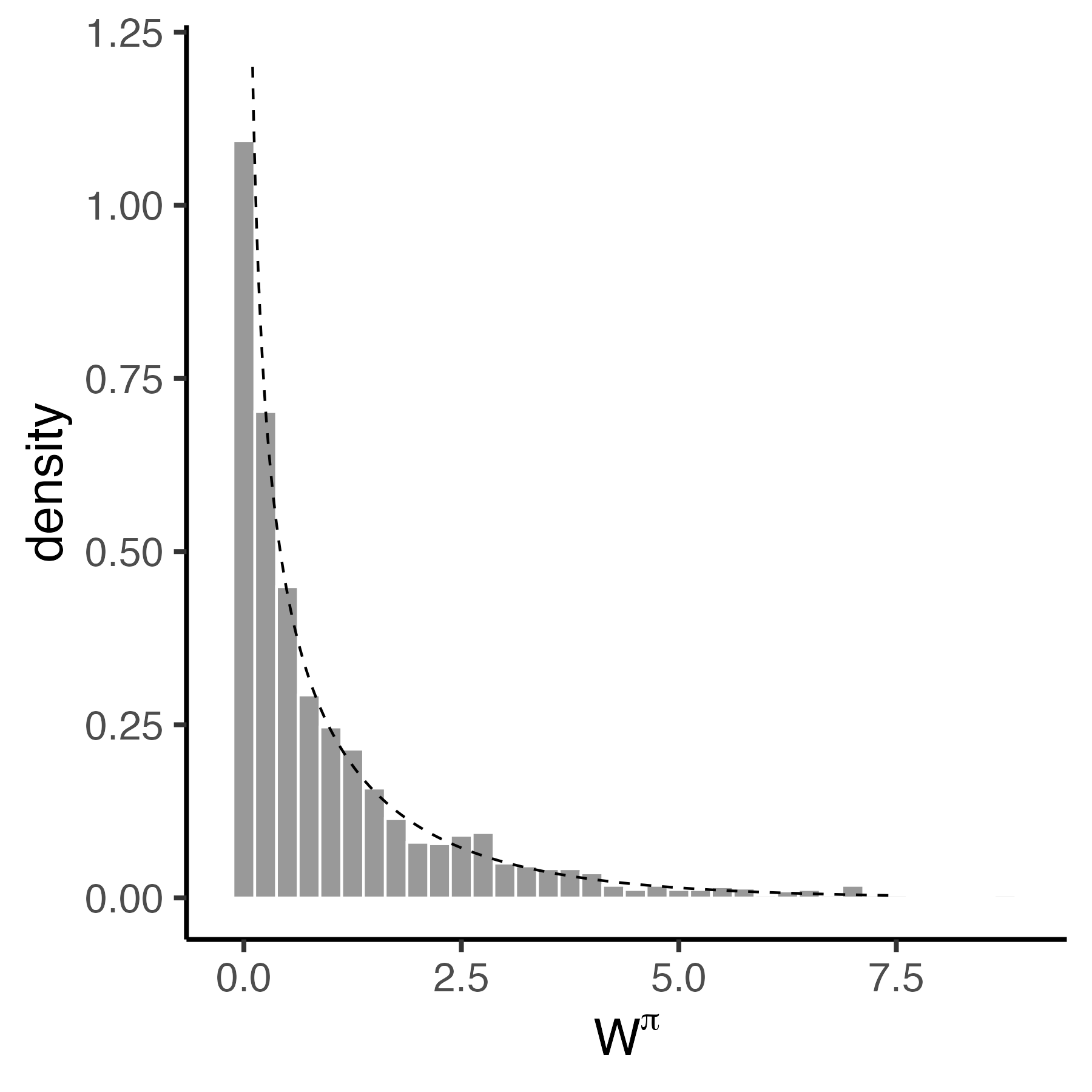}
\end{subfigure}
\caption{Comparing the permutation distribution and sampling distribution with permuting-$b$ strategy under the weak null hypothesis \eqref{eqn:null-mrqap-partial}. The histograms are permutation distributions and the dashed lines are asymptotic approximations. (i) Left: asymptotic sampling and permutation distributions of the non-studentized statistic $\sqrt{n}\hvartheta_1$; (ii) Right: asymptotic sampling and permutation distributions of the Wald statistic $W$. }
\label{fig:test-weak-null-MRQAP}
\end{figure}

We simulate two correlated unit-specific features $T$ and $S$, from a bivariate normal distribution with mean zero, variance one, and correlation coefficient $0.5$.
The noise feature, $R$ is generated by $R = T \cdot Z$,
where $Z$ is a standard normal variable that is independent of $(T, S)$. The kernels $\beta$, $\gamma$, and $\epsilon$ are generated as pairwise averages of $S$, $T$ and $R$, respectively, as in the first case of Example \ref{exp:walsh-average}.
Generate the outcome network $A=(a_{ij})$ from
\begin{align}\label{eqn:linear-setup}
    a_{ij} = 
    \vartheta_0 + 
    \vartheta_1 \cdot b_{ij} 
    + 
    \varrho\cdot c_{ij}
    +
    e_{ij}. 
\end{align}
We can check that Assumption \ref{asp:plm} holds. Set the coefficients in \eqref{eqn:linear-setup} as
$\vartheta_0 = \vartheta_1 = 0$, $ \varrho = 1
$, so that the weak null hypothesis in \eqref{eqn:null-mrqap-partial} holds. The results are summarized in Figure \ref{fig:test-weak-null-MRQAP}, which shows a good match between the sampling distribution and permutation distribution with studentization and a mismatch without studentization. 

\section{Case study based on MRQAP}\label{sec:case}

In this section, we revisit the analysis conducted by \cite{elmer2020depressive}, which studies how depressive symptoms are associated with social isolation in face-to-face interaction networks. Their study involves a group of students spending a weekend together in a remote camp house, during which the time of how long two individuals interact with each other is tracked by a special device. This social interaction dyadic matrix $A = (a_{ij})$ serves as the dependent variable. The predictors of interest are pairwise depressive means and depression similarity, which are dyadic data defined by a depression score based on the German version of the Center for Epidemiologic Studies Depression Scale. Meanwhile, several covariates are adjusted, including pairwise age mean, age similarity, gender difference, and friendship ties. 

The original study consisted of two separate dyadic datasets. We use one dataset with $n = 73$ students and $n(n-1)/2 = 2628$ dyadic observations to illustrate the proposed methods. We apply MRQAP in Definition \ref{def:permute-b} with least-squares coefficients studentized by the cluster-robust standard error multiplied by $2$ (recalling Section \ref{sec:cluster-var-est}). When $b_{ij}$ is a one-dimensional measure, this is equivalent to the Wald statistic. Table \ref{tab:case} reports the coefficients and $p$-values.
From Table \ref{tab:case}, a higher level of depression leads to less interaction between two individuals. With the studentized regression coefficient as the test statistic, it is significant by both the $p$-value based on the asymptotic approximation and the $p$-value based on MRQAP. In contrast, the $p$-value based on MRQAP without studentization is not significant. Our theory does not recommend using it because it is not robust for testing the weak null hypothesis. This example highlights the importance of studentization in MRQAP. We have generated the permutation distributions of the studentized least-squares coefficients. They are close to the standard normal distribution, which provides evidence about the validity of normal approximation. To save space, we omit them in the main paper but provide code to generate them in the replication package. We do not find a significant association between social interaction and depression similarity based on all $p$-values. We also include the interaction term between the depression mean level and depression similarity, as well as between depression level and friendship, as suggested by \cite{elmer2020depressive}. They are not significant and thus omitted in the main paper. See our replication package for more details. 
\begin{table}[ht!]
\centering
\caption{MVQAP to test the relation between social interaction and depressive symptoms}
\label{tab:case}
\begin{tabular}{P{3.8cm}P{1.5cm}P{1.5cm}P{2.7cm}P{2.5cm}}
\toprule
    & \textbf{\footnotesize Coefficient} & \textbf{\footnotesize p-value (normal)} & \textbf{\footnotesize QAP p-value (non-studentized)} & \textbf{\footnotesize QAP p-value (studentized)} \\ \midrule
\textbf{\footnotesize Depression mean}  & -0.0671               & 0.0049$^*$       & 0.1445                         & 0.0475$^*$                     \\
\textbf{\footnotesize Depression similarity}  & 0.0693                & 0.0189       & 0.2695                         & 0.1695                     \\ \bottomrule
\end{tabular}
\end{table}


\section{Discussion}\label{sec:discussion}

QAP and MRQAP are popular tools for analyzing dyadic data. We provide a rigorous formulation for a hypothesis testing problem in dyadic network applications and establish the asymptotic theory for QAP and MRQAP. Our asymptotic theory suggests that properly studentized statistics in QAP and MRQAP can guarantee finite-sample exactness under the strong null hypothesis and control the asymptotic type one error rate under the weak null hypothesis. From a technical side, we leverage the theory of U-statistics to derive the asymptotic sampling distributions and the theory of double-indexed permutation statistics to derive the asymptotic permutation distributions, respectively. We also derive some new results on permutation-related statistics, which are of independent interest to other problems involving permutations.

There are several remaining issues to explore. First, our asymptotic analysis relies on non-degeneracy conditions (such as $\eta_{1,\phi} > 0$ in Theorem \ref{thm:asp-super}) on the kernels. It is of interest to generalize the results to incorporate degenerate kernels to accommodate more applications. This extension is potentially useful for many problems and is nontrivial as it relies on more delicate limiting theorems for degenerate U-statistics and double-indexed permutation statistics \citep[e.g.][]{shi2022distribution}. Second, many applications involve multiple separate dyadic networks, which motivate a multi-group extension of MRQAP \citep{burnett2015relationship, elmer2020depressive}. In particular, the double permutation should be conducted independently across the separate networks. It is interesting to discuss the theoretical foundation for such an extension. Third, it is a natural question to extend the multiple linear regression model on dyadic data to generalized linear models \citep{graham2020dyadic}. Fourth, many real-world applications involve large dyadic networks, which needs computationally efficient algorithms for estimation and inference. We leave these as future research. 

\bibliographystyle{apalike}
\bibliography{ref}

\newpage 

\renewcommand{\thesection}{\Alph{section}}
\renewcommand{\thetheorem}{\Alph{section}.\arabic{theorem}}
\setcounter{theorem}{0}
\renewcommand{\thelemma}{\Alph{section}.\arabic{lemma}}
\setcounter{lemma}{0}
\renewcommand{\theproposition}{\Alph{section}.\arabic{proposition}}
\setcounter{proposition}{0}
\renewcommand{\thecorollary}{\Alph{section}.\arabic{corollary}}
\setcounter{corollary}{0}
\renewcommand{\thedefinition}{\Alph{section}.\arabic{definition}}
\setcounter{definition}{0}
\renewcommand{\thepage}{S\arabic{page}}
\setcounter{page}{1}
\renewcommand{\theequation}{\Alph{section}.\arabic{equation}}
\setcounter{equation}{0}
\renewcommand{\thetable}{S\arabic{table}}
\setcounter{table}{0}

\begin{center}
\Huge 
Supplementary material
\end{center}

\appendix

In the supplementary material, we provide additional technical details. For ease of reference, we provide a table of standard acronyms:
\begin{table}[ht!]
    \centering
    \caption{Summary of acronyms}
    \label{tab:acronym}
    \begin{tabular}{cc}
    \toprule
    {Acronym} & {Full Name}\\
    \midrule
        CLT & central limit theorem\\
        DIPS & double-indexed permutation statistic\\
        MRQAP & multiple regression quadratic assignment procedure \\
        OLS & ordinary least squares\\
        QAP & quadratic assignment procedure \\
        SIPS & single-indexed permutation statistic \\
        SLLN & strong law of large numbers\\
        WLLN & weak law of large numbers\\
        \bottomrule
    \end{tabular}
\end{table}

The supplementary material is organized as follows. Section \ref{sec:more-mrqap} presents more results for MRQAP. Section \ref{sec:U} reviews probabilistic results for one- and multi-dimensional U-statistics of degree $2$, which are useful for the analysis of the sampling distributions. Section \ref{sec:DIPS} reviews the technical results of DIPS, which are useful for the analysis of the permutation distributions. Section \ref{sec:pf-main} presents the proofs.


\section{More results on MRQAP}\label{sec:more-mrqap}

In this section, we present more results on MRQAP. Section \ref{sec:hatvartheta} presents the technical details for the asymptotic distribution of $\hat{\vartheta}$ in Section \ref{sec:MRQAP-model} of the main paper. Section \ref{sec:compare-permutation} compares different permutation strategies for MRQAP, which are briefly mentioned in Section \ref{sec:MRQAP-other} of the main paper. Section \ref{sec:additional-results} provides additional technical results that support some statements in the main paper. 

\subsection{Asymptotic distribution of $\hat{\vartheta}$}\label{sec:hatvartheta}

Introduce the kernel vector $\phi = (\phi_b,\phi_c)$ where $\phi_b = (\phi_b^{(1:p)})$ and $\phi_c = (\phi_c^{(1:q)})$ with
\begin{align*}
    \phi^{(k)}_b(r,s,t;r',s',t') = \epsilon(r,r')\tbeta^{(k)}(s,s'), 
    \quad 
    \phi^{(l)}_c(r,s,t;r',s',t') = \epsilon(r,r')\tgamma^{(l)}(t,t'). 
\end{align*}
Following the discussion in Section \ref{sec:QAP}, let $H_{1,\phi}$ be the covariance matrix of the first-order projected kernels of $\phi$. 
Based on Table \ref{tab:notation-phi}, recall the definitions for the centered kernels:
\begin{align}
    \tphi_b^{(k)}(r,s,t;r',s',t') = \phi_b^{(k)}(r,s,t;r',s',t') - \phi_{b0}^{(k)}, 
    \quad 
    \phi_{b0}^{(k)} = \E{\phi_b^{(k)}(R,S,T;R',S',T')}
\end{align}
and 
\begin{align}
    \tphi_c^{(k)}(r,s,t;r',s',t') = \phi_c^{(k)}(r,s,t;r',s',t') - \phi^{(k)}_{c0}, 
    \quad 
    \phi_{c0}^{(k)} = \E{\phi_c^{(k)}(R,S,T;R',S',T')}.
\end{align}
Also, recall the first-order projection kernels
    \begin{align}
        \tphi^{(l)}_{b1}(r,s,t) = \E{\tphi_b^{(l)}(r,s,t;R',S',T')}, \quad 
        \tphi^{(l)}_{c1}(r,s,t) = \E{\tphi_c^{(l)}(r,s,t;R',S',T')}.
    \end{align}
Define the covariance matrix of the first-order projection kernels:
\begin{align*}
    H_{1,\phi} 
    =
    \begin{pmatrix}
        H_{1,\phi_b} & H_{1,\phi_b, \phi_c}\\
        H_{1,\phi_b, \phi_c} & H_{1,\phi_c}
    \end{pmatrix},
\end{align*}
where 
\begin{gather*}
    H_{1,\phi_b} = \lt( \eta^{(kk')}_{1,\phi_b} \rt)_{k,k'\in[p]},\quad \eta^{(kk')}_{1,\phi_b} = \E{\tphi^{(k)}_{b1}(R,S,T) \tphi^{(k')}_{b1}(R,S,T)},\\
    H_{1,\phi_c} = \lt( \eta^{(ll')}_{1,\phi_c} \rt)_{l,l'\in[p]},\quad \eta^{(ll')}_{1,\phi_c} = \E{\tphi^{(l)}_{c1}(R,S,T) \tphi^{(l')}_{c1}(R,S,T)},\\
    H_{1,\phi_b, \phi_c} = \lt( \eta^{(kl)}_{1,\phi_b, \phi_c} \rt)_{k\in[p],l\in[q]},\quad \eta^{(kl)}_{1,\phi_b} = \E{\tphi^{(k)}_{b1}(R,S,T) \tphi^{(l)}_{c1}(R,S,T)}.
\end{gather*}
The following Theorem \ref{thm:MR-super} gives the asymptotic sampling distribution of ${\hvartheta}$. This is a direct application of Theorem \ref{thm:asp-super-ns-p}. The only modification is to use a subset of the OLS coefficient and a submatrix of the covariance estimator. 

Recall the definition of $F = (I_p, 0_{p\times q})^\top \in \bbR^{(p+q)\times p}$. Also, recall the definition of the covariance matrix $\Sigma$ in \eqref{eqn:mrqap-Sigma}. 
\begin{theorem}\label{thm:MR-super}
Assume Assumption \ref{asp:plm}. Assume
\begin{align}
    \E{\zeta^2}<\infty, \quad \E{\beta^{(k)}(S, S')^2}<\infty, \quad  \E{\gamma^{(l)}(T, T')^2}<\infty, \quad \E{\epsilon(R, R')^2}<\infty.
\end{align}
Assume the kernels $\phi_b^{(k)}$ and $\phi_c^{(l)}$ satisfy:
\begin{align*}
    \eta_{1,\phi_b}^{(kk)} > 0 , ~ \E{\phi_b^{(k)}(R,S,T;R',S',T')^2} < \infty;\quad 
    \eta_{1,\phi_c}^{(ll)} > 0 , ~ \E{\phi_c^{(l)}(R,S,T;R',S',T')^2} < \infty.
\end{align*} 
\begin{enumerate}
    \item Under $\mathrm{H}_{0\textsc{w}}: \vartheta = 0$ in \eqref{eqn:null-mrqap-partial}, we have
    \begin{align}
       \sqrt{n} \hvartheta = \sqrt{n} F^\top \hw \rightsquigarrow \cN(0, 4 F^\top \Sigma^{-1}H_{1,\phi}\Sigma^{-1} F). 
    \end{align}

    \item Under $\mathrm{H}_{0\textsc{s}}: (R,T)\indep S$ in \eqref{eqn:null-mrqap-partial}, we have
   \begin{align}
     \sqrt{n} \hvartheta = \sqrt{n} F^\top \hw \rightsquigarrow \cN(0, 4 \eta_{1,\epsilon} \Sigma_{bb}^{-1} H_{1,\beta} \Sigma_{bb}^{-1}).
   \end{align}
\end{enumerate}
\end{theorem}

The variance estimator for $\hvartheta$ is given by $F^\top \hV F$, with:
\begin{align}\label{eqn:hV-supp}
    \hV = 4 \hSigma^{-1} \hH_{1,\phi} \hSigma^{-1}.
\end{align}
Here, we have the covariance estimator
\begin{align}
\hSigma = 
   \begin{pmatrix}
       \hSigma_{bb} ~ \hSigma_{bc} \\
       \hSigma_{bc}^\top ~ \hSigma_{cc}
   \end{pmatrix} 
\end{align}
as defined in \eqref{eqn:hw}. The covariance of the first-order projection kernel can be estimated by
\begin{align}
\hH_{1,\phi} = \frac{1}{n} \sum_{i=1}^n \hat{\phi}_{1,i}\hat{\phi}_{1,i}^\top,
\end{align}
with the estimated projected kernels
\begin{align}
    \hat{\phi}_{1,i} = 
    \frac{1}{n-1}\sum_{j: j\neq i} \hat{e}_{ij} \begin{pmatrix}
        \bsb_{ij} - \overline{\bsb} \\
        \bc_{ij} - \overline{\bc}
    \end{pmatrix},
\end{align}
where $\widehat{e}_{ij}$ is the residual from the OLS fit: 
\begin{align}
    \hat{e}_{ij} = y_{ij} - \oy 
    - 
    (\bsb_{ij} - \overline{\bsb})^\top \hvartheta
    -
    (\bc_{ij} - \overline{\bc})^\top \hat{\varrho}
    . 
\end{align}
Theorem \ref{thm:var-plm} below gives the results on variance estimation and the Wald statistic $W$ in \eqref{eqn:plm-W}. 

\begin{theorem}\label{thm:var-plm}
Under the same conditions as Theorem \ref{thm:MR-super}, we have $\hV \rightarrow 4\Sigma^{-1}H_{1,\phi}\Sigma^{-1}$, $\bbP\text{-a.s.}$
Therefore, the Wald statistic \eqref{eqn:plm-W} follows a $\chi^2_p$ distribution asymptotically: $W \rightsquigarrow \chi^2_p$. 

\end{theorem}
Following the discussion in Section \ref{sec:cluster-var-est}, the variance estimator in Theorem \ref{thm:var-plm} also has a regression implementation via the cluster-robust variance estimation.

\subsection{Other permutation strategies for MRQAP}\label{sec:compare-permutation}
This section provides a discussion of the properties of MRQAP, with different permutation strategies reviewed in Section \ref{sec:MRQAP-other} in the main paper. It also serves as a theoretical justification of the simulation studies in \cite{dekker2007sensitivity}. Consider two statistics. The first one is a smooth function of the partial regression coefficients, such as the squared Euclidean norm:
\begin{align}\label{eqn:T}
    L_n = f(\sqrt{n}\hvartheta) = f(\sqrt{n} F^\top \hw). 
\end{align}
The other one is the Wald statistic: 
\begin{align}
    W = 
    n 
    \cdot 
    \{F^\top \hw\}^\top
    \{F^\top \hV F\}^{-1}
    \{F^\top \hw\}.
\end{align}
We compare four permutation strategies:
\begin{enumerate}
        \item Permuting outcome: permute the $(a_{ij})$ matrix; 
        \item Permuting covariates: permute the $(b_{kij})$ matrices, as introduced in Definition \ref{def:permute-b};
        \item Permuting $\hepsilon_b$: permute the residuals from regressing each of the $(b_{kij})$'s on $1$ and the $(c_{lij})$'s, as proposed by \cite{dekker2007sensitivity};
        \item Permuting $\hepsilon$: permute the residuals from regressing the outcomes $(a_{ij})$ on $1$ and the $(c_{lij})$'s, which extends the Freedman--Lane approach under the classic OLS setting \citep{freedman1983nonstochastic, dekker2007sensitivity}.
\end{enumerate}

\begin{theorem}\label{thm:mrqap-p-ns}
   Assume Assumption \ref{asp:plm} with $\E{\zeta^4} <\infty$. Assume there are univariate functions $\kappa_\epsilon, \kappa_\beta^{(k)}, \kappa_\gamma^{(l)}$ such that 
\begin{align}
    |\epsilon(r,r')| \le \kappa_\epsilon(r) + \kappa_\epsilon(r'), 
    ~ 
    |\beta^{(k)}(s,s')| \le \kappa_\beta^{(k)}(s) + \kappa_\beta^{(k)}(s'),
    ~ 
    |\gamma^{(l)}(t,t')| \le \kappa_\gamma^{(l)}(t) + \kappa_\gamma^{(l)}(t'),
\end{align}
where $\E{\kappa_\epsilon(R)^4} < \infty$, $\bbE\{\kappa_\beta^{(k)}(S)^4\} < \infty$ and $\bbE\{\kappa_\gamma^{(l)}(T)^4\} < \infty$.
Also, assume non-degeneracy of the kernels $\epsilon$ and $\beta^{(k)}$'s in the sense that 
$\eta_{1,\epsilon} > 0$ and $H_{1,(\beta,\gamma)} = (\eta_{1, (\beta, \gamma)}^{(kl)})_{k,l\in[p]}$ is positive definite.
    The partial regression coefficient has the following asymptotic distributions with the four permutation strategies:
    \begin{enumerate}
        \item Permuting outcomes: $\lim_{n\to\infty}\sup_{t\in\bbR} |\cL(t;\sqrt{n}\hvartheta^{\pi}) - \cL(t;\cN(0, F^\top V_\alpha F))| = 0 ~ \bbP\text{-a.s.}$, where $V_\alpha = 4\eta_{1,\alpha} \Sigma^{-1} H_{1,(\beta,\gamma)} \Sigma^{-1}$, 
        and $\alpha$ is the kernel
        \begin{align}
            \alpha(r,t;r',t') 
            = \sum_{l=1}^q \varrho_l \gamma^{(l)}(t,t') + \epsilon(r,r').
        \end{align}
        It matches the asymptotic sampling distribution under $\mathrm{H}_{0\textsc{s}}$ in \eqref{eqn:null-mrqap-partial} if $\varrho = 0$. 
        
        \item Permuting $b$: $\lim_{n\to\infty}\sup_{t\in\bbR} |\cL(t;\sqrt{n}\hvartheta^{\pi}) - \cL(t;\cN(0, V_b))| = 0 ~ \bbP\text{-a.s.}$, where $V_b = 4\eta_{1,\epsilon} \Sigma_{bb}^{-1} H_{1,\beta} \Sigma_{bb}^{-1}$. 
        It matches the asymptotic sampling distribution under $\mathrm{H}_{0\textsc{s}}$ in \eqref{eqn:null-mrqap-partial}. 

        \item Permuting $\hepsilon_b$: $\lim_{n\to\infty}\sup_{t\in\bbR} |\cL(t;\sqrt{n}\hvartheta^{\pi}) - \cL(t;\cN(0, F^\top V_{eb} F))| = 0 ~ \bbP\text{-a.s.}$, where $V_{eb} = 4\eta_{1,\epsilon} \Sigma^{-1} H_{1,(\beta,\gamma)} \Sigma^{-1}$. 
        It matches the asymptotic sampling distribution under $\mathrm{H}_{0\textsc{s}}$ in \eqref{eqn:null-mrqap-partial}. 

        \item Permuting $\hepsilon$: $\lim_{n\to\infty}\sup_{t\in\bbR} |\cL(t;\sqrt{n}\hvartheta^{\pi}) - \cL(t;\cN(0, F^\top V_{e} F))| = 0 ~ \bbP\text{-a.s.}$, where $V_{e} = 4\eta_{1,\epsilon} \Sigma^{-1} H_{1,(\beta,\gamma)} \Sigma^{-1}$. It matches the asymptotic sampling distribution under $\mathrm{H}_{0\textsc{s}}$ in \eqref{eqn:null-mrqap-partial}. 
    \end{enumerate}
    
\end{theorem}

Based on Theorem \ref{thm:mrqap-p-ns} and the continuous mapping theorem, the asymptotic distribution of $L_n$ defined in \eqref{eqn:T} is given by applying $f$ to the asymptotic permutation distribution under each strategy. From Theorem \ref{thm:mrqap-p-ns}, for the non-studentized statistics, none of the asymptotic permutation distributions match the asymptotic distribution of $\sqrt{n}\hvartheta$ under $\mathrm{H}_{0\textsc{w}}$. If the permuting outcome strategy is adopted, the asymptotic permutation distribution can even fail to match the sampling asymptotic distribution under the strong null hypothesis.





\begin{theorem}\label{thm:mrqap-p-s}
    Under the conditions in Theorem \ref{thm:mrqap-p-ns}, the Wald statistic $W$ in \eqref{eqn:plm-W} has an asymptotic distribution $\chi^2_p$ under all four permutation regimes:
    \begin{align}
        \lim_{n\to\infty}\sup_{t\in\bbR}|\cL(t; W^\pi) - \cL(t;\chi^2_p)| = 0 \quad \bbP\text{-a.s.}
    \end{align}
    It matches the asymptotic sampling distribution of $W$ under both the weak and strong null hypotheses in \eqref{eqn:null-mrqap-partial}.
\end{theorem}

Theorem \ref{thm:mrqap-p-s} states that all four permutation tests are asymptotically valid upon studentization. 


\subsection{Additional results}\label{sec:additional-results}

\subsubsection{Justification of the non-degeneracy assumption in Theorem \ref{thm:asp-super}}
The following Proposition \ref{prop:non-degeneracy-normal} provides a large class of examples for the non-degeneracy assumption $\eta_{1,\phi} > 0$ in Theorem \ref{thm:asp-super}, echoing our discussion after Theorem \ref{thm:asp-super}.

\begin{proposition}\label{prop:non-degeneracy-normal}
    Let $R$ and $S$ be random vectors in $\bbR^p$ and $\bbR^q$, respectively, and have a joint normal distribution with mean zero and covariance $\Sigma$. Let $\alpha(r,r') = \|r-r'\|_\alpha$ and $\beta(s,s') = \|s-s'\|_\beta$ be any two norm-induced metrics in $\bbR^p$ and $\bbR^q$, respectively. Define the kernel $\phi(r,s;r',s') = \alpha(r,r')\beta(s,s')$. Then 
    \begin{align}
        \eta_{1,\alpha} > 0, \quad \eta_{1,\beta} > 0, \quad \eta_{1,\phi} > 0. 
    \end{align}
\end{proposition}
\begin{proof}[Proof of Proposition \ref{prop:non-degeneracy-normal}]
If $\eta_{1,\alpha} = 0$, then the first-order projection
\begin{align}
    \E{\|R-R'\|_\alpha \mid R' = r} = \E{\|R - r\|_\alpha}
\end{align}
must be a constant function in $r$. By the triangular inequality, 
\begin{align}
    \E{\|R - r\|_\alpha} \ge \E{\|r\|_\alpha - \|R\|_\alpha} = \|r\|_\alpha - \E{\|R\|_\alpha}.
\end{align}
Therefore, $\E{\|R - r\|_\alpha} \to \infty$ as $\|r\|_\alpha\to\infty$ and can not be any finite constant. Similar argument applies to $ \E{\|S - s\|_\beta}$ and $\E{\|R - r\|_\alpha \|S - s\|_\beta}$.

\end{proof}

\subsubsection{Equivalence of $\widehat{V}$ and the cluster-robust variance estimator}\label{sec:LZ}
In this section, we justify the use of cluster-robust variance estimators multiplied by $4$ to recover $\widehat{V}$ as discussed in Section \ref{sec:cluster-var-est}.

By Theorem 3 of \cite{ding2021frisch}, running OLS with intercept gives the same point estimator and cluster-robust variance estimator as those by running OLS with centered outcome and covariates. Therefore, we can formulate the regression as follows:
\begin{align}
    \ty_g = 
    \begin{pmatrix}
        a_{1g} - \oa\\
        \vdots\\
        a_{ig} - \oa\\
        \vdots\\
        a_{ng} - \oa
    \end{pmatrix} \in \bbR^n,
    \quad 
    \tX_g = 
    \begin{pmatrix}
        b_{11g} - \ob_1 & \cdots & b_{k1g} - \ob_k & \cdots & b_{p1g} - \ob_p\\
        \vdots & \vdots & \vdots & \vdots & \vdots\\
        b_{1ig} - \ob_1 & \cdots & b_{kig} - \ob_k & \cdots & b_{pig} - \ob_p\\
        \vdots & \vdots & \vdots & \vdots & \vdots\\
        b_{1ng} - \ob_1 & \cdots & b_{kng} - \ob_k & \cdots & b_{png} - \ob_p
    \end{pmatrix} \in \bbR^{n\times p}.
\end{align}
The stacked observations across different clusters correspond to the vectorized dyadic matrices $\tA = (a_{ij} - \oa)_{i,j\in[n]}$ and $\tB_k = (b_{kij} - \ob_k)_{i,j\in[n]}$: 
\begin{align}\label{eqn:stack-y-X}
    \begin{pmatrix}
        \ty_1 \\
        \vdots \\
        \ty_G
    \end{pmatrix}
    =
    \operatorname{vec}(\tA), 
    \quad 
    \begin{pmatrix}
        \tX_1 \\
        \vdots \\
        \tX_G
    \end{pmatrix}
    =
    (\operatorname{vec}(\tB_1), \dots,  \operatorname{vec}(\tB_p)).
\end{align}
Therefore, running OLS of the stacked $y_g$'s on the stacked $X_g$'s recovers the estimators for the regression coefficients $\hvartheta$. 

For the covariance estimator, based on \eqref{eqn:stack-y-X}, we first have
\begin{align}
    \frac{1}{n(n-1)}\sum_{g=1}^n \tX_g^\top \tX_g = \hSigma_{bb}. 
\end{align}
Meanwhile, we can compute
\begin{align}
    \tX_g^\top \hat{e}_g 
    = 
    (\tilde{b}_{\cdot 1g}, \dots, \tilde{b}_{\cdot ng})
    \cdot
    \begin{pmatrix}
        \hat{e}_{1g}\\
        \vdots\\
        \hat{e}_{ng}
    \end{pmatrix}
    =
    \sum_{j} \hat{e}_{jg}\tilde{b}_{\cdot jg}
    =
    \sum_{j\neq g} \hat{e}_{jg}\tilde{b}_{\cdot jg},
\end{align}
where $\tilde{b}_{\cdot ig} = (b_{1ng} - \ob_1,\  
    \dots,\  
    b_{kng} - \ob_k,\  
    \dots,\ 
    b_{png} - \ob_p)^\top\in\bbR^p$ 
is the transpose of the $i$-th row of the matrix $\tilde{X}_g$. 

Therefore, 
\begin{align}
    \sum_{g=1}^G \tX^\top_g\hat{e}_g\hat{e}_g^\top \tX_g
    =
    \sum_{g=1}^n (\sum_{j\neq g} \hat{e}_{jg}\tilde{\bsb}_{jg})(\sum_{j\neq g} \hat{e}_{jg}\tilde{\bsb}_{jg})^\top
    =
    (n-1)(n-2)(n-4)\hH_{1,\phi} 
\end{align}
and the cluster-robust variance estimator equals
\begin{align}
    \hV_{\text{LZ}}
    = 
    \frac{(n-2)(n-4)}{n^2(n-1)} \hSigma_{bb}^{-1} \hH_{1,\phi} \hSigma_{bb}^{-1} 
    = 
    \frac{(n-2)(n-4)}{4n^2(n-1)} \hV
    \approx \frac{1}{4} \widehat{V}. 
\end{align}
This also verifies the correction factor proposed.







\section{Review of some results on U-statistics}\label{sec:U}

This section reviews some basic results on U-statistics and exchangeably dissociated arrays, which are useful for proving the asymptotic properties of the test statistics.

\subsection{U-statistics of degree $2$}\label{sec:U2}
Below, we review some useful results of U-statistics for proving Theorems \ref{thm:asp-super} and \ref{thm:asp-super-s}. Let $\{X_i\}_{i=1}^n$ be an i.i.d. sequence of random variables. A U-statistic of order $2$ is defined as 
\begin{align*}
    U_n = \frac{2}{n(n-1)} \sum_{i<j}\Phi(X_i, X_j),
\end{align*}
where $\Phi$ is a symmetric function (kernel) of two arguments. 

Recall the notation related to the kernel function $\Phi$ in Table \ref{tab:notation-phi}. $\Phi_0 = \E{\Phi(X,X')}$ is the expectation. $\tPhi_1(x)$ and $\tPhi_2(x,x')$ are the centered kernels:
\begin{gather*}
    \tPhi_1(x) = \E{\Phi(x,X')} - \Phi_0, \quad \tPhi_2(x,x') = \Phi(x,x') - \Phi_0.
\end{gather*}
$\eta_{1}$ and $\eta_{2}$ are the variances of the centered kernels:
\begin{align*}
    \eta_{1} = \E{\tPhi_1^2(X)}, \quad \eta_{2} = \E{\tPhi_2^2(X,X')}.
\end{align*}
The following lemma gives the mean and variance of $U_n$:
\begin{lemma}[Mean and variance of $U_n$, \cite{hoeffding1948class}]\label{lem:mean-var-u}
    We have
    \begin{align}
        \E{U_n} = \Phi_0, \quad \Var{U_n} = \frac{4(n-2)}{n(n-1)}\eta_1 + \frac{2}{n(n-1)}\eta_2.
    \end{align}
\end{lemma}

The following lemma gives a SLLN for $U_n$, which will be used to prove the consistency of the point and variance estimators:
\begin{lemma}[SLLN for $U_n$, \cite{korolyuk2013theory}]\label{lem:slln-u}
    Assume that the kernel $\Phi$ satisfies $\E{|\Phi(X,X')|} < \infty$. Then as $n\to\infty$,  $U_n \rightarrow \Phi_0$, $\bbP\text{-a.s.}$
    and $\E{|U_n - \Phi_0|} \rightarrow 0$. 
\end{lemma}

The following lemma gives a CLT for $U_n$, which will be used to prove the asymptotic distributions of the test statistics:
\begin{lemma}[CLT for $U_n$, \cite{hoeffding1948class}]\label{lem:clt-u}
    If $\eta_1 > 0$ and $ \E{\Phi^2(X,X')} < \infty $, then
    \begin{align}
        {\sqrt{n}(U_n - \Phi_0)} \rightsquigarrow \cN(0,4\eta_1). 
    \end{align}
\end{lemma}
The following lemma further provides a consistent variance estimator.
\begin{lemma}[Variance estimation, Proposition 1 of \cite{sen1960some}]\label{lem:var-est-1-dim}
    Assume $\E{\Phi(X,X')^2}<\infty$. Let 
    \begin{align}
        \heta_1 = \frac{n-1}{(n-2)(n-4)}\sum_{i=1}^n \lt\{\frac{1}{n-1}\sum_{j:j\neq i} (\Phi(X_i, X_j) - U_n)\rt\}^2. 
    \end{align}
    We have
    \begin{align}
        \E{\heta_1} = \eta_1 + \frac{1}{n-4}\eta_2. 
    \end{align}
    As $n\to\infty$, we have $
    \heta_1 \rightarrow \eta_1$, $ \bbP\text{-a.s.}$
\end{lemma}

\subsection{Multi-dimensional U-statistics of degree $2$}
In this section, we present a generalization of the one-dimensional U-statistic theory to the multi-dimensional setting, which is useful for proving the theory of QAP and MRQAP including Theorems \ref{thm:asp-super-ns-p} and \ref{thm:asp-super-s-p} in the main paper, as well as Theorems \ref{thm:MR-super} and \ref{thm:var-plm} in the Supplementary material. Consider a $p$-dimensional U-statistic of the form $U_n = (U_n^{(1)}, \dots, U_n^{(p)})^\top$, where
\begin{align*}
    U_n^{(k)} = \frac{2}{n(n-1)}\sum_{i<j}\Phi^{(k)}(X_i, X_j), \quad k = 1,\dots,p.
\end{align*}
Let $\Phi_0^{(k)} = \E{\Phi^{(k)}(X,X')}$, $k = 1,\dots, p$. 
Define the centered projection kernels:
\begin{align*}
    \tPhi_1^{(k)}(x) = \E{\Phi^{(k)}(x, X')} - \Phi_0^{(k)}, \quad \tPhi_2^{(k)}(x, x') = \Phi^{(k)}(x, x') - \Phi_0^{(k)}.
\end{align*}
Define
\begin{align}
    \eta_1^{(kl)} = \E{\tPhi_1^{(k)}(X)\tPhi_1^{(l)}(X)}, \quad 
    \eta_2^{(kl)} = \E{\tPhi_2^{(k)}(X, X')\tPhi_2^{(l)}(X, X')}.
\end{align}

The following four lemmas extend Lemmas \ref{lem:mean-var-u}--\ref{lem:var-est-1-dim}. Their proofs are straightforward and thus omitted.
\begin{lemma}
The mean and variance of $U_n^{(k)}$ is 
\begin{align}
    \E{U_n^{(k)}} = \Phi_0^{(k)}, 
    \quad 
    \Var{U_n^{(k)}} = 
    \frac{4(n-2)}{n(n-1)}\eta_1^{(kk)} 
    + 
    \frac{2}{n(n-1)} \eta_2^{(kk)}. 
\end{align}
The covariance between $U_n^{(k)}$ and $U_n^{(l)}$ is 
\begin{align}
    \Cov{U_n^{(k)}}{U_n^{(l)}} = 
    \frac{4(n-2)}{n(n-1)}\eta_1^{(kl)} 
    + 
    \frac{2}{n(n-1)} \eta_2^{(kl)}. 
\end{align}
\end{lemma}

\begin{lemma}[SLLN for $U_n$]\label{lem:slln-u-md}
    Assume that the kernel $\Phi$ satisfies $\E{|\Phi(X,X')|} < \infty$. Then for $k = 1,\dots,p$, as $n\to\infty$, we have
    $U_n^{(k)} \rightarrow \Phi_0^{(k)} \quad \bbP\text{-a.s.}
    $
    and 
    \begin{align}
        \E{|U_n^{(k)} - \Phi_0^{(k)}|} \rightarrow 0. 
    \end{align}
\end{lemma}

\begin{lemma}
    Assume each kernel $\Phi^{(k)}$ satisfies:
    \begin{align*}
        \eta_1^{(kk)} > 0 , \quad \E{(\Phi^{(k)}(X,X'))^2} < \infty. 
    \end{align*}
    Then as $n\to\infty$, ${\sqrt{n}}(U_n - \Phi_0)\rightsquigarrow \cN(0, 4H_1)$, 
    where $H_1 = (\eta_1^{(kl)})_{k,l\in[p]}$.
\end{lemma}

\begin{lemma}\label{lem:var-est-p-dim}
Assume $\E{\Phi^{(k)}(X,X')^2}<\infty$ for $k\in[p]$. Let $\hH_1 = (\heta_1^{(kl)})_{k,l\in[p]}$, where
    \begin{align}
        \heta_1^{(kl)} = \frac{n-1}{(n-2)(n-4)}\sum_{i=1}^n \lt\{\frac{1}{n-1}\sum_{j:j\neq i} (\Phi^{(k)}(X_i, X_j) - U_n^{(k)})\rt\} \lt\{\frac{1}{n-1}\sum_{j:j\neq i} (\Phi^{(l)}(X_i, X_j) - U_n^{(l)})\rt\}. 
    \end{align}
    We have $\bbE\{\heta_1^{(kl)}\} = \eta_1^{(kl)} + {(n-4)^{-1}}\eta_2^{(kl)}$. 
    As $n\to\infty$, we have $\heta_1^{(kl)} \rightarrow \eta_1^{(kl)}$, $\bbP\text{-a.s.}$
\end{lemma}

\subsection{Exchangeably disassociated arrays} 
U-statistics are special cases of the sample average of the \textit{exchangeably disassociated arrays}, which we will review below.  Suppose that an $m$-tuple $J$ is defined to be an ordered set of $m$ distinct positive integers $j_1, \dots, j_m$. Let $P(m)$ be the set of all $m$-tuples and $P(m, n)$ be the set of $m$-tuples, all of whose elements are in $[n]$. Let $\bbN$ be the set of all positive numbers. Let $\pi(J)$ denote the $m$-tuple $(\pi(j_1), \dots, \pi(j_m))$, where $\pi:\bbN\to\bbN$ is a permutation of the positive integers. Let $\fX_m$ denote an array of random variables $X_J$, indexed by all $J$ in $P(m)$. For example, when $m=1$, $\fX_m$ includes a sequence of random variables $\{X_1,X_2,\dots,X_n,\dots\}$. When $m=2$, $\fX_m$ is a triangular array
\begin{align}
    \begin{matrix}
        X_{12} &        &        & \\
        X_{13} & X_{23} &        & \\
        X_{14} & X_{24} & X_{34} & \\
        \vdots & \vdots & \vdots & \ddots
    \end{matrix}
\end{align}
and the definition of each $X_{ij}$ can depend on the concrete problems.

We call $\mathcal{J},  \mathcal{K} \subset P(m)$ as two disjoint subsets of $P(m)$ if the sets of integers $\{j:j\in J \text{ for some }J\in\cJ\}$ and $\{k: k \in \cK \text{ for some } K \in \cK\}$ are disjoint. The exchangeable dissociated array is defined below:
\begin{definition}[Exchangeably dissociated array]\label{def:ed}
$\fX_m$ is called an {\em exchangeably dissociated array} if the following two conditions hold:
\begin{enumerate}
    \item Independence. $\{X_J: J\in\cJ\}$ and $\{X_K: K\in\cK\}$ are independent for any disjoint pair of subsets $\cJ$ and $\cK$ of $P(m)$.
    
    \item Exchangeability. For any positive integer $l=1,2,\dots$, any finite sequence $J_1,\dots, J_l$ in $P(m)$ and any permutation $\pi:\bbN\to\bbN$ of the positive integers, $(X_{\pi(J_1)}, \dots, X_{\pi(J_l)})$ has the same distribution as $(X_{J_1}, \dots, X_{J_l})$. 
\end{enumerate}
\end{definition}
As a direct consequence of exchangeability in Definition \ref{def:ed}, $X_J$ and $X_{J'}$ have the same distribution for any $J, J' \in P(m,n)$.  
Define the average
\begin{align}
    T_n = \lt\{m! {n\choose m} \rt\}^{-1} \sum_{J\in P(m,n)} X_J. 
\end{align}
The averaging factor is the number of elements in $P(m,n)$.
$n\choose m$ is the number of different combinations of integers to choose from $1,\dots,n$, and $m!$ counts the number of permutations with any $m$ integers. 
The U-statistics of degree $2$, $U_n$, in Section \ref{sec:U2} are special cases of $T_n$ if we take $m=2$ and define $X_J$ as the $\Phi(X_i, X_j)$. 
If $\fX_m$ is an exchangeably dissociated array, we have the following SLLN for the sample average of $T_n$, which is established with a general SLLN in \cite{eagleson1978limit} applied to $\fX_m$.  
\begin{lemma}[SLLN for $T_n$, {\citet[][Theorem 3]{eagleson1978limit}}]\label{lem:slln-ed}
    If $\E{|X_J|} < \infty$, then $T_n \rightarrow \E{X_J}~ {\text{a.s.}}$
\end{lemma}
We will use Lemma \ref{lem:slln-ed} later to prove Theorem \ref{thm:asp-mantel-s-p}.


\section{Review of some theory on DIPS and additional results on multiple-indexed permutation statistics }\label{sec:DIPS}
In this section, we review theory of one dimensional DIPS (Section \ref{sec:DIPS-1d}), multi-dimensional DIPS (Section \ref{sec:dips-md}), and $K$-indexed permutation statistics (Section \ref{sec:KIPS}) as a generalization of DIPS. For convenience, we introduce a set of quantities defined by a generic matrix $A = (a_{ij}) \in \bbR^{n\times n}$.  Define the off-diagonal average $\oa$, demeaned row-wise averages $\oa_i$'s, and demeaned elements $\ta_{ij}$'s as follows:
\begin{gather}\label{eqn:matrix-quantities}
    \overline{a} = \frac{1}{n(n-1)} \sum_{i\neq j}a_{ij}, \quad 
    \overline{a}_i = \frac{1}{n-2} \sum_{j:j\neq i} (a_{ij} - \overline{a}),\quad 
    \tilde{a}_{ij} = a_{ij} - \overline{a}_i - \overline{a}_j - \overline{a}. 
\end{gather}
We can verify that
\begin{align*}
    \sum_{i=1}^n \overline{a}_i = 0;\quad \text{ for all } i, ~
    \sum_{j:i\neq j}\tilde{a}_{ij} = 0.
\end{align*}

Besides, define the moments as follows:
\begin{align}\label{eqn:dips-moments}
    m_{1k}(a) = \frac{1}{n}\sum_{i=1}^n |\overline{a}_i|^k,  \quad m_{2k}(a) =  \frac{1}{n(n-1)} \sum_{i\neq j}|\tilde{a}_{ij}|^k.
\end{align}
Here the index ``1'' indicates moments for the row-wise (or column-wise) means, $\overline{a}_{i}$'s, while the index ``2'' indicates moments for the demeaned elements, $\ta_{ij}$'s. 

Furthermore, for a set of matrices, $A^{(k)} = (a^{(k)}_{ij})$, $k=1,\dots,p$, we use $m_{12}^{(kl)}(a)$ to denote the covariance between the demeaned row-wise averages of $A^{(k)}$ and $A^{(l)}$:
\begin{align}
    m_{12}^{(kl)} = \frac{1}{n}\sum_{i=1}^n \overline{a}_i^{(k)}\overline{a}_i^{(l)}.
\end{align}
\subsection{One dimensional DIPS}\label{sec:DIPS-1d}
In this section, we review some theoretical results on DIPS, which are useful for proving Theorems \ref{thm:asp-mantel-ns} and \ref{thm:asp-mantel-s}. Consider two symmetric matrices $A = (a_{ij})$ and $B = (b_{ij})$. Besides, we assume the diagonals of $A$ and $B$ are all zero: $a_{ii} = 0$, $b_{ii} = 0$. Consider the following DIPS:
\begin{align*}
    W = \frac{1}{n(n-1)}\sum_{i\neq j} a_{ij} b_{\pi(i)\pi(j)}.
\end{align*}
This is a special case of the general DIPS discussed in \cite{zhao1997error}, which is also called the matrix correlation statistics \citep{barbour1986random, barbour2005stein}. 



Recall the quantities defined in \eqref{eqn:matrix-quantities} and \eqref{eqn:dips-moments} regarding a general matrix. We apply these definitions to $A$ and $B$ here and further introduce the following two random components:
\begin{align*}
    V = \frac{2(n-2)}{n(n-1)} \sum_{i=1}^n \overline{a}_i \overline{b}_{\pi(i)}, 
    \quad 
    \Delta = \frac{1}{n(n-1)}\sum_{i\neq j} \tilde{a}_{ij}\tilde{b}_{\pi(i)\pi(j)},
\end{align*}
where $V$ is a SIPS and $\Delta$ is still a DIPS. 
Following the discussion in \cite{barbour2005stein} and \cite{chen2011normal}, we provide the lemma below:
\begin{lemma}[Decomposition of $W$]\label{lem:decomp-dips}
The following decomposition holds for $W$:
   \begin{align*}
    W = \E{W} + {V} + {\Delta}.
   \end{align*} 
   Moreover, we have
   \begin{align*}
       \E{W} = \overline{a}\cdot \overline{b},
       \quad \E{V} = 0, 
       \quad \E{\Delta} = 0,
       \quad \E{V\Delta} = 0,
   \end{align*}
   and
   \begin{align*}
       \Var{V} = \frac{4(n-2)^2}{(n-1)^3}m_{12}(a) m_{12}(b), 
       \quad 
       \Var{\Delta} = \frac{2}{n(n-3)}m_{22}(a) m_{22}(b).
   \end{align*}
\end{lemma}

We also prove a CLT for $W$. We will focus on the regime which assumes that the SIPS term $V$ is dominant in the decomposition of Lemma \ref{lem:decomp-dips}. The CLT is then derived from the following CLT of the SIPS.
\begin{lemma}[CLT for SIPS, Lemma S.3.3 of \cite{diciccio2017robust}]\label{lem:clt-sips}
    Let $a = (a_1, \dots, a_n)$ and $b^{(1)} = (b_1^{(1)}, \dots, b_n^{(1)})$, $\dots$, $b^{(p)} = (b_1^{(p)}, \dots, b_n^{(p)})$ be vectors whose components are real numbers, possibly depending on \( n \), satisfying
\begin{align}
    \sum_{i=1}^{n} a_i &= \sum_{i=1}^{n} b_i^{(k)} = 0 \quad \text{for each } k = 1, \dots, p,
\end{align}
and
\begin{align}
    \lim_{n \to \infty} \sum_{i=1}^{n} a_i^2 = \sigma < \infty,\quad 
    \lim_{n \to \infty} \frac{1}{n} \sum_{i=1}^{n} b_i^{(r)} b_i^{(s)} = \sigma_{r,s} < \infty \text{ for each } r,s \in [p]. 
\end{align}
Let $\pi$ be a uniformly chosen permutation of \( \{ 1, \dots, n \} \). Then
\begin{align}
    S_n & = \left[ \sum_{i=1}^{n} a_i b_{\pi(i)}^{(1)}, \dots, \sum_{i=1}^{n} a_i b_{\pi(i)}^{(p)} \right]
\end{align}
is asymptotically normal with mean zero and covariance matrix \( \Sigma \) with \( \Sigma_{i,j} = \sigma \cdot \sigma_{i,j} \) provided that for each \( \epsilon > 0 \), there exists a constant \( d > 0 \) such that
\begin{align}
    \sum_{i=1}^{n} a_i^2 I \left\{ \sqrt{n} |a_i| \geq d \right\} < \epsilon, \quad 
    \max_{1 \leq i \leq n} \frac{b_i^{(r)}}{\sqrt{n}} \to 0
\end{align}
as $n \to \infty$ for $r = 1, \dots, p$.
\end{lemma}

With Lemma \ref{lem:clt-sips}, we can establish the following CLT for DIPS: 
\begin{lemma}[CLT for DIPS] \label{lem:clt-dips}
Assume that as $n\to\infty$, we have
\begin{align*} 
    \frac{\max_{i} \overline{a}_i^2 }{n\cdot m_{12}(a) } \to 0, \quad \lim_{d\to\infty}\limsup_{n\to\infty} \frac{1}{n}\sum_{i=1}^n \frac{\overline{b}_i^2}{m_{12}(b)} \ind{\frac{|\overline{b}_i|}{\sqrt{m_{12}(b)}} > d} = 0, 
    \quad \frac{\Var{\Delta}}{\Var{V}} 
    \to 0.
\end{align*}
Then
\begin{align}
    \frac{W-\E{W}}{\sqrt{\Var{W}}} \rightsquigarrow \cN(0,1).
\end{align}
\end{lemma}
\begin{proof}[Proof of Lemma \ref{lem:clt-dips-md}]
    By Lemma \ref{lem:decomp-dips}, 
    \begin{align}
        W =  \E{W} + V + \Delta.
    \end{align}
    The condition ${\Var{\Delta}}/{\Var{V}} 
    \to 0$ implies that the dominant term is $V$. By Lemma \ref{lem:clt-sips}, 
    \begin{align*}
        \frac{W-\E{W}}{\sqrt{\Var{W}}} \rightsquigarrow \cN(0,1).
    \end{align*}
    Hence we conclude the proof. 
\end{proof}

\subsection{Multi-dimensional DIPS}\label{sec:dips-md}
Now we present an extension to multi-dimensional DIPS, which will be used to prove Theorem \ref{thm:asp-mantel-s-p} for QAP as well as Theorem \ref{thm:permute-b-W} for MRQAP. Consider symmetric matrices $A^{(k)} = (a_{ij}^{(k)})$ ($k=1,\dots,p$) and $B = (b_{ij})$ with zero diagonals: $a_{ii}^{(k)} = 0$, $b_{ii} = 0$.

Analogous to the one-dimensional case, for $A^{(k)}$, define the quantities $\oa^{(k)}$, $\oa_{i}^{(k)}$ and $\ta_{ij}^{(k)}$, as well as moments $m_{12}^{(kl)}(a)$ and $m_{22}^{(k)}(a)$ following \eqref{eqn:matrix-quantities} and \eqref{eqn:dips-moments}. Also, define the similar quantities corresponding to the matrix $B$. 
We introduce the following random quantities: 
\begin{align*}
    V^{(k)} = \frac{2(n-2)}{n(n-1)} \sum_{i=1}^n \overline{a}^{(k)}_i \overline{b}_{\pi(i)}, 
    \quad 
    \Delta^{(k)} = \frac{1}{n(n-1)}\sum_{i\neq j} \tilde{a}_{ij}^{(k)}\tilde{b}_{\pi(i)\pi(j)}.
\end{align*}

Consider the following multi-dimensional DIPS:
\begin{align}\label{eqn:mv-dips}
    W = (W^{(k)})_{k\in[p]} = \lt(\frac{1}{n(n-1)}\sum_{i\neq j} a^{(k)}_{ij} b_{\pi(i)\pi(j)}\rt)_{k \in [p]}.
\end{align}
By Lemma \ref{lem:decomp-dips}, we have the following decomposition for $W^{(k)}$:
\begin{align*}
 W^{(k)} = \E{W^{(k)}} + {V^{(k)}} + {\Delta^{(k)}}.
\end{align*} 
Moreover, we have
\begin{align*}
    \E{W^{(k)}} = \overline{a}^{(k)}\cdot \overline{b},
       \quad \E{V^{(k)}} = 0, 
       \quad \E{\Delta^{(k)}} = 0,
       \quad \E{V^{(k)}\Delta^{(k)}} = 0,
\end{align*}
and
\begin{gather*}
    \Var{V^{(k)}} = \frac{4(n-2)^2}{(n-1)^3}m_{12}^{(kk)}(a) m_{12}(b), 
    \quad 
    \Var{\Delta^{(k)}} = \frac{2}{n(n-3)}m_{22}^{(k)}(a) m_{22}(b),\\
    \Cov{V^{(k)}}{V^{(l)}} = \frac{4(n-2)^2}{(n-1)^3}m^{(kl)}_{12}(a)m_{12}(b).
\end{gather*}

We establish a multivariate CLT for $W$ in \eqref{eqn:mv-dips}:
\begin{lemma}[CLT for multi-dimensional DIPS] \label{lem:clt-dips-md}
Assume that as $n\to\infty$, we have
\begin{enumerate}
    \item $m_{12}(b)\rightarrow \xi_{12}(b)$, $m_{22}(b) \rightarrow \xi_{22}(b)$. For $k,l\in[p]$, 
    $m_{12}^{(kl)}(a) \rightarrow \xi_{12}^{(kl)}(a)$,  $m_{22}^{(k)}(a) \rightarrow \xi_{22}^{(k)}(a)$. Moreover, $\xi_{12}(b)>0$ and the matrix $\Xi_{12}(a) = \lt(\xi_{12}^{(kl)}(a)\rt)_{k,l\in[p]}$ is positive definite;
    
    \item For $k\in[p]$, $ {\max_{i} |\overline{a}_i^{(k)}| }/{\sqrt{n}} \to 0 $;
    
    \item $
    \lim_{d\to\infty}\limsup_{n\to\infty} {n^{-1}}\sum_{i=1}^n {\overline{b}_i^2}  \ind{ {|\overline{b}_i|} > d} = 0
    $.
\end{enumerate}
Then
\begin{align*}
    \sqrt{n}(W - \E{W}) \rightsquigarrow \cN(0, 4\xi_{12}(b)\cdot \Xi_{12}(a)).
\end{align*}
\end{lemma}
Lemma \ref{lem:clt-dips-md} is a direct application of the Cramer--Wold device, but we give the detailed proof below for completeness.
\begin{proof}[Proof of Lemma \ref{lem:clt-dips-md}]
    For any $c\in\bbR^p$ with $\|c\|_2 = 1$, we have
    \begin{align}
        \sum_{k=1}^p c_k W^{(k)} = \sum_{k=1}^p c_k \E{W^{(k)}} + \sum_{k=1}^p c_k V^{(k)} + \sum_{k=1}^p c_k \Delta^{(k)}.
    \end{align}
    Then we can compute
    \begin{align*}
        \Var{\sum_{k=1}^p c_k \Delta^{(k)}} \le p\sum_{k=1}^p c_k^2 \Var{\Delta^{(k)}} \le p\max_{k\in[p]} \Var{\Delta^{(k)}} = O(n^{-2}),
    \end{align*}
    and 
    \begin{align*}
        \Var{\sum_{k=1}^p c_k V^{(k)}} \asymp \frac{4(n-2)^2}{(n-1)^3} c^\top \Xi_{12}(a)c \asymp n^{-1}.
    \end{align*}
    Therefore, the dominant term of $W$ is $(V^{(k)})_{k\in[p]}$. By Lemma \ref{lem:clt-dips}, 
    \begin{align*}
        \sqrt{n}V \rightsquigarrow \cN(0, 4\xi_{12}(b)\cdot \Xi_{12}(a))
    \end{align*}
    Hence we conclude the proof. 
\end{proof}

\subsection{Some results on $K$-indexed permutation statistics}\label{sec:KIPS}
In this section, we discuss $K$-indexed permutation statistics, as an extension of DIPS. The results are used to prove theorems in both the main paper and supplementary materials (such as Theorem \ref{thm:asp-mantel-s} and Theorem \ref{thm:mrqap-p-ns}). Besides, they may also be of independent interest. 

Let $D = (d(i_1 \dots i_K; j_1 \dots j_K))$ be a multi-dimensional array indexed by 
\begin{align}
  (i_1 \dots i_K; j_1 \dots j_K)\in [n]^{2K}.
\end{align}
Consider the following $K$-indexed permutation statistic
\begin{align}\label{eqn:Gamma-K}
    \Gamma = \sum_{i_1\neq \dots\neq i_K} d(i_1 \dots i_K; \pi(i_1) \dots \pi(i_K)).
\end{align}
The following lemma gives the expectation and an upper bound on the variance of $\Gamma$, which
generalizes \citet[][Lemma 2.3]{barbour2005stein} on DIPS.  
\begin{lemma}[Variance bound for $K$-indexed permutation statistics]\label{lem:var-array}
The expectation of $\Gamma$ defined in \eqref{eqn:Gamma-K} equals
\begin{align}
    \E{\Gamma} = \frac{1}{(n)_K} \sum_{i_1\neq \dots\neq i_K}\sum_{j_1\neq \dots\neq j_K} d(i_1 \dots i_K; j_1 \dots j_K), 
\end{align}
and the variance of $\Gamma$ is upper bounded by 
\begin{gather}
    \Var{\Gamma} \le 
    \frac{C_K n^{2K - 1}}{(n)_K^2}\sum_{i_1\neq \dots\neq i_K} \sum_{j_1\neq \dots\neq j_K} \td(i_1 \dots i_K; j_1 \dots j_K)^2,
\end{gather}
where $C_K$ is a constant that only depends on $K$,  $(n)_K$ denotes the falling factorial: $(n)_K = \prod_{l=1}^K (n-l+1)$, and $\td(\cdot;\cdot)$ is the demeaned array
\begin{align}
& \td(i_1 \dots i_K; j_1 \dots j_K)\\
= 
    &d(i_1 \dots i_K; j_1 \dots j_K)
    - 
    \frac{1}{(n)_K} \sum_{i_1'\neq \dots\neq i_K'}d(i_1' \dots i_K'; j_1 \dots j_K)
    & \\
    & 
    - \frac{1}{(n)_K} \sum_{j_1'\neq \dots\neq j_K'}d(i_1 \dots i_K; j_1' \dots j_K') 
    + \frac{1}{(n)_K^2} \sum_{i_1'\neq \dots\neq i_K'}\sum_{j_1'\neq \dots\neq j_K'}d(i_1' \dots i_K'; j_1' \dots j_K').
\end{align}
\end{lemma}
As a special case of the variance bound in Lemma \ref{lem:var-array}, when $K=1$, $\Gamma$ is a SIPS with variance  
\begin{align}
    \Var{\Gamma} = \frac{1}{n-1}\sum_{i=1}^n\sum_{j=1}^n \td(i;j)^2. 
\end{align}
See, for example, \citet[][Theorem 2]{hoeffding1951combinatorial}. This aligns with the variance bound in Lemma \ref{lem:var-array} with $C_1 = 2$. 

The following lemma gives a WLLN based on Lemma \ref{lem:var-array}.

\begin{lemma}\label{lem:Ppi-converge}
    Under the condition
    \begin{align}
        \frac{1}{(n)_K^2}\sum_{i_1 \neq \cdots \neq i_K} \sum_{j_1 \neq \cdots \neq j_K} \td(i_1 \dots i_K; j_1 \dots j_K)^2 = o(n), 
    \end{align}
    $\Gamma$ defined in \eqref{eqn:Gamma-K} satisfies
    \begin{align}
        \frac{1}{(n)_K}(\Gamma-\E{\Gamma}) = o_{\bbP}(1). 
    \end{align}
    In particular, let $\{a(i_1,\dots,i_K)\}$ and $\{b(i_1,\dots,i_K)\}$ be two arrays indexed by $(i_1,\dots,i_K)\in[n]^K$, and $\{\ta(i_1,\dots,i_K)\}$ and $\{\tb(i_1,\dots,i_K)\}$ be their centered versions:
    \begin{align}
        \ta(i_1,\dots,i_K) = 
        a(i_1,\dots,i_K)
        -
        \frac{1}{(n)_K}\sum_{i_1\neq\cdots\neq i_K}a(i_1,\dots,i_K), \\
        \tb(i_1,\dots,i_K) = 
        b(i_1,\dots,i_K)
        -
        \frac{1}{(n)_K}\sum_{i_1\neq\cdots\neq i_K}b(i_1,\dots,i_K).
    \end{align}

    Consider the permutational statistics:
    \begin{align*}
        U_n = \frac{1}{(n)_K}\sum_{i_1\neq \cdots \neq i_K} a(i_1,\dots,i_K)b(\pi(i_1), \dots, \pi(i_K)),
    \end{align*}
    which is a special case of $\Gamma$ in \eqref{eqn:Gamma-K}. 
    Under the condition 
    \begin{align}\label{eqn:var-to-0}
        \frac{1}{n} {\lt\{\frac{1}{(n)_K}\sum_{i_1\neq \dots \neq i_K} \ta(i_1,\dots,i_K)^2\rt\} \lt\{\frac{1}{(n)_K}\sum_{i_1\neq \dots \neq i_K} \tb(i_1,\dots,i_K)^2\rt\}} \rightarrow 0,
    \end{align}
    we have ${U_n} - {\E{U_n}} = o_{\bbP}(1)$. 
\end{lemma}

\subsection{Proofs}

\subsubsection{Proof of Lemma \ref{lem:var-array}}
\begin{proof}[Proof of Lemma \ref{lem:var-array}]
\textbf{Compute the expectation.}  Because $\pi$ is a uniform random permutation, $(\pi(i_1) \cdots \pi(i_K))$ will run through all $(n)_K$ distinct $K$-tuples $(i_1'\cdots i_K')$ with $i_1' \neq \cdots \neq i_K'$. Therefore, we have
\begin{align}
    \E{\Gamma} = \frac{1}{(n)_K} \sum_{i_1\neq\cdots\neq i_K}\sum_{i_1'\neq \cdots\neq i_K'} d(i_1 \dots i_K; i_1' \dots i_K').
\end{align}

\textbf{Bound the variance.} We divide this part into several steps.

\textit{Step I. Reformulate $\Gamma$ with a demeaned array.} By the formula of the expectation, we can verify
\begin{align}
    &\Gamma - \E{\Gamma} 
    \\
    = &
    \sum_{i_1\neq\cdots\neq i_K}
    \Bigg\{
    d(i_1 \dots i_K; \pi(i_1) \dots \pi(i_K))
    - 
    \frac{1}{(n)_K} \sum_{i_1'\neq \dots\neq i_K'}d(i_1' \dots i_K'; \pi(i_1) \dots \pi(i_K))
    & \\
    &\quad 
    - \frac{1}{(n)_K} \sum_{j_1'\neq \dots\neq j_K'}d(i_1 \dots i_K; j_1' \dots j_K') 
    + \frac{1}{(n)_K^2} \sum_{i_1'\neq \dots\neq i_K'}\sum_{j_1'\neq \dots\neq j_K'}d(i_1' \dots i_K'; j_1' \dots j_K')\Bigg\}\\
    = &
    \sum_{i_1\neq\cdots\neq i_K}
    \td(i_1 \dots i_K; \pi(i_1) \dots \pi(i_K)).
\end{align}
The summands $\tilde{d}(i_1 \cdots i_K; i_1' \cdots i_K')$'s satisfy that,  for any distinct tuples $i_1 \neq \cdots \neq i_K$ and $i_1' \neq \cdots \neq i_K'$,
\begin{align}\label{eqn:margin-zero}
    \sum_{i_1'\neq \cdots\neq i_K'} \td(i_1 \dots i_K; i_1' \dots i_K') = 0, \quad \sum_{i_1\neq \cdots\neq i_K} \td(i_1 \dots i_K; i_1' \dots i_K') = 0.
\end{align}
In other words, $\td$ has zero marginal sums.

The variance can be expressed using the demeaned array:
\begin{align}\label{eqn:var-Gamma-mid}
    \Var{\Gamma} = \sum_{\substack{i_1\neq\cdots\neq i_K, \\ j_1\neq \dots\neq j_K}} \E{\td(i_1 \cdots i_K; \pi(i_1) \cdots \pi(i_K))\td(j_1 \cdots j_K; \pi(j_1) \cdots \pi(j_K))}.
\end{align}

\textit{Step II. Partition the summation of the variance formula.} The key now is to bound the expectation
\begin{align}
    \E{\td(i_1 \cdots i_K; \pi(i_1) \cdots \pi(i_K))\td(j_1 \cdots j_K; \pi(j_1) \cdots \pi(j_K))}. 
\end{align}
Because $\pi$ is a random permutation over $[n]$, this expectation depends crucially on how many indices overlap between $(i_1 \dots i_K)$ and $(j_1 \dots j_K)$. For example, when $i_1 \neq \cdots \neq i_K \neq j_1 \neq \cdots \neq j_K$, the permuted indices are also all distinct: $\pi(i_1) \neq \cdots \neq \pi(i_K) \neq \pi(j_1) \neq \cdots \neq \pi(j_K)$. Hence, the expectation is averaging
\begin{align}
    \td(i_1 \cdots i_K; i_1' \cdots i_K')\td(j_1 \cdots j_K; j_1' \cdots j_K')
\end{align}
over all distinct indices $i_1' \neq \cdots \neq i_K' \neq j_1' \neq \cdots \neq j_K'$, which contain $(n)_{2K}$ combinations. Generally, when there are $p$ overlapping indices, the expectation is averaging over $(n)_{2K-p}$ combinations. For simplicity, we consider the overlapping indices to be the first $p$ ones because we can handle the rest cases similarly. If we denote
\begin{align*}
    \mathcal{E}_p = \{(i_1\dots i_K), (j_1\dots j_K): |&\{i_1 , i_2, \dots , i_K\}| = K; |\{j_1, j_2 , \dots , j_K\}| = K; \\ &
    |\{i_1 , i_2, \dots , i_K\} \cup \{j_1, j_2 , \dots , j_K\}| = 2K - p\},
\end{align*}
then \eqref{eqn:var-Gamma-mid} becomes
\begin{align}
    \Var{\Gamma} = \Big(\sum_{\mathcal{E}_0} + \cdots + \sum_{\mathcal{E}_K} \Big) \E{\td(i_1 \cdots i_K; \pi(i_1) \cdots \pi(i_K))\td(j_1 \cdots j_K; \pi(j_1) \cdots \pi(j_K))}. 
\end{align}

\textit{Step III. Bound the expectation in the variance formula.} 

We consider three cases:

\textbf{Case 1: all indices are distinct, i.e., a summation over $\mathcal{E}_0$. } For this case, we have
\begin{align*}
    &\E{\td(i_1 \dots i_K; \pi(i_1) \dots \pi(i_K))\td(j_1 \dots j_K; \pi(j_1) \dots \pi(j_K))}\\
    = & \frac{1}{(n)_{2K}} \sum_{i_1' \neq \cdots \neq i_K' \neq j_1' \neq \cdots \neq j_K'}\td(i_1 \dots i_K; i'_1 \dots i'_K)\td(j_1 \dots j_K; j'_1 \dots j'_K)\\
    = & -\frac{1}{(n)_{2K}} \sum_{i_1' \neq \cdots \neq i_K'} \td(i_1 \dots i_K; i'_1 \dots i'_K)\sum_{\substack{j_1' \neq \cdots \neq j_K',\\ j'_k = i'_l \text{ for some $k,l$}}} \td(j_1 \dots j_K; j'_1 \dots j'_K),
\end{align*}
where the last equality is due to \eqref{eqn:margin-zero}:
\begin{align}
    0 = 
    \sum_{j_1'\neq \cdots\neq j_K'} \td(j_1 \dots j_K; j_1' \dots j_K') 
    =
    \lt(\sum_{\substack{j_1' \neq \cdots \neq j_K',\\ j'_k = i'_l \text{ for some $k,l$}}} + \sum_{\substack{j_1' \neq \cdots \neq j_K',\\ j'_k \neq i'_l \text{ for all $k,l$}}}\rt) \td(j_1 \dots j_K; j_1' \dots j_K').
\end{align}

Hence
\begin{align*}
    &\lt|\E{\td(i_1 \dots i_K; \pi(i_1) \dots \pi(i_K))\td(j_1 \dots j_K; \pi(j_1) \dots \pi(j_K))}\rt|\\
    \le & \frac{1}{(n)_{2K}} \Bigg| \sum_{\substack{i_1' \neq \cdots \neq i_K',\\j_1' \neq \cdots \neq j_K',\\ j_k' = i_l' \text{ for some $k,l$}}} \td(i_1 \dots i_K; i'_1 \dots i'_K)\td(j_1 \dots j_K; j'_1 \dots j'_K)\Bigg|\\
    \le & \frac{1}{(n)_{2K}}\sum_{\substack{i_1' \neq \cdots \neq i_K',\\j_1' \neq \cdots \neq j_K',\\ j_k' = i_l' \text{ for some $k,l$}}}\Bigg|\td(i_1 \dots i_K; i'_1 \dots i'_K)\td(j_1 \dots j_K; j'_1 \dots j'_K)\Bigg|\\
    \le & \frac{1}{(n)_{2K}}\sum_{\substack{i_1' \neq \cdots \neq i_K',\\j_1' \neq \cdots \neq j_K',\\ j_k' = i_l' \text{ for some $k,l$}}} \lt(\frac{\td(i_1 \dots i_K; i'_1 \dots i'_K)^2}{2} +  \frac{\td(j_1 \dots j_K; j'_1 \dots j'_K)^2}{2}\rt)\\
    \le & \frac{1}{(n)_{2K}} \sum_{\substack{i_1' \neq \cdots \neq i_K',\\j_1' \neq \cdots \neq j_K',\\ j_k' = i_l' \text{ for some $k,l$}}}\frac{\td(i_1 \dots i_K; i'_1 \dots i'_K)^2}{2}
    + \frac{1}{(n)_{2K}} \sum_{\substack{i_1' \neq \cdots \neq i_K',\\j_1' \neq \cdots \neq j_K',\\ j_k' = i_l' \text{ for some $k,l$}}}  \frac{\td(j_1 \dots j_K; j'_1 \dots j'_K)^2}{2}\\
    \triangleq& \mathrm{I}(i_1 \dots i_K) + \mathrm{II}(j_1 \dots j_K). 
\end{align*}
For the term I, the index $(j_1' \dots j_K')$ does not appear in the summands. The total number of such $K$-tuples $(j_1' \dots j_K')$ that has overlap with $(i_1' \dots i_K')$ is $(n)_K - (n-K)_K$, i.e., the total number of $K$-tuples minus the number of $K$-tuples with no overlap. Therefore, we have
\begin{align}
    \mathrm{I}(i_1 \dots i_K)
    = & \frac{(n)_K - (n-K)_K}{(n)_{2K}} \sum_{\substack{i_1' \neq \cdots \neq i_K'}}\frac{\td(i_1 \dots i_K; i'_1 \dots i'_K)^2}{2} \notag\\
    \le & \frac{K(n-1)_{K-1}}{{2}(n)_{2K}} \sum_{\substack{i_1' \neq \cdots \neq i_K'}}{\td(i_1 \dots i_K; i'_1 \dots i'_K)^2}. \label{eqn:I-bound}
\end{align}
By symmetry, we bound the term II by
\begin{align}
    \mathrm{II}(j_1 \dots j_K)
    \le & \frac{K(n-1)_{K-1}}{{2}(n)_{2K}} \sum_{\substack{j_1' \neq \cdots \neq j_K'}}{\td(j_1 \dots j_K; j'_1 \dots j'_K)^2}. \label{eqn:II-bound}
\end{align}
Therefore, in Case 1, 
\begin{align*}
    &\Bigg|\sum_{\substack{i_1 \neq \cdots \neq i_K \neq j_1 \neq \cdots \neq j_K}} \E{\td(i_1 \dots i_K; \pi(i_1) \dots \pi(i_K))\td(j_1 \dots j_K; \pi(j_1) \dots \pi(j_K))}\Bigg|\\
    \le & \sum_{\substack{i_1 \neq \cdots \neq i_K \neq j_1 \neq \cdots \neq j_K}} \{\mathrm{I}(i_1 \dots i_K) + \mathrm{II}(j_1 \dots j_K)\} \\
    = & (n-K)_{K}\sum_{\substack{i_1 \neq \cdots \neq i_K}} \mathrm{I}(i_1 \dots i_K) + (n-K)_{K}\sum_{\substack{j_1 \neq \cdots \neq j_K}}\mathrm{II}(j_1 \dots j_K)\\
    \le & \frac{K(n-1)_{K-1}(n-K)_{K}}{(n)_{2K}} \sum_{\substack{i_1 \neq \cdots \neq i_K, \\j_1 \neq \cdots \neq j_K}}{\td(i_1 \dots i_K; j_1 \dots j_K)^2}~
    \see{using \eqref{eqn:I-bound} and \eqref{eqn:II-bound}}\\
    = & \frac{K(n-1)_{K-1}(n-K)_{K}(n)_K^2}{(n)_{2K}} \cdot \sum_{\substack{i_1 \neq \cdots \neq i_K, \\j_1 \neq \cdots \neq j_K}}{\td(i_1 \dots i_K; j_1 \dots j_K)^2}\\
    \le & C_K n^{2K - 1} \sum_{\substack{i_1 \neq \cdots \neq i_K, \\j_1 \neq \cdots \neq j_K}}{\td(i_1 \dots i_K; j_1 \dots j_K)^2}. 
\end{align*}

\textbf{Case 2: $i_k = j_l$ for exactly one pair of $k,l\in[K]$, i.e., the summation over $\mathcal{E}_1$. } For simplicity, we consider the case where overlapping occurs at the first index with 
$i_1 = j_1$, and the rest are all different because we can handle the rest cases similarly. We have $|\{i_1,i_2\dots i_K,j_2,\dots,j_K\}| = 2K - 1$. Then
\begin{align*}
    &\E{\td(i_1i_2 \dots i_K; \pi(i_1)\pi(i_2) \dots \pi(i_K))\td(i_1j_2 \dots j_K; \pi(i_1) \pi(j_2) \dots \pi(j_K))}\\
    = & \frac{1}{(n)_{2K-1}} \sum_{\substack{|\{i_1' , i_2', \dots , i_K' , j_2' , \dots , j_K'\}|=2K-1}}\td(i_1i_2 \dots i_K; i'_1i'_2 \dots i'_K)\td(i_1j_2 \dots j_K; i'_1j_2' \dots j'_K)\\
    \le & \frac{1}{2(n)_{2K-1}} \sum_{\substack{|\{i_1' , i_2', \dots , i_K' , j_2' , \dots , j_K'\}|=2K-1}} 
    \{\td(i_1i_2 \dots i_K; i'_1i'_2 \dots i'_K)^2 + \td(i_1j_2 \dots j_K; i'_1j_2' \dots j'_K)^2\}\\
    = & \frac{(n-K)_{K-1}}{(n)_{2K-1}} \sum_{\substack{|\{i_1' , i_2', \dots , i_K'\}|=K}}\td(i_1i_2 \dots i_K; i'_1i'_2 \dots i'_K)^2.
\end{align*}
The last equality holds because, for $\td(i_1i_2 \dots i_K; i'_1i'_2 \dots i'_K)$, fixing a $K$-tuple $(i_1', \dots, i_K')$, the summation has $K-1$ indices $j_2', \dots, j_K'$ that do not appear in the summands. Such $j_2', \dots, j_K'$ can take $K-1$ different values from $[n]\backslash\{i_1', \dots, i_K'\}$, which gives a total of $(n-K)_{K-1}$ combinations. Similar argument also applies to $\td(i_1j_2 \dots j_K; i'_1j_2' \dots j'_K)$. Therefore,
\begin{align*}
    &\sum_{\substack{|\{i_1 , i_2, \dots , i_K , j_2 , \dots , j_K\}|=2K-1}}\lt|\E{\td(i_1i_2 \dots i_K; \pi(i_1)\pi(i_2) \dots \pi(i_K))\td(i_1j_2 \dots j_K; \pi(i_1)\pi(j_2) \dots \pi(j_K))}\rt|\\
    \le & \frac{(n-K)_{K-1}}{(n)_{2K-1}}\sum_{\substack{|\{i_1 , i_2, \dots , i_K , j_2 , \dots , j_K\}|=2K-1}}\sum_{\substack{|\{i_1' , i_2', \dots , i_K'\}|=K}}\td(i_1i_2 \dots i_K; i'_1i'_2 \dots i'_K)^2\\
    \le & 
    \frac{(n-K)_{K-1}^2}{(n)_{2K-1}}\sum_{\substack{|\{i_1 , i_2, \dots , i_K\}|=K, \\|\{i_1' , i_2', \dots , i_K'\}|=K }} \td(i_1i_2 \dots i_K; i'_1i'_2 \dots i'_K)^2\\
    &\see{since given any $i_1,\dots,i_K$, the indices $j_2,\dots,j_K$ did not appear in the summands}.
\end{align*}

\textbf{Case 3: $i_k = j_l$ for more than one pairs of $k,l\in[K]$. } In general, when there are exactly $p$ ($p\ge1$) pairs of identical $(i_k,j_l)$, we can take one representative set of indices:
\begin{align*}
    \mathcal{E}_p' = \{|&\{i_1 , i_2, \dots , i_K\}| = K; |\{j_1, j_2 , \dots , j_K\}| = K; \\ &i_k = j_k, k=1,\dots,p; i_k \neq j_k, k=p+1,\dots,K\}.
\end{align*}
Following a similar argument as Case 2, we can prove
\begin{align}\label{eqn:bound-E-p}
    &\sum_{(i_1\dots i_Kj_1\dots j_K)\in\mathcal{E}_p'}\lt|\E{\td(i_1i_2 \dots i_K; \pi(i_1)\pi(i_2) \dots \pi(i_K))\td(i_1j_2 \dots j_K; \pi(i_1)\pi(j_2) \dots \pi(j_K))}\rt|\\
    \le & C_K n^{2K-p} \sum_{\substack{|\{i_1 , i_2, \dots , i_K\}|=K, \\|
    \{i_1' , i_2', \dots , i_K'\}|=K }} \td(i_1i_2 \dots i_K; i'_1i'_2 \dots i'_K)^2. 
\end{align}
As a remark, \eqref{eqn:bound-E-p} only applies to $p\ge 1$ and cannot cover $p = 0$, because we used a different technique in Case 1 to bound the case with $p = 0$. In general, the cases with $p>1$ contribute smaller order terms to the variance.

Summarizing Cases 1--3, we conclude that
\begin{align*}
    \Var{\Gamma} \le C_K \frac{n^{2K - 1}}{(n)_K^2} \sum_{\substack{|i_1 , i_2, \dots , i_K|=K, \\|i_1' , i_2', \dots , i_K'|=K }} \td(i_1i_2 \dots i_K; i'_1i'_2 \dots i'_K)^2. 
\end{align*}

\end{proof}

\subsubsection{Proof of Lemma \ref{lem:Ppi-converge}}
\begin{proof}[Proof of Lemma \ref{lem:Ppi-converge}]
The first part is a direct application of Chebyshev's inequality. We only prove the second part below.

For variance, we can verify that the demeaned array $\td(i_1\dots i_K;j_1\dots j_K)$ is 
\begin{align}
    \td(i_1\dots i_K;j_1\dots j_K) = \ta(i_1\dots i_K) \tb(j_1\dots j_K).
\end{align}
Applying Lemma \ref{lem:var-array}, we have
\begin{align}
        \Var{U_n} &\le \frac{C_K n^{2K-1}}{(n)_K^2}\lt\{\frac{1}{(n)_K}\sum_{i_1\neq \dots \neq i_K} \ta(i_1,\dots,i_K)^2\rt\} \lt\{\frac{1}{(n)_K}\sum_{i_1\neq \dots \neq i_K} \tb(i_1,\dots,i_K)^2\rt\} \\
        &\le \frac{C_K}{n} \lt\{\frac{1}{(n)_K}\sum_{i_1\neq \dots \neq i_K} \ta(i_1,\dots,i_K)^2\rt\} \lt\{\frac{1}{(n)_K}\sum_{i_1\neq \dots \neq i_K} \tb(i_1,\dots,i_K)^2\rt\}.
\end{align}
    
    Hence, under Condition \eqref{eqn:var-to-0}, Chebyshev's inequality implies
    \begin{align}
        {U_n} - {\E{U_n}} = O_{\bbP}\lt(\sqrt{\Var{{U_n}}}\rt) =  o_{\bbP}(1). 
    \end{align}
\end{proof}

\section{Proofs of the main results}\label{sec:pf-main}

In this section, we prove the theoretical results. We will deal with two different probability measures: one is on the random population data, and the other is the permutation probability measure conditional on the population data. For the population level, let $\bbP$ denote the probability measure and $\bbE$ denote the expectation over $\bbP$. For the permutation level, let $\bbP^\pi$ denote the probability measure over permutations conditional on the population data and $\bbE^\pi$ denote the expectation taken over $\bbP^\pi$. 

\subsection{Lemmas}

\paragraph{Covergence of maxima}
The following lemma establishes the convergence of the maxima of a sequence of random variables, which will be useful for proving results for the asymptotic permutation distributions, including Theorems \ref{thm:asp-mantel-ns}, \ref{thm:asp-mantel-s} and \ref{thm:asp-mantel-s-p}.

\begin{lemma}[Almost sure convergence of maxima, \cite{einmahl2005general}]\label{lem:maxima-as}
Consider a sequence of random variables $\{X_n, n \ge 1\}$. Let ${m(n), n \ge 1}$ be a sequence of monotonically increasing positive constants that diverge to $\infty$. Then
\begin{align}\label{eqn:maxima-condition}
\frac{\max_{1\le i\le n} |X_i|}{m(n)} \to 0 \text{ a.s. } \text{ if and only if } ~ \frac{X_n}{m(n)} \to 0 \text{ a.s. }
\end{align}
A sufficient condition for \eqref{eqn:maxima-condition} is 
\begin{align*}
    \sum_{n=1}^\infty \Prob{|X_n| > \epsilon m(n)} < \infty \quad \text{ for any } \epsilon > 0. 
\end{align*}
Specially, if $m(n) = n$, then a sufficient condition for \eqref{eqn:maxima-condition} is
\begin{align}
    \sup_{n\ge 1}~\E{|X_n|^p} <  \infty, \quad \text{ for some } p > 1.
\end{align}

\end{lemma}

\paragraph{Decomposition of moments.} The following lemma gives a decomposition of the moment $m_{22}(a)$, introduced in \eqref{eqn:dips-moments}, which will be useful for proving Theorem \ref{thm:asp-mantel-s}. 
\begin{lemma}\label{lem:m22a-decomp}
    The following decomposition holds for $m_{22}(a)$:
    \begin{align}
        m_{22}(a) = \frac{1}{n(n-1)}\sum_{i\neq j} (a_{ij} - \oa)^2 
        - 
        \frac{(n-2)}{n(n-1)}\sum_{i} \oa_i^2 
        - 
        \frac{(n-2)}{n(n-1)}\sum_{j} \oa_j^2.
    \end{align}
\end{lemma}
\begin{proof}[Proof of Lemma \ref{lem:m22a-decomp}]
    We can directly verify
    \begin{align}
        m_{22}(a)=&\frac{1}{n(n-1)}\sum_{i\neq j} (a_{ij} - \oa_i - \oa_j - \oa)^2 \\
        = & \frac{1}{n(n-1)}\sum_{i\neq j} (a_{ij} - \oa)^2 + 
        \frac{1}{n(n-1)}\sum_{i\neq j} \oa_i^2 + 
        \frac{1}{n(n-1)}\sum_{i\neq j} \oa_j^2 \\
        - &
        \frac{2}{n(n-1)}\sum_{i\neq j} (a_{ij} - \oa)\oa_i - 
        \frac{2}{n(n-1)}\sum_{i\neq j} (a_{ij} - \oa)\oa_j +  
        \frac{2}{n(n-1)}\sum_{i\neq j} \oa_i\oa_j \\
        = & 
        \frac{1}{n(n-1)}\sum_{i\neq j} (a_{ij} - \oa)^2 
        + 
        \frac{1}{n}\sum_{i} \oa_i^2 
        + 
        \frac{1}{n}\sum_{j} \oa_j^2 \\
        - &
        \frac{2(n-2)}{n(n-1)}\sum_{i} \oa_i^2 
        - 
        \frac{2(n-2)}{n(n-1)}\sum_{j} \oa_j^2 
        -  
        \frac{2}{n(n-1)}\sum_{i} \oa_i^2 \\
        = & 
        \frac{1}{n(n-1)}\sum_{i\neq j} (a_{ij} - \oa)^2 
        - 
        \frac{(n-2)}{n(n-1)}\sum_{i} \oa_i^2 
        - 
        \frac{(n-2)}{n(n-1)}\sum_{j} \oa_j^2.\label{eqn:decomp-xi22}
    \end{align}
\end{proof}

\subsection{Proof of Theorem \ref{thm:asp-super}}
\begin{proof}[Proof of Theorem \ref{thm:asp-super}]
    Recall $\alpha_0 = \E{\alpha(R, R')}$ and $\beta_0 = \E{\beta(S, S')}$. Decompose $\widehat{\phi}_0$ into the following two terms:
\begin{align*}
\hphi_0 &= \frac{1}{n(n-1) - 1} \sum_{i\neq j} (a_{ij} - \alpha_0)(b_{ij} - \beta_0)  - \frac{n(n-1)}{n(n-1) - 1} (\overline{a} - \alpha_0)(\overline{b} - \beta_0).
\end{align*}
We now prove the result using the U-statistic CLT and SLLN in Lemma \ref{lem:clt-u} and Lemma \ref{lem:slln-u}. By the CLT for U-statistics (Lemma \ref{lem:clt-u}), 
\begin{align}
    \frac{\frac{1}{n(n-1) - 1} \sum_{i\neq j} (a_{ij} - \alpha_0)(b_{ij} - \beta_0) - \phi_0}{\sqrt{4\eta_{1,\phi}/n}} \rightsquigarrow \cN(0,1).
\end{align}
By the variance formula of U-statistics (Lemma \ref{lem:mean-var-u}), 
\begin{align}
    \oa - \alpha_0 = O_\bbP\lt(\sqrt{\frac{\eta_{1,\alpha}}{n}}\rt), \quad 
    \ob - \beta_0 = O_\bbP\lt(\sqrt{\frac{\eta_{1,\beta}}{n}}\rt). 
\end{align}
Therefore, we have ${\sqrt{n}(\hphi_0 - \phi_0)} \rightsquigarrow \cN(0, {4{\eta_{1,\phi}}})$. 

Now we prove the asymptotic distribution of $\widehat{\rho}$. Under $\mathrm{H}_{0\textsc{w}}$ in \eqref{eqn:null-qap}, $\phi_0 = 0$. 
By the SLLN for U-statistics (Lemma \ref{lem:slln-u}), the sample variance of $a_{ij}$ and $b_{ij}$ converge almost surely: $\heta_{2,\alpha} \rightarrow \eta_{2,\alpha}$, $\heta_{2,\beta} \rightarrow \eta_{2,\beta}$, $\bbP\text{-a.s.}$ By Slutsky's Theorem, we have
\begin{align}
    \sqrt{n}\hrho = \frac{\sqrt{n}\hphi_0}{\sqrt{\heta_{2,\alpha}}\cdot\sqrt{\heta_{2,\beta}}}
    \rightsquigarrow
    \cN(0, \frac{4\eta_{1,\phi}}{\eta_{2,\alpha}\eta_{2,\beta}}).
\end{align}

Under $\mathrm{H}_{0\textsc{s}}$, we have $R\indep S$, which implies $\eta_{1,\phi} = \eta_{1,\alpha} \eta_{1,\beta}$. 

\end{proof}

\subsection{Proof of Theorem \ref{thm:asp-super-s}}
\begin{proof}[Proof of Theorem \ref{thm:asp-super-s}]
    This is a direct result of Theorem \ref{thm:asp-super} and the variance estimation consistency given by Lemma \ref{lem:var-est-1-dim}. 
\end{proof}

\subsection{Proof of Theorem \ref{thm:asp-mantel-ns}}\label{sec:pf-thm-asp-mantel-ns}
\begin{proof}[Proof of Theorem \ref{thm:asp-mantel-ns}]
The permuted Pearson correlation coefficient is given by
\begin{align}
    \hrho^\pi = \frac{\hphi_0^{\pi}}{(\heta_{2,\alpha}^\pi)^{1/2} (\heta_{2,\beta}^\pi)^{1/2}}.
\end{align}
The proof is divided into several steps.

\textbf{Step I. Convergence of the denominator $\heta_{2,\alpha}$ and $\heta_{2,\beta}$.} First, by the definition of $\heta_{2,\alpha}$ and $\heta_{2,\beta}$ in \eqref{eqn:heta-alpha} and \eqref{eqn:heta-beta}, we can verify that they are invariant under any permutation $\pi\in\bbS_n$:  
\begin{align}\label{eqn:sigma-pi}
    \heta_{2,\alpha}^\pi = \heta_{2,\alpha},\quad \heta_{2,\beta}^\pi = \heta_{2,\beta}.
\end{align}
Based on the SLLN in Lemma \ref{lem:slln-u}, we have 
\begin{align}\label{eqn:eta-as-converge}
    \heta_{2,\alpha} \to \eta_{2,\alpha}, \quad 
    \heta_{2,\beta} \to \eta_{2,\beta} \quad \bbP\text{-a.s.}
\end{align}

\textbf{Step II. Convergence of the numerator $\hphi_0^{\pi}$. }
For $\hphi_0^\pi$, we have
\begin{align}
    \hphi_0^\pi & = \frac{1}{n(n-1)-1}\sum_{i \neq j}(a_{ij} - \overline{a})(b_{\pi(i)\pi(j)} - \overline{b}) \\
    & = \frac{n(n-1)}{n(n-1)-1}\sum_{i \neq j}(a_{ij} - \alpha_0)(b_{\pi(i)\pi(j)} - \beta_0) - \frac{n(n-1)}{n(n-1)-1} (\overline{a} - \alpha_0)(\overline{b} - \beta_0).
\end{align}
Introduce the following quantities based on Lemma \ref{lem:decomp-dips}:
\begin{gather*}
    \oa_{i} = \frac{1}{n-2}\sum_{j:j\neq i}(a_{ij} - \oa), \quad \ta_{ij} = a_{ij} - \oa_i - \oa_j - \oa, \\
    \ob_{i} = \frac{1}{n-2}\sum_{j:j\neq i}(b_{ij} - \ob), \quad \tb_{ij} = b_{ij} - \ob_i - \ob_j - \ob.
\end{gather*}
Further, recall the definition of moments from \eqref{eqn:dips-moments}:
\begin{gather}
    m_{1k}(a) = \frac{1}{n}\sum_{i=1}^n |\overline{a}_i|^k,  \quad m_{2k}(a) =  \frac{1}{n(n-1)} \sum_{i\neq j}|\tilde{a}_{ij}|^k; \\
    m_{1k}(b) = \frac{1}{n}\sum_{i=1}^n |\overline{b}_i|^k,  \quad m_{2k}(b) =  \frac{1}{n(n-1)} \sum_{i\neq j}|\tilde{b}_{ij}|^k; \\
    s_1^2 = \frac{4(n-2)^2}{(n-1)^3}m_{12}(a)m_{12}(b), 
    \quad 
    s_2^2 = \frac{2}{n(n-3)}m_{22}(a) m_{22}(b).
\end{gather}
Based on Lemma \ref{lem:clt-dips}, conditional on $(R_{i}, S_i)_{i=1}^n$, 
\begin{align}\label{eqn:hphi-pi-asp}
    \frac{\sqrt{n}\hphi_0^\pi}{2(\xi_{12}\gamma_{12})^{1/2}} \rightsquigarrow \cN(0,1), 
\end{align}
when the following holds:
\begin{align} \label{eqn:ab-conditions}
    \frac{\max_{i} \overline{a}_i^2 }{n\cdot m_{12}(a) } \to 0, 
    ~ \lim_{d\to\infty}\limsup_{n\to\infty} \frac{1}{n}\sum_{i=1}^n \frac{\overline{b}_i^2}{m_{12}(b)} \ind{\frac{|\overline{b}_i|}{\sqrt{m_{12}(b)}} > d} = 0, 
    ~ \frac{s_1^2}{s_2^2} 
    \to 0.
\end{align}

\textbf{Step III. Verify conditions for the convergence of the numerator.} Now we check that \eqref{eqn:ab-conditions} holds almost surely. 

(i)  We check the first condition in \eqref{eqn:ab-conditions}. This is shown by two results:
    \begin{align}\label{eqn:a-part}
        \frac{\max_{i} \oa_i^2}{n} \rightarrow 0  \quad \bbP\text{-a.s.}
    \end{align}
    and
    \begin{align}\label{eqn:xi-part}
         m_{12}(a) \rightarrow \eta_{1,\alpha}  \quad \bbP\text{-a.s.}
    \end{align}
    To prove the first result \eqref{eqn:a-part}, we notice the bound 
    \begin{align}
        |\oa_i| 
        & = \lt|\frac{1}{n-2}\sum_{j:j\neq i}(a_{ij} - \oa)\rt| \\
        & \le \frac{1}{n-2}\sum_{j:j\neq i}|a_{ij}| + \frac{1}{n-2}|\oa| \\
        & \le \frac{1}{n-2}\sum_{j:j\neq i}\{\kappa_\alpha(S_i) + \kappa_\alpha(S_j)\} + \frac{1}{n-2}\cdot\frac{1}{n(n-1)}\sum_{j\neq k}\{\kappa_\alpha(S_j) + \kappa_\alpha(S_k)\} \see{by \eqref{eqn:kernel-bd}}\\
        & = \kappa_{\alpha}(S_i) + \frac{n+2}{n(n-2)}\sum_{j=1}^n \kappa_\alpha(S_j) \\
        & \le \kappa_{\alpha}(S_i) + \frac{3}{n-2}\sum_{j=1}^n \kappa_\alpha(S_j).
    \end{align}
    Therefore, 
    \begin{align}\label{eqn:a-bound}
        \max_{i} |\oa_i| \le \max_{i}\kappa_{\alpha}(S_i) + \frac{3}{n-2}\sum_{j=1}^n \kappa_\alpha(S_j).
    \end{align}
    Here $\kappa_\alpha(S_i)$'s are i.i.d. with 
    $ \E{\kappa_\alpha(S_i)^2} < \infty $. By the SLLN for the average of i.i.d. variables, 
    \begin{align}\label{eqn:avg-kappa-S}
        \frac{1}{n-2}\sum_{j=1}^n\kappa_\alpha(S_j)\rightarrow \E{\kappa_\alpha(S)} \quad \bbP\text{-a.s.}
    \end{align}
    Hence we have
    \begin{align}
        \frac{\max_i |\oa_i|^2}{n} 
        \le
        \frac{2\max_i |\kappa_\alpha(S_i)|^2}{n} 
        + 
        \frac{18}{n} 
        \lt\{\frac{1}{n-2}\sum_{j=1}^n \kappa_\alpha(S_j)\rt\}^2.
    \end{align}
    Applying Lemma \ref{lem:maxima-as}, the first term above converges to zero. By \eqref{eqn:avg-kappa-S}, the second term also converges to zero. Thus we conclude that 
    \begin{align}\label{eqn:max-oa-as-0}
        \frac{\max_{i} \oa_i^2}{n} \rightarrow 0  \quad \bbP\text{-a.s.}
    \end{align}
    To prove the second result \eqref{eqn:xi-part}, we have decomposition
    \begin{align}\label{eqn:xi12}
        m_{12}(a) = \frac{1}{n(n-1)^2}\sum_{i=1}\lt(\sum_{j\neq i}(a_{ij} - \alpha_0)\rt)^2 - \frac{(n-1)^2}{(n-2)^2} (\oa - \alpha_0)^2  = \ostar_1 + \ostar_2.
    \end{align}
    By the SLLN of U-statistics (Lemma \ref{lem:slln-u}), 
    \begin{align}\label{eqn:ostar2-as}
    \ostar_2 \to 0 \quad \bbP\text{-a.s.} 
    \end{align}
    For $\ostar_1$, we further have
    \begin{align*}
        \ostar_1 &= \frac{1}{n(n-1)^2}\sum_{i=1}\lt(\sum_{j\neq i}(a_{ij} - \alpha_0)\rt)^2\\
        &= \frac{1}{n(n-1)^2}\lt\{\sum_{j\neq i}(a_{ij} - \alpha_0)^2 + \sum_{j\neq k \neq i} (a_{ij} - \alpha_0)(a_{ik} - \alpha_0)\rt\}
    \end{align*}
     which, by the SLLN (Lemma \ref{lem:slln-u}), has
     \begin{align}\label{eqn:ostar1-as}
         \ostar_1 \rightarrow \eta_{1,\alpha} > 0 \quad \bbP\text{-a.s.} 
     \end{align}
    Therefore, by \eqref{eqn:ostar2-as} and  \eqref{eqn:ostar1-as}, 
    \begin{align}\label{eqn:xi-12-as}
        m_{12}(a) \rightarrow \eta_{1,\alpha}  \quad \bbP\text{-a.s.}
    \end{align} 
    Hence we have proved that the first condition in \eqref{eqn:ab-conditions} holds $\bbP\text{-a.s}$.
    
    
(ii) We now prove the second condition in \eqref{eqn:ab-conditions}. Similar to \eqref{eqn:ostar1-as}, we can prove
    \begin{align}\label{eqn:gamma-12-as}
        m_{12}(b)\to \eta_{1,\beta} > 0 \quad \bbP\text{-a.s.}
    \end{align}
    Similar to \eqref{eqn:a-bound}, we have
    \begin{align}\label{eqn:upper-bi}
        |\ob_i| \le \kappa_\beta(S_i) + \frac{3}{n-2}\sum_{j = 1}^n \kappa_\beta(S_j). 
    \end{align}
    
    To show the second condition in \eqref{eqn:ab-conditions}, we have
    \begin{align}
        &\frac{1}{n}\sum_{i=1}^n \frac{\overline{b}_i^2}{m_{12}(b)} \ind{\frac{|\overline{b}_i|}{\sqrt{m_{12}(b)}} > d}\\
        \le & 
        \frac{1}{n}\sum_{i=1}^n \frac{\overline{b}_i^2}{m_{12}(b)} \lt(\ind{\frac{|\overline{b}_i|}{\sqrt{\eta_{1,\beta}}} > d, m_{12}(b) \ge \eta_{1,\beta}} + \ind{m_{12}(b) < \eta_{1,\beta}}\rt)\\
        &\see{because if ${|\overline{b}_i|}/{\sqrt{m_{12}(b)}} > d$, at least one of the two indicators must be 1}\\
        \le &
        \frac{2}{n}\sum_{i=1}^n \frac{2\kappa_\beta^2(S_i) + 18(n-2)^{-2}\{\sum_{j=1}^n \kappa_\beta(S_j)\}^2}{\eta_{1,\beta}} \ind{\frac{|\overline{b}_i|}{\sqrt{\eta_{1,\beta}}} > d}\\
        + &
        \frac{2}{n}\sum_{i=1}^n \frac{2\kappa_\beta^2(S_i) + 18(n-2)^{-2}\{\sum_{j=1}^n \kappa_\beta(S_j)\}^2}{m_{12}(b)}\ind{m_{12}(b) > \eta_{1,\beta}}
        \see{using \eqref{eqn:upper-bi}} \\
        = & \ostar_1 + \ostar_2.
    \end{align}
    Now, by the SLLN for the average of i.i.d. variables, the following results hold:
    \begin{gather}
        \frac{1}{n}\sum_{i=1}^n \kappa_\beta(S_i) \rightarrow \E{\kappa_\beta(S)}, \quad
        \frac{1}{n}\sum_{i=1}^n \kappa_\beta^2(S_i)\rightarrow \E{\kappa_\beta^2(S)} \quad \bbP\text{-a.s.}
    \end{gather}
    Together with \eqref{eqn:gamma-12-as}, we get $\ostar_2 \to 0$ a.s. 

    We now bound $\ostar_1$:  
    \begin{align}    
        \ostar_1\le &
        \frac{2}{n}\sum_{i=1}^n \frac{2\kappa_\beta^2(S_i) + 18(n-2)^{-2}\{\sum_{j=1}^n \kappa_\beta(S_j)\}^2}{\eta_{1,\beta}} 
        \ind{\frac{\kappa_\beta(S_i)}{\sqrt{\eta_{1,\beta}}} > \frac{d}{2}} \\
        + &\frac{2}{n}\sum_{i=1}^n \frac{2\kappa_\beta^2(S_i) + 18(n-2)^{-2}\{\sum_{j=1}^n \kappa_\beta(S_j)\}^2}{\eta_{1,\beta}}\ind{\frac{3\sum_{j=1}^n\kappa_\beta(S_j)}{(n-2)\sqrt{\eta_{1,\beta}}} > \frac{d}{2}}\\
        & \see{because if ${|\overline{b}_i|}/{\sqrt{\eta_{1,\beta}}} > d$, at least one of the two indicators above must be one} \\
        = & \ostar_{1.1} + \ostar_{1.2}. 
    \end{align}
    Now, by the SLLN for the average of i.i.d. variables, the following results hold:
    \begin{gather}
        \frac{1}{n}\sum_{i=1}^n \ind{\frac{\kappa_\beta(S_i)}{\sqrt{\eta_{1,\beta}}} > d/2}\rightarrow \Prob{\frac{\kappa_\beta(S)}{\sqrt{\eta_{1,\beta}}} > d/2} \quad \bbP\text{-a.s.}, \\
        \frac{1}{n}\sum_{i=1}^n \frac{\kappa_\beta^2(S_i)}{\eta_{1,\beta}}\ind{\frac{\kappa_\beta(S_i)}{\sqrt{\eta_{1,\beta}}} > d/2}\rightarrow \E{\frac{\kappa_\beta^2(S)}{\eta_{1,\beta}}\ind{\frac{\kappa_\beta(S)}{\sqrt{\eta_{1,\beta}}} > d/2}}\quad \bbP\text{-a.s.}
    \end{gather}
    Therefore, 
    \begin{align}
        \ostar_{1.1}\rightarrow 4\E{\frac{\kappa_\beta^2(S)}{\eta_{1,\beta}}\ind{\frac{\kappa_\beta(S)}{\sqrt{\eta_{1,\beta}}} > d/2}}
        +
        36\eta_{1,\beta}^{-1} \E{\kappa_\beta^2(S)}\Prob{\frac{\kappa_\beta(S)}{\sqrt{\eta_{1,\beta}}} > d/2} \quad \bbP\text{-a.s.}
    \end{align}
    We can see that $ \lim_{d\to\infty}\lim_{n\to\infty}
    \ostar_{1.1}= 0 $. 
    Moreover,
    \begin{align}
        \ostar_{1.2} \rightarrow 
        \lt(4\E{\frac{\kappa_\beta^2(S)}{\eta_{1,\beta}}}
        +
        36\eta_{1,\beta}^{-1} \E{\kappa_\beta^2(S)}\rt)\ind{\frac{3\E{\kappa_\beta(S)}}{\sqrt{\eta_{1,\beta}}} > d/2} \quad \bbP\text{-a.s.}
    \end{align}
    Hence, $ \lim_{d\to\infty}\lim_{n\to\infty}
    \ostar_{1.2}= 0 $. 
    Therefore, 
    \begin{align}
        \lim_{d\to\infty}\limsup_{n\to\infty}\frac{1}{n}\sum_{i=1}^n \frac{\overline{b}_i^2}{m_{12}(b)} \ind{\frac{|\overline{b}_i|}{\sqrt{m_{12}(b)}} > d}
        \le 
        \lim_{d\to\infty}\lim_{n\to\infty}
        (\ostar_{1.1} + \ostar_{1.2} + \ostar_{2}) = 0.
    \end{align}
    This verifies the second condition in \eqref{eqn:ab-conditions}. 

    (iii) Now we prove the third condition in \eqref{eqn:ab-conditions}. 
    By the decomposition in Lemma \ref{lem:m22a-decomp} and the SLLN of U-statistic in Lemma \ref{lem:slln-u}, the following limit holds for \eqref{eqn:decomp-xi22}:
    \begin{align}\label{eqn:xi-gamma-asp}
        m_{22}(a)\to \eta_{2,\alpha} - 2\eta_{1,\alpha} <\infty, \quad m_{22}(b)\to \eta_{2,\beta} - 2\eta_{1,\beta} <\infty \quad\bbP\text{-a.s.}
    \end{align}
    Hence for the third part of \eqref{eqn:ab-conditions}, we have
    \begin{align*}
        \frac{s_2^2}{s_1^2} = \frac{\frac{2}{n(n-3)}m_{22}(a)m_{22}(b)}{\frac{4(n-2)^2}{(n-1)^3}m_{12}(a)m_{12}(b)} \asymp \frac{1}{2n}\frac{(\eta_{2,\alpha} - 2\eta_{1,\alpha})(\eta_{2,\beta} - 2\eta_{1,\beta})}{\eta_{1,\alpha}\eta_{1,\beta}} \rightarrow 0 \quad\bbP\text{-a.s.}
    \end{align*}

\textbf{Step IV. Summarizing the results.} Combining \eqref{eqn:sigma-pi}, \eqref{eqn:eta-as-converge} and the fact that \eqref{eqn:ab-conditions} holds almost surely, we obtain that, under $\mathrm{H}_{0\textsc{w}}$ in \eqref{eqn:null-qap}, 
\begin{align*}
    \frac{\sqrt{n}\hphi_0^\pi}{({\heta_{2,\alpha}^\pi})^{1/2} \cdot ({\heta_{2,\beta}^\pi})^{1/2}} \rightsquigarrow \cN\lt(0, \frac{4\eta_{1,\alpha}\eta_{1,\beta}}{\eta_{2,\alpha}\eta_{2,\beta}}\rt) \quad\bbP\text{-a.s.}
\end{align*}
By Polya's theorem \citep{bickel1992uniform}, we have
\begin{align*}
    \lim_{n\to\infty} \sup_{t\in\bbR} \lt|\cL(t;{\sqrt{n}\hrho^\pi}) - \cL(t; \cN(0,v_\textsc{s}))\rt| = 0 \quad\bbP\text{-a.s.}
\end{align*}

\end{proof}

\subsection{Proof of Theorem \ref{thm:asp-mantel-s}}

\begin{proof}[Proof of Theorem \ref{thm:asp-mantel-s}]

To analyze the permutation distribution of the studentized $\widehat{\rho}$, we only need to further analyze $\hv^\pi=4\heta_{1,\phi}^\pi/(\heta^\pi_{2,\alpha} \heta^\pi_{2,\alpha})$ because \eqref{eqn:hphi-pi-asp}  has established the asymptotic normality of $\hphi_0^\pi$, and \eqref{eqn:sigma-pi} has established the invariance of $\heta_{2,\alpha}^\pi$ and $\heta_{2,\beta}^\pi$ under permutation.

Now we analyze the limit of $\heta_{1,\phi}^\pi$.

\textbf{Step 1. Decomposing $\heta_{1,\phi}^\pi$.} Recall that we have
    \begin{align*}
      \heta_{1,\phi}^\pi & = \frac{n-1}{(n-2)(n-4)}\sum_{i=1}^n \lt(\frac{1}{n-1}\sum_{j\neq i}^n ({a}_{ij} - \overline{a})({b}_{\pi(i)\pi(j)} - \overline{b})\rt)^2\\
      & = \frac{n(n-1)}{(n-1)(n-2)(n-4)}\cdot \underbrace{\frac{1}{n(n-1)} \sum_{i\neq j}({a}_{ij} - \overline{a})^2({b}_{\pi(i)\pi(j)} - \overline{b})^2}_{\text{Term I}} \\
      & + \frac{n(n-1)(n-2)}{(n-1)(n-2)(n-4)}\cdot\underbrace{\frac{1}{n(n-1)(n-2)}\sum_{i\neq j\neq k} ({a}_{ij} - \overline{a})({a}_{ik} - \overline{a})({b}_{\pi(i)\pi(j)} - \overline{b}) ({b}_{\pi(i)\pi(k)} - \overline{b})}_{\text{Term II}},
    \end{align*}
    with
    \begin{gather}
        \E[][\pi]{\text{Term I}} = \frac{1}{n(n-1)} \sum_{i\neq j} ({a}_{ij} - \overline{a})^2 \cdot \frac{1}{n(n-1)}\sum_{i\neq j}({b}_{ij} - \overline{b})^2, \\
        \E[][\pi]{\text{Term II}} = \frac{1}{n(n-1)(n-2)} \sum_{i\neq j\neq k} ({a}_{ij} - \overline{a})({a}_{ik} - \overline{a}) \cdot \frac{1}{n(n-1)(n-2)}\sum_{i\neq j\neq k}({b}_{ij} - \overline{b}) ({b}_{ik} - \overline{b}).
    \end{gather}
    By the SLLN of U-statistics (Lemma \ref{lem:slln-u}), the following results hold almost surely: 
    \begin{gather*}
        \frac{1}{n(n-1)}\sum_{i\neq j} (a_{ij} - \oa)^2 \rightarrow \E{\tilde{\alpha}_2(S,S')^2} = \eta_{2,\alpha} > 0, \\
        \frac{1}{n(n-1)}\sum_{i\neq j} (b_{ij} - \ob)^2 \rightarrow \E{\tilde{\beta}_2(R,R')^2} = \eta_{2,\beta} > 0,\\
        \frac{1}{n(n-1)(n-2)}\sum_{i\neq j\neq k} ({a}_{ij} - \overline{a})({a}_{ik} - \overline{a}) \rightarrow \E{\tilde{\alpha}_2(S,S')\tilde{\alpha}_2(S,S'')} = \eta_{1,\alpha} > 0, \\
        \frac{1}{n(n-1)(n-2)}  \sum_{i\neq j\neq k} ({b}_{ij} - \overline{b})({b}_{ik} - \overline{b})
        \rightarrow \E{\tilde{\beta}_2(R,R')\tilde{\beta}_2(R,R'')} = \eta_{1,\beta} > 0.
    \end{gather*}
    Hence, it holds almost surely that
    \begin{align}\label{eqn:Epi-I-II-as}
        \E[][\pi]{\text{Term I}} \rightarrow \eta_{2,\alpha}\eta_{2,\beta}, \quad 
        \E[][\pi]{\text{Term II}} \rightarrow \eta_{1,\alpha}\eta_{1,\beta}.
    \end{align}
    
    \textbf{Step 2. Limit of  $\heta^\pi_{1,\phi}$.} Now we show that
    \begin{align}\label{eqn:hsigma1P-P}
    {\heta^\pi_{1,\phi}} - \eta_{1,\alpha}\eta_{1,\beta} = o_{\bbP^\pi}(1) \quad\bbP\text{-a.s.}
    \end{align}
    
    For Term I, Lemma \ref{lem:Ppi-converge} suggests that
    \begin{align}\label{eqn:term-I-asp}
        \lt|\text{Term I} - \E[][\pi]{\text{Term I}}\rt| = o_{\bbP^\pi}(1)
    \end{align}
    under the condition
    \begin{gather}\label{eqn:suff-term-I}
    \frac{1}{n} { \lt\{\frac{1}{n(n-1)}\sum_{i\neq j} (a_{ij} - \oa)^4\cdot \frac{1}{n(n-1)}\sum_{i\neq j} (b_{ij} - \ob)^4\rt\}} \rightarrow 0.
    \end{gather}
    \eqref{eqn:suff-term-I} holds almost surely because the SLLN for U-statistic ensures
    \begin{align*}
        \frac{1}{n(n-1)}\sum_{i\neq j} (a_{ij} - \oa)^4 \rightarrow \E{\tilde{\alpha}_2(S,S')^4}, \quad
        \frac{1}{n(n-1)}\sum_{i\neq j} (b_{ij} - \ob)^4 \rightarrow \E{\tilde{\beta}_2(R,R')^4}.
    \end{align*}
    
    For Term II, Lemma \ref{lem:Ppi-converge} ensures that 
    \begin{align}\label{eqn:term-II-asp}
        {\text{Term II} - \E[][\pi]{\text{Term II}}} = o_{\bbP^\pi}(1),
    \end{align}
    if 
    \begin{gather}\label{eqn:suff-term-II}
    \frac{1}{n} {\lt\{ \frac{1}{(n)_3}\sum_{i\neq j\neq k} ({a}_{ij} - \overline{a})^2({a}_{ik} - \overline{a})^2 \cdot \frac{1}{(n)_3}\sum_{i\neq j\neq k}  ({b}_{ij} - \overline{b})^2({b}_{ik} - \overline{b})^2 \rt\}} \rightarrow 0.
    \end{gather}
    \eqref{eqn:suff-term-II} holds almost surely because the following limits are true by the SLLN for exchangeably dissociated arrays (Lemma \ref{lem:slln-ed}):
    \begin{gather}
        \frac{1}{n(n-1)(n-2)}\sum_{i\neq j\neq k} ({a}_{ij} - \overline{a})^2({a}_{ik} - \overline{a})^2 \rightarrow \E{\tilde{\alpha}_2(S,S')^2\tilde{\alpha}_2(S,S'')^2}, \\
        \frac{1}{n(n-1)(n-2)}\sum_{i\neq j\neq k}  ({b}_{ij} - \overline{b})^2({b}_{ik} - \overline{b})^2\rightarrow \E{\tilde{\beta}_2(R,R')^2\tilde{\beta}_2(R,R'')^2}. 
    \end{gather}
    To see this, we use the first result for an illustration. It has the same order as 
    \begin{align}
        \frac{1}{n(n-1)(n-2)}\sum_{i\neq j\neq k} \talpha(S_i,S_j)^2\talpha(S_i,S_k)^2. 
    \end{align}
    We can check that $\talpha(S_i,S_j)^2\talpha(S_i,S_k)^2$'s form an exchangeably dissociated array:
\begin{itemize}
    \item \textit{Independence}. Consider a pair of disjoint subsets $\cJ$ and $\cK$. For $J\in\cJ$ and $K\in\cK$, they involve different indices, which means they are defined based on two different subsets of $S_i$'s. Hence they are independent. 

    \item \textit{Exchangeability.} The $S_i$'s are i.i.d. variables, thus are exchangeable. Therefore, under any permutation $\pi:\bbN_+\to\bbN_+$, $(S_i)_{i=1}^\infty$ always has the same distribution as $(S_{\pi(i)})_{i=1}^\infty$. Hence, $\talpha(S_i,S_j)^2\talpha(S_i,S_k)^2$'s are also exchangeable. 
\end{itemize}
Therefore, by Lemma \ref{lem:slln-ed}, we have
\begin{align}
    \frac{1}{n(n-1)(n-2)}\sum_{i\neq j\neq k} ({a}_{ij} - \overline{a})^2({a}_{ik} - \overline{a})^2 \rightarrow \E{\tilde{\alpha}_2(S,S')^2\tilde{\alpha}_2(S,S'')^2}. 
\end{align}

    By \eqref{eqn:Epi-I-II-as}, \eqref{eqn:suff-term-I} and \eqref{eqn:suff-term-II}, we conclude that \eqref{eqn:hsigma1P-P} is correct. 

    \textbf{Step 3. Asymptotic distribution of $\hrho^\pi$.} Using \eqref{eqn:Epi-I-II-as}, as well as \eqref{eqn:xi-12-as}, \eqref{eqn:gamma-12-as}, 
    \begin{align}\label{eqn:delta-asp}
        \frac{\E[][\pi]{\text{Term II}}}{m_{12}(a)m_{12}(b)} \rightarrow 1, \quad\bbP\text{-a.s.}
    \end{align}
    By \eqref{eqn:hsigma1P-P}, \eqref{eqn:delta-asp} and Slutsky's Theorem, conditional on $(R_i,S_i)_{i=1}^n$,
    \begin{align}\label{eqn:student-asp}
    \frac{\sqrt{n}\hrho^\pi}{(\hv^\pi)^{1/2}} 
    = 
    \frac{\sqrt{n}\hphi_0^\pi}{2(\heta_{1,\phi}^\pi)^{1/2}}
    =
    \frac{\frac{\sqrt{n}\hphi_0^\pi}{2(m_{12}(a)m_{12}(b))^{1/2}}}{\frac{2(\heta_{1,\phi}^\pi)^{1/2}}{2(\E[][\pi]{\text{Term II}})^{1/2}}}\cdot \frac{2(m_{12}(a)m_{12}(b))^{1/2}}{2(\E[][\pi]{\text{Term II}})^{1/2}}
    \rightsquigarrow \cN(0,1). 
    \end{align}

    Thus, 
    \begin{align*}
        \frac{\sqrt{n}\hrho^\pi}{(\hv^\pi)^{1/2}} \rightsquigarrow \cN(0,1) \quad \bbP\text{-a.s.}
    \end{align*}

\end{proof}

\subsection{Proof of Theorem \ref{thm:asp-super-ns-p}}
\begin{proof}[Proof of Theorem \ref{thm:asp-super-ns-p}]

Under Assumption \ref{asp:noise-lm}, the OLS estimator satisfies
    \begin{align*}
        \hvartheta - \vartheta = \hSigma_{bb}^{-1}\hSigma_{be}. 
    \end{align*}
    Applying the SLLN for U-statistics in Lemma \ref{lem:slln-u} and the continuous mapping theorem, we have 
    \begin{align}\label{eqn:hSigma-bb-as}
        \hSigma_{bb}^{-1} \rightarrow \Sigma_{bb}^{-1}, \quad \hSigma_{be} \rightarrow 0 \quad\bbP\text{-a.s.}
    \end{align}
    We need a more delicate decomposition to analyze the asymptotic distribution of $\hvartheta$. Decompose $\hSigma_{be}$ into
    \begin{align*}
        \hSigma_{be} = \hSigma_{b\epsilon} + \hSigma_{b\zeta},
    \end{align*}
    where
    \begin{gather*}
        \hSigma_{b\epsilon} = \lt(\frac{1}{n(n-1)}\sum_{i\neq j} (\epsilon_{ij} - \oepsilon)(b_{kij} - \ob_k)\rt)_{k\in[p]}, \\
        \hSigma_{b\zeta} = \lt(\frac{1}{n(n-1)}\sum_{i\neq j} (\zeta_{ij} - \ozeta)(b_{kij} - \ob_k)\rt)_{k\in[p]}.
    \end{gather*}
    Notice that
    \begin{align*}
        \hSigma_{b\zeta}^{(k)} = \frac{1}{n(n-1)}\sum_{i\neq j} (\zeta_{ij} - \ozeta)(b_{kij} - \ob_k) = \underbrace{\frac{1}{n(n-1)}\sum_{i\neq j} \zeta_{ij}(b_{kij} - \beta_{k0})}_{{\text{Term I}}} - \underbrace{\ozeta (\ob_k - \beta^{(k)}_{0}) \vphantom{\sum_{i\neq j}}}_{\text{Term II}}.
    \end{align*}
    For Term I, we compute its variance:
    \begin{align}
        \Var{\text{Term I}}
        = &
        \frac{1}{n^2(n-1)^2}\E{\sum_{i\neq j}\sum_{u\neq v} \zeta_{ij}\zeta_{uv}(b_{kij} - \beta_{k0})(b_{kuv} - \beta_{k0})}\\
        = &
        \frac{1}{n^2(n-1)^2}\sum_{i\neq j} \E{\zeta_{ij}^2}\E{(b_{kij} - \beta_{k0})^2}\\
        \asymp & O(n^{-2}).
    \end{align}
    By Chebyshev's inequality, we have
    \begin{align}
        \text{Term I} = O_\bbP(n^{-1}).
    \end{align}
    Analogously, because $ \ozeta = O_\bbP(n^{-1}) $ and $ \ob_k - \beta^{(k)}_{0} = O_\bbP(n^{-1/2})$, we have
    \begin{align}
        \text{Term II} = O_\bbP(n^{-3/2}).
    \end{align}

Under $\mathrm{H}_{0\textsc{w}}$ in \eqref{eqn:null-mrqap-full}, applying the CLT for multi-dimensional U-statistic ({Lemma \ref{lem:clt-dips-md}}), we have
    \begin{align}\label{eqn:hSigma-ab-clt}
        \sqrt{n}\hSigma_{be} \rightsquigarrow \cN(0, 4H_{1,\phi})
    \end{align}

    Hence by \eqref{eqn:hSigma-bb-as} and \eqref{eqn:hSigma-ab-clt}, we have
    \begin{align}
        \sqrt{n} \hvartheta \rightsquigarrow \cN(0, 4\Sigma_{bb}^{-1}H_{1,\phi}\Sigma_{bb}^{-1}).
    \end{align}

The second part holds because under the strong null hypothesis that $(R, G) \indep S$, we have $H_{1,\phi} = \eta_{1,\epsilon} H_{1,\beta}$.
\end{proof}

\subsection{Proof of Theorem \ref{thm:asp-super-s-p}}
\begin{proof}[Proof of Theorem \ref{thm:asp-super-s-p}]
Recall that $\hV = \hSigma_{bb}^{-1} \hH_{1,\phi} \hSigma_{bb}^{-1} $ and that $\hSigma_{bb} \to \Sigma_{bb}$ by SLLN of U-statistics (Lemma \ref{lem:slln-u}). It remains to show that $\hH_{1,\phi}\to H_{1,\phi}$. Under $\mathrm{H}_{0\textsc{w}} $,
we have the following decomposition:
\begin{align}
    \hat{\phi}_{1,i} 
    =& 
    \frac{1}{n-1}\sum_{j: j\neq i} \{y_{ij} - \oy - (\bsb_{ij} - \overline{\bsb})^\top \hvartheta\} (\bsb_{ij} - \overline{\bsb})\\
    =&
    \frac{1}{n-1}\sum_{j: j\neq i} \{e_{ij} - (\bsb_{ij} - \overline{\bsb})^\top \hvartheta\} (\bsb_{ij} - \overline{\bsb}).
\end{align}
Then
\begin{align}
    \hH_{1,\phi} =& \frac{n-1}{(n-2)(n-4)} \sum_{i=1}^n \hat{\phi}_{1,i}\hat{\phi}_{1,i}^\top\\
    =& \frac{1}{(n-1)(n-2)(n-4)}
    \sum_{i=1}^n 
    \lt\{\sum_{j:j\neq i} e_{ij}(\bsb_{ij} - \overline{\bsb})\rt\}
    \lt\{\sum_{j:j\neq i} e_{ij}(\bsb_{ij} - \overline{\bsb})\rt\}^\top \\
    -& \frac{2}{(n-1)(n-2)(n-4)}
    \sum_{i=1}^n 
    \lt\{\sum_{j:j\neq i} e_{ij}(\bsb_{ij} - \overline{\bsb})\rt\} \hvartheta^\top 
    \lt\{\sum_{j: j\neq i} (\bsb_{ij} - \overline{\bsb})(\bsb_{ij} - \overline{\bsb})^\top \rt\}\\
    +& \frac{2}{(n-1)(n-2)(n-4)}
    \sum_{i=1}^n 
    \lt\{\sum_{j: j\neq i} (\bsb_{ij} - \overline{\bsb})(\bsb_{ij} - \overline{\bsb})^\top\rt\} \hvartheta \hvartheta^\top
    \lt\{\sum_{j: j\neq i} (\bsb_{ij} - \overline{\bsb})(\bsb_{ij} - \overline{\bsb})^\top \rt\} \\
    =& \ostar_1 - \ostar_2 + \ostar_3 . 
\end{align}
Consider the vector kernel:
\begin{align}
\Phi(s,r;s',r') = 
    \begin{pmatrix}
        \epsilon(r,r')\boldsymbol{\tbeta}(s,s') \\
        \tbeta^{(1)}(s,s')\boldsymbol{\tbeta}(s,s') \\
        \vdots \\
        \tbeta^{(p)}(s,s')\boldsymbol{\tbeta}(s,s') \\
    \end{pmatrix} \in \bbR^{p+p^2}. 
\end{align}
We also write $\Phi = (\phi, ~\boldsymbol{\tbeta}_{\otimes})$, because the first $p$ elements of $\Phi$ corresponds to the kernel $\phi$, while the rest $p^2$ corresponds to a Kronecker product of two $\boldsymbol{\tbeta}$ kernels. Now we partition the first-order covariance of $\Phi$ as follows:
\begin{align}
    H_{1,\Phi} = 
    \begin{pmatrix}
        H_{1, \phi} & H_{1, (\phi\boldsymbol{\tbeta}_{\otimes})} \\
        H_{1, (\boldsymbol{\tbeta}_{\otimes} \phi)} & H_{1, \boldsymbol{\tbeta}_{\otimes}}
    \end{pmatrix}
\end{align}
For $\ostar_1$, following a similar argument as the proof for Theorem \ref{thm:asp-super-s}, we can obtain
\begin{align}\label{eqn:prob-limit}
    \ostar_1 \to H_{1,\phi}.
\end{align}
Now we analyze $\ostar_2$ and $\ostar_3$. Let $\operatorname{vec}(\cdot)$ denotes the vectorization of a matrix by stacking its columns into a vector. By the property of Kronecker product, we can obtain 
\begin{gather}
    \operatorname{vec}(\ostar_2) 
    =
    \underbrace{\frac{2}{(n-1)(n-2)(n-4)}
    \sum_{i=1}^n 
    \lt\{\sum_{j: j\neq i} (\bsb_{ij} - \overline{\bsb})(\bsb_{ij} - \overline{\bsb})^\top \otimes 
    \sum_{j:j\neq i} e_{ij}(\bsb_{ij} - \overline{\bsb})\rt\}}_{\ostar_{2.1}} \hvartheta, \\
    \operatorname{vec}(\ostar_3) 
    =
    \underbrace{\frac{2}{(n-1)(n-2)(n-4)}
    \sum_{i=1}^n 
    \lt\{\sum_{j: j\neq i} (\bsb_{ij} - \overline{\bsb})(\bsb_{ij} - \overline{\bsb})^\top \otimes 
   \sum_{j: j\neq i} (\bsb_{ij} - \overline{\bsb})(\bsb_{ij} - \overline{\bsb})^\top\rt\}}_{\ostar_{3.1}} \hvartheta \hvartheta^\top.
\end{gather}
Following a similar argument as Theorem \ref{thm:asp-super-s}, we can obtain 
\begin{align}
    \ostar_{2.1} \rightarrow H_{1,\phi\boldsymbol{\tbeta}_{\otimes}},  
    \quad 
    \ostar_{3.1} \rightarrow 
    H_{1,\boldsymbol{\tbeta}_{\otimes}} \quad \quad \bbP\text{-a.s.}
\end{align}
Meanwhile, under $\mathrm{H}_{0\textsc{w}}$, as we stated in \eqref{eqn:hSigma-bb-as} by SLLN, $\hvartheta \to 0~~\bbP\text{-a.s.}$
Therefore, combining \eqref{eqn:prob-limit}, we conclude $\hH_{1,\phi} \to H_{1,\phi}$, $\bbP\text{-a.s.}$ 

Under $\mathrm{H}_{0\textsc{w}}$ in \eqref{eqn:null-mrqap-full}, this further establishes the convergence of the variance estimator: $\hV \to V_{\textsc{w}} ~~ \bbP\text{-a.s.}$ Combining this with the asymptotic distribution for $\hvartheta$ in Theorem \ref{thm:asp-super-ns-p}, we use the continuous mapping theorem to conclude that the Wald statistic converges to a $\chi^2_p$ distribution.

\end{proof}

\subsection{Proof of Theorem \ref{thm:asp-mantel-s-p}}
The proof is conducted in two steps. First, we establish the asymptotic permutation distribution of $\sqrt{n} \hvartheta^\pi$. Second, we establish the limit of $\hV^\pi$ under the permutation distribution. Combining these two steps concludes the proof. 

\begin{proof}[Proof of Theorem \ref{thm:asp-mantel-s-p}]
  \textbf{Limiting distribution of $\sqrt{n} \hvartheta^\pi$.} We first show
    \begin{align}\label{eqn:asp-hu-pi}
    \lim_{n\to\infty}\sup_{t\in\bbR}|\cL(t;\sqrt{n}\hvartheta^\pi) - \cL(t;\cN(0, 4 \eta_{1,\epsilon} \Sigma_{bb}^{-1} H_{1,\beta} \Sigma_{bb}^{-1}))|=0
    \quad\bbP\text{-a.s.}
    \end{align}

    Conditional on the data, $\sqrt{n} \hvartheta^\pi= \sqrt{n}(\hSigma_{bb}^\pi)^{-1}\hSigma_{be}^\pi$, where 
    \begin{align*}
        \hSigma_{bb}^\pi = \hSigma_{bb}, \quad \hSigma_{be}^\pi = \hSigma_{b\epsilon}^\pi + \hSigma_{b\zeta}^\pi.
    \end{align*}
\textbf{Step I.}    We will apply the CLT for multi-dimensional DIPS (Lemma \ref{lem:clt-dips-md}) to $\sqrt{n}\hvartheta^\pi$. We define $y_{ij} = \epsilon_{ij}$, $x_{ij}^{(k)} = b_{kij} - \beta_{k0}$.
    Following the notation and definition in Section \ref{sec:dips-md}, we also define the moments $m_{12}(y)$, $m_{22}(y)$,  $m_{12}^{(kl)}(x)$, $m_{22}^{(k)}(x)$.
    The following facts hold almost surely as $n\to\infty$:
\begin{enumerate}
    \item $m_{12}(y)\rightarrow \xi_{12}(y) = \eta_{1,\epsilon} > 0$, $m_{22}(y)\rightarrow \xi_{22}(y) = \eta_{2,\epsilon} - 2\eta_{1,\epsilon}$. For $k,l\in[p]$, 
    $m_{12}^{(kl)}(x) \rightarrow \xi_{12}^{(kl)}(x) = \eta_{1,\beta}^{(kl)}$,  $m_{22}^{(k)}(x)\rightarrow \xi_{22}^{(k)}(x) = \eta_{2,\beta}^{(kk)} - 2 \eta_{1,\beta}^{(kk)}$. 
    
    \item For $k\in[p]$, $ {\max_{i} |\overline{x}_i^{(k)}| }/{\sqrt{n}} \to 0 $;
    
    \item $
    \lim_{d\to\infty}\limsup_{n\to\infty} {n^{-1}}\sum_{i=1}^n {\overline{y}_i^2}  \ind{ {|\overline{y}_i|} > d} = 0
    $.
\end{enumerate}
The justification of these three facts is similar to the proof for Theorem \ref{thm:asp-mantel-ns} in Section \ref{sec:pf-thm-asp-mantel-ns}. See the proof following the argument \eqref{eqn:ab-conditions}. 

\textbf{Step II.} Besides, we need to show that
\begin{align}
    \sqrt{n}\hSigma_{b\zeta}^\pi = o_{\bbP^\pi}(1), \quad  \bbP\text{-a.s.}
\end{align}
Notice that
\begin{align*}
    (\hSigma_{b\zeta}^{(k)})^\pi &= \frac{1}{n(n-1)}\sum_{i\neq j} (\zeta_{\pi(i)\pi(j)} - \ozeta)(b_{kij} - \ob_k) \\
    &= \underbrace{\frac{1}{n(n-1)}\sum_{i\neq j} \zeta_{\pi(i)\pi(j)}(b_{kij} - \beta_{k0})}_{{\text{Term I}}} - \underbrace{\ozeta (\ob_k - \beta^{(k)}_{0}) \vphantom{\sum_{i\neq j}}}_{\text{Term II}}.
\end{align*}
For Term I, by Lemma \ref{lem:decomp-dips}, we can compute
\begin{align}\label{eqn:varpi-zeta-b}
    \Varpi{\text{Term I}} = \underbrace{\frac{4(n-2)^2}{(n-1)^3}m_{12}(\zeta)m^{(k)}_{12}(x)}_{ s_1^2(n) } + 
      \underbrace{\frac{2}{n(n-3)}m_{22}(\zeta)m^{(k)}_{22}(x) }_{ s_2^2(n) }, 
\end{align}
Recalling that $m_{12}(\zeta)$ and $m_{22}(\zeta)$ are the moments:
\begin{align}
    m_{12}(\zeta) = \frac{1}{n}\sum_{i=1}^n \ozeta_{i}^2, \quad m_{22}(\zeta) = \frac{1}{n(n-1)} \sum_{i\neq j}\tzeta_{ij}^2.
\end{align}
Similar to the proofs of \eqref{eqn:xi-12-as}, \eqref{eqn:gamma-12-as} and \eqref{eqn:xi-gamma-asp}, by the SLLN of U-statistics (Lemma \ref{lem:slln-u}), we have 
\begin{align}
    m^{(k)}_{12}(x) \rightarrow \eta^{(kk)}_{1,\beta}, \quad m^{(k)}_{22}(x) \rightarrow \eta_{2,\beta}^{(kk)} - 2\eta^{(kk)}_{1,\beta} \quad \bbP\text{-a.s.}
\end{align}

For $\delta_{12}$ and $\delta_{22}$, we can show that
\begin{align}
    \sqrt{n}\delta_{12} \rightarrow 0, \quad \delta_{22} \rightarrow \E{\zeta^2} \quad \bbP\text{-a.s.}
\end{align}
The second convergence result above is a direct result of the SLLN for the average of i.i.d. random variables. The first convergence result above is derived based on the variance computation
\begin{align}
     &\Var{\frac{1}{\sqrt{n}}\sum_{i=1}^n \lt(\frac{1}{n-2}\sum_{j:j\neq i}\zeta_{ij}\rt)^2} \\
    \le & \frac{1}{n}\sum_{i=1}^n \E{\lt(\frac{1}{n-2}\sum_{j:j\neq i}\zeta_{ij}\rt)^4} \\
    =  &\frac{(n-1)\E{\zeta^4} + 6(n-1)(n-2)(\E{\zeta^2})^2}{n^2(n-2)^2} = O(\frac{1}{n^2}). 
\end{align}
By the Borel--Cantelli Lemma, 
\begin{align}
    \frac{1}{\sqrt{n}}\sum_{i=1}^n \lt(\frac{1}{n-2}\sum_{j:j\neq i}\zeta_{ij}\rt)^2 \rightarrow 0 \quad \bbP\text{-a.s.}
\end{align}
Going back to \eqref{eqn:varpi-zeta-b}, we now have
\begin{align*}
    \Varpi{\sqrt{n}\cdot \text{Term I}} \rightarrow 0 \quad \bbP\text{-a.s.}
\end{align*}
Similarly, we can show that for Term II, 
\begin{align}
    \sqrt{n} \cdot \text{Term II} \rightarrow 0 \quad \bbP\text{-a.s.} 
\end{align}

Now applying Lemma \ref{lem:clt-dips-md}, under $\mathrm{H}_{0\textsc{w}}$ in \eqref{eqn:null-mrqap-full}, we can prove that the following holds:
    \begin{align*}
        \lim_{n\to\infty}\sup_{t\in\bbR}|\cL(t; \sqrt{n} \hvartheta^\pi) - \cL(t;\cN(0, V_\textsc{s}))|
        \to 0 \quad \bbP\text{-a.s.}
    \end{align*}
 

  
\textbf{Limit of the variance estimator $\hV^\pi$ under permutation.} Now we verify that conditional on the data, the permutation variance estimator, $\hV^\pi$, almost surely satisfies
\begin{align}
    \hV^\pi - 4\eta_{1,\epsilon} H_{1,\beta} = o_{\bbP^\pi}(1) \quad \bbP\text{-a.s.}
\end{align}

Recall the definition \eqref{eqn:hV}. Now we have
\begin{align}\label{eqn:hV-pi-limit}
    \hSigma_{bb}^\pi = \hSigma_{bb}.
\end{align}
It remains to show the convergence of
\begin{align}
    \hH_{1,\phi}^\pi &= \frac{n-1}{(n-2)(n-4)} \sum_{i=1}^n \hat{\phi}_{1,i}^\pi(\hat{\phi}_{1,i}^\pi)^\top \\
    & = \frac{n-1}{(n-2)(n-4)} \sum_{i=1}^n \lt\{\frac{1}{n-1}\sum_{j: j\neq i} e_{\pi(i)\pi(j)} (\bsb_{ij} - \overline{\bsb})\rt\} \lt\{\frac{1}{n-1}\sum_{j: j\neq i} e_{\pi(i)\pi(j)} (\bsb_{ij} - \overline{\bsb})\rt\}^\top \\
    & = \frac{1}{(n-1)(n-2)(n-4)} \sum_{i\neq j} e_{\pi(i)\pi(j)}^2 (\bsb_{ij} - \overline{\bsb})(\bsb_{ij} - \overline{\bsb})^\top \\
    & + 
    \frac{1}{(n-1)(n-2)(n-4)}  \sum_{i\neq j\neq k} e_{\pi(i)\pi(j)}e_{\pi(i)\pi(k)} (\bsb_{ij} - \overline{\bsb})(\bsb_{ik} - \overline{\bsb})^\top.\label{eqn:hH-pi}
\end{align}

By looking at each entry of $\hH_{1,\phi}^\pi$ separately, the rest of the proof is analogous to the proof of Theorem \ref{thm:asp-mantel-s}. The extension of the proof relies on the fact that the following results hold $\bbP\text{-a.s.}$:
\begin{eqnarray}
    \frac{1}{n(n-1)}\sum_{i\neq j}e_{ij}^2 &\rightarrow& \E{\epsilon(R,R')^2} + \E{\zeta^2},\\
    \frac{1}{n(n-1)}\sum_{i\neq j}e_{ij}^4 &\rightarrow& \E{\epsilon(R,R')^4} + \E{\zeta^4} + 6\E{\epsilon(R,R')^2}\E{\zeta^2}, \\
    \frac{1}{n(n-1)(n-2)}\sum_{i\neq j\neq k} e_{ij}e_{ik} &\rightarrow& \E{\epsilon(R,R')\epsilon(R,R'')} = \eta_{1,\epsilon}, \label{eqn:key-part-eta} \\
    \frac{1}{n(n-1)(n-2)}\sum_{i\neq j\neq k} e_{ij}^2e_{ik}^2 &\rightarrow& \E{\epsilon(R,R')^2\epsilon(R,R'')^2} + (\E{\zeta^2})^2 + 2\E{\epsilon(R,R')^2}\E{\zeta^2}.
\end{eqnarray}
The proof of the above applies the SLLN for exchangeable dissociated arrays (Lemma \ref{lem:slln-ed}). We use the first result for illustration. Recall the definition of $e_{ij}$ from Assumption \ref{asp:noise-lm}:
\begin{align}
    e_{ij} = \epsilon(R_i, R_j) + \zeta_{ij}. 
\end{align}
We can check that $e_{ij}^2$'s form an exchangeably dissociated array:
\begin{itemize}
    \item \textit{Independence}. Consider a pair of disjoint subsets $\cJ$ and $\cK$. For $J\in\cJ$ and $K\in\cK$, they involve different indices, which means they are defined based on the $R_i$'s and $\zeta_{ij}$'s from two different subsets of units. Hence they are independent. 

    \item \textit{Exchangeability.} The $R_i$'s and $\zeta_{ij}$'s are both i.i.d. variables, thus are exchangeable. Besides, $R_i$'s are independent of $\zeta_{ij}$'s. Therefore, under any permutation $\pi:\bbN_+\to\bbN_+$, $\{(R_i)_{i=1}^\infty, (\zeta_{ij})_{i,j=1}^\infty\}$ always has the same distribution as $\{(R_{\pi(i)})_{i=1}^\infty, (\zeta_{\pi(i)\pi(j)})_{i,j=1}^\infty\}$. Hence, $e_{ij}^2$'s are also exchangeable. 
\end{itemize}
Therefore, by Lemma \ref{lem:slln-ed}, we have
\begin{align}
    \frac{1}{n(n-1)}\sum_{i\neq j}e_{ij}^2 \to 
    \E{e_{ij}^2} = \E{\epsilon(R, R')^2} + \E{\zeta^2} \quad \bbP\text{-a.s.} 
\end{align}
The rest can be justified similarly. 

Moreover, using Lemma \ref{lem:slln-ed}, we can also justify that, for any $l,m\in[p]$, the following results hold $\bbP\text{-a.s.}$:
\begin{eqnarray}
    \frac{1}{n(n-1)}\sum_{i\neq j}  ({b}^{(l)}_{ij} - \overline{b})({b}^{(m)}_{ij} - \overline{b}) &\rightarrow& \E{\tilde{\beta}_2^{(l)}(R,R')\tilde{\beta}_2^{(m)}(R,R')} = \eta^{(lm)}_{2,\beta}, \\
    \frac{1}{n(n-1)}\sum_{i\neq j}  ({b}^{(l)}_{ij} - \overline{b})^2({b}^{(m)}_{ij} - \overline{b})^2&\rightarrow& \E{\tilde{\beta}^{(l)}_2(R,R')^2\tilde{\beta}^{(m)}_2(R,R')^2},\\
    \frac{1}{n(n-1)(n-2)}\sum_{i\neq j\neq k}  ({b}^{(l)}_{ij} - \overline{b})({b}^{(m)}_{ik} - \overline{b})&\rightarrow& \E{\tilde{\beta}_2^{(l)}(R,R')\tilde{\beta}_2^{(m)}(R,R'')} = \eta^{(lm)}_{1,\beta}, \label{eqn:key-part-beta}\\
    \frac{1}{n(n-1)(n-2)}\sum_{i\neq j\neq k}  ({b}^{(l)}_{ij} - \overline{b})^2({b}^{(m)}_{ik} - \overline{b})^2&\rightarrow& \E{\tilde{\beta}^{(l)}_2(R,R')^2\tilde{\beta}^{(m)}_2(R,R'')^2}.
\end{eqnarray}
Now using \eqref{eqn:hH-pi}, \eqref{eqn:key-part-eta} and \eqref{eqn:key-part-beta},  we conclude that 
\begin{align*}
    \hH_{1,\phi}^{\pi}- \eta_{1,\epsilon} H_{1,\beta} = o_{\bbP^\pi}(1) \quad \bbP\text{-a.s.}
\end{align*}
Combining \eqref{eqn:asp-hu-pi} and \eqref{eqn:hV-pi-limit}, we finish the proof. 
\end{proof}

\subsection{Proof of Theorem \ref{thm:mrqap-p-ns}}
\begin{proof}[Proof of Theorem \ref{thm:mrqap-p-ns}]
    \begin{enumerate}
        \item Permuting outcomes:
        \begin{align}
            \sqrt{n}\hvartheta^{(\pi)} &= \sqrt{n}F^\top
            \begin{pmatrix}
                \hSigma_{b b} & \hSigma_{bc}\\
                \hSigma_{c b} & \hSigma_{cc}
            \end{pmatrix}^{-1}
            \begin{pmatrix}
                \hSigma_{ba_\pi }\\
                \hSigma_{ca_\pi }
            \end{pmatrix}, 
        \end{align}
        where $\hSigma_{ba_\pi }$ and $ \hSigma_{ca_\pi }$ are the empirical covariance between $b,c$ and the permuted outcome $a_\pi$, respectively. 
        We can treat the $ (\gamma_{lij}) $ covariates and the noise terms $ (\epsilon_{ij}) $ as the pseudo noise, and repeat the proof in Theorem \ref{thm:asp-mantel-s-p} again. We omit the details here.

        \item Permuting covariates: let $\hSigma_{b_\pi b_\pi}$ and  $\hSigma_{b_\pi c}$ be the sample covariance matrices defined by the permuted covariates $b_\pi$ and the original $c$. Then
        \begin{align}
            \sqrt{n}\hvartheta^{(\pi)} &= 
            \sqrt{n}F^\top
            \begin{pmatrix}
                \hSigma_{b_\pi b_\pi} & \hSigma_{b_\pi c}\\
                \hSigma_{c b_\pi } & \hSigma_{cc}
            \end{pmatrix}^{-1}
            \begin{pmatrix}
                \hSigma_{b_\pi a}\\
                \hSigma_{ca}
            \end{pmatrix}\\
            & = \sqrt{n}F^\top
            \begin{pmatrix}
                \hSigma_{b_\pi b_\pi} & \hSigma_{b_\pi c}\\
                \hSigma_{c b_\pi } & \hSigma_{cc}
            \end{pmatrix}^{-1}
            \lt\{
            \begin{pmatrix}
                \hSigma_{b_\pi b_\pi}\\
                \hSigma_{c b_\pi}
            \end{pmatrix} \cdot 0
            +
            \begin{pmatrix}
                \hSigma_{b_\pi c}\\
                \hSigma_{c c}
            \end{pmatrix} \cdot v
            +
            \begin{pmatrix}
                \hSigma_{b_\pi e}\\
                \hSigma_{c e}
            \end{pmatrix}
            \rt\} \\
            & = \sqrt{n}F^\top
            \begin{pmatrix}
                \hSigma_{b_\pi b_\pi} & \hSigma_{b_\pi c}\\
                \hSigma_{c b_\pi } & \hSigma_{cc}
            \end{pmatrix}^{-1}
            \begin{pmatrix}
                \hSigma_{b_\pi e}\\
                \hSigma_{c e}
            \end{pmatrix}.
        \end{align}
        First, by Lemma \ref{lem:slln-u-md},  we have $\hSigma_{b_\pi b_\pi} = \hSigma_{bb} \to \Sigma_{bb}$, $\bbP{\text{-a.s.}}$
        
        By the WLLN in Lemma \ref{lem:Ppi-converge}, we can show that, we have almost surely 
        \begin{gather}\label{eqn:bpi-c-0}
            \hSigma_{b_\pi c} = o_{\bbP^\pi}(1) \quad \bbP\text{-a.s.}. 
        \end{gather}
        Therefore, the permuted inverse covariance matrix has a block diagonal limit in a large sample. 
        Meanwhile, following a similar argument to Theorem \ref{thm:asp-mantel-s-p}, we can prove that 
        \begin{align}
            \lim_{n\to\infty}\sup_{t\in\bbR}|\cL(t;\sqrt{n} \hSigma_{b_\pi e}) - \cL(t;\cN(0, 4 \eta_{1,\epsilon} H_{1,\beta}))|
            = 0
            \quad 
            \bbP\text{-a.s.}
        \end{align}
        Therefore, 
        \begin{align}
         \lim_{n\to\infty}\sup_{t\in\bbR}|\cL(t;\sqrt{n}\hvartheta^{(\pi)}) - \cL(t;\cN(0, 4\eta_{1,\epsilon} \Sigma_{bb}^{-1} H_{1,\beta} \Sigma_{bb}^{-1}))|
            = 0
            \quad 
            \bbP\text{-a.s.} 
        \end{align}

        \item Permuting $ \hepsilon_{b} $. If we permute the residual of the $ (b_{kij}) $ after regressing on $(c_{lij})$, according to part 2 of this theorem (the permuting covariates part), we can obtain an asymptotic normal distribution: 
        \begin{align}
         \lim_{n\to\infty}\sup_{t\in\bbR}|\cL(t;\sqrt{n}\hvartheta^{(\pi)}) - \cL(t;\cN(0, 4\eta_{1,\epsilon} \Sigma_{\epsilon_b \epsilon_b}^{-1} H_{1,\epsilon_b} \Sigma_{\epsilon_b \epsilon_b}^{-1}))|
            = 0
            \quad 
            \bbP\text{-a.s.} 
        \end{align} 
        where $\epsilon_b$ is the population residual kernel by regressing the $(\tbeta_k)$ kernels on $(\tgamma_l)$ kernels. 
        We now show that the following holds:
        \begin{align}
            \Sigma_{\epsilon_b \epsilon_b}^{-1} H_{1,\epsilon_b} \Sigma_{\epsilon_b \epsilon_b}^{-1}
            =
            F^\top \Sigma^{-1} H_{1,(\beta,\gamma)} \Sigma^{-1} F. 
        \end{align}
        For simplicity, denote
        \begin{align}
            \Psi = \begin{pmatrix}
                I_p & \Sigma_{bc}\Sigma_{cc}^{-1}\\
                0_{q\times p } & I_q
            \end{pmatrix}.
        \end{align}
        Recall the definition
        \begin{align}
            \Sigma 
            &= 
            \E{
            \begin{pmatrix}
                \tbeta\\
                \tgamma
            \end{pmatrix} 
            \begin{pmatrix}
                \tbeta^\top & 
                \tgamma^\top 
            \end{pmatrix} 
            }
            =
            \E{
            \Psi
            \begin{pmatrix}
                \epsilon_b\\
                \tgamma
            \end{pmatrix} 
            \begin{pmatrix}
                \epsilon_b^\top & 
                \tgamma^\top 
            \end{pmatrix} 
            \Psi^\top}
            = \Psi 
            \begin{pmatrix}
                \Sigma_{\epsilon_b\epsilon_b} & 0_{p\times q} \\
                 0_{q \times p} & \Sigma_{cc}.
            \end{pmatrix}
            \Psi^\top.
        \end{align}
        Similarly, noting that $H_{1,(\beta,\gamma)}$ is the covariance for the first-order kernels $\tbeta_1$ and $\tgamma_1$, we can show that, 
        \begin{align}
            H_{1,(\beta, \gamma)} 
            = 
            \Psi
            \begin{pmatrix}
                H_{1,\epsilon_b} & H_{1,\epsilon_b\tgamma} \\
                H_{1,\tgamma\epsilon_b} & H_{1,\gamma}
            \end{pmatrix}
            \Psi^{\top}
            . 
        \end{align}
        Now plugging in the above decomposition, we have 
        \begin{align}
            & F^\top \Sigma^{-1} H_{1,(\beta,\gamma)} \Sigma^{-1} F\\
            = & F^\top 
            \Psi^{-\top} 
            \begin{pmatrix}
                \Sigma_{\epsilon_b\epsilon_b}^{-1} & 0_{p\times q} \\
                 0_{q \times p} & \Sigma_{cc}^{-1}.
            \end{pmatrix} 
            \begin{pmatrix}
                H_{1,\epsilon_b} & H_{1,\epsilon_b\tgamma} \\
                H_{1,\tgamma\epsilon_b} & H_{1,\gamma}
            \end{pmatrix}
            \begin{pmatrix}
                \Sigma_{\epsilon_b\epsilon_b}^{-1} & 0_{p\times q} \\
                 0_{q \times p} & \Sigma_{cc}^{-1}.
            \end{pmatrix} 
            \Psi^{-1} 
            F,
        \end{align}
        and noting that $\Psi^{-1} F = F$, 
        we conclude that 
        \begin{align*}
            F^\top \Sigma^{-1} H_{1,(\beta,\gamma)} \Sigma^{-1} F = \Sigma_{\epsilon_b \epsilon_b}^{-1} H_{1,\epsilon_b} \Sigma_{\epsilon_b \epsilon_b}^{-1}. 
        \end{align*}

        \item Permuting residuals $\hat{e}$: 
        \begin{align}
            \sqrt{n}\hvartheta^{(\pi)} &= \sqrt{n}F^\top
            \begin{pmatrix}
                \hSigma_{b b} & \hSigma_{b c}\\
                \hSigma_{c b} & \hSigma_{cc}
            \end{pmatrix}^{-1}
            \begin{pmatrix}
                \hSigma_{ba}\\
                \hSigma_{ca}
            \end{pmatrix}\\
            & = \sqrt{n}F^\top
            \begin{pmatrix}
                \hSigma_{b b} & \hSigma_{bc}\\
                \hSigma_{c b} & \hSigma_{cc}
            \end{pmatrix}^{-1}
            \lt\{
            \begin{pmatrix}
                \hSigma_{b c}\\
                \hSigma_{c c}
            \end{pmatrix} \cdot \hat{v}
            +
            \begin{pmatrix}
                \hSigma_{b \hat{e}^\pi}\\
                \hSigma_{c \hat{e}^\pi} 
            \end{pmatrix}
            \rt\} \\
            & = \sqrt{n}F^\top
            \begin{pmatrix}
                \hSigma_{b b} & \hSigma_{bc}\\
                \hSigma_{c b} & \hSigma_{cc}
            \end{pmatrix}^{-1}
            \begin{pmatrix}
                \hSigma_{b \hat{e}^\pi}\\
                \hSigma_{c \hat{e}^\pi}
            \end{pmatrix}.
        \end{align}
        Again, following a similar argument as Proof of Theorem \ref{thm:asp-mantel-s-p}, we can show that 
        \begin{align}
            \begin{pmatrix}
                \hSigma_{b b} & \hSigma_{bc}\\
                \hSigma_{c b} & \hSigma_{cc}
            \end{pmatrix}
            \to 
            \begin{pmatrix}
                \Sigma_{b b} & \Sigma_{bc}\\
                \Sigma_{c b} & \Sigma_{cc}
            \end{pmatrix} 
            \quad 
            \bbP\text{-a.s.}
        \end{align}
        and 
        \begin{align}
            \lim_{n\to\infty}\sup_{t\in\bbR}|\cL(t; \sqrt{n} 
            (\hSigma_{b \hat{e}_\pi} \quad 
            \hSigma_{c \hat{e}_\pi})^\top)
            - 
            \cL(t;\cN(0, 4\eta_{1\epsilon} H_{1,(\beta,\gamma)}))| = 0
            \quad 
            \bbP\text{-a.s.}
        \end{align}
        Therefore, we conclude that 
        \begin{align}
            \lim_{n\to\infty}\sup_{t\in\bbR}|\cL(t;\sqrt{n}\hvartheta^{(\pi)}) - \cL(t;\cN(0, 4\eta_{1\epsilon} F^\top \Sigma^{-1} H_{1,(\beta,\gamma)} \Sigma^{-1} F ))| = 0 \quad \bbP\text{-a.s.}
        \end{align}
    \end{enumerate}
\end{proof}

\subsection{Proof of Theorem \ref{thm:mrqap-p-s}}
\begin{proof}[Proof of Theorem \ref{thm:mrqap-p-s}]
The permuted Wald statistic can be expressed as
\begin{align}
    W^{(\pi)} = 
    n 
    \cdot 
    \{F^\top \hw^{(\pi)}\}^\top
    \{F^\top \hV^{(\pi)} F\}^{-1}
    \{F^\top \hw^{(\pi)}\}.
\end{align}
We have shown the distributional convergence of $\sqrt{n} F^\top \hw^{(\pi)}$ in Theorem \ref{thm:mrqap-p-ns}. Now we prove the convergence of the variance estimator under different permutation regimes. 

\begin{enumerate}
    \item Permuting outcome: this is equivalent to treating the $ (\gamma_{lij}) $ and $\epsilon_{ij}$ as the pseudo noise. Similar to the proof in Theorem \ref{thm:asp-mantel-s-p}, we can show that 
    \begin{align*}
        \hH_{1,\phi}^{(\pi)} - \eta_{1\alpha} H_{1,(\beta,\gamma)} = o_{\bbP^\pi}(1) \quad \bbP\text{-a.s.}
    \end{align*}
    Therefore, 
    \begin{align*}
        \hV^{(\pi)} - 4\eta_{1\alpha} \Sigma^{-1} H_{1,(\beta,\gamma)} \Sigma^{-1} = o_{\bbP^\pi}(1) \quad \bbP\text{-a.s.}
    \end{align*}

    \item Permuting covariates: in this case, when constructing $\hV^{(\pi)}$, the regression coefficients $\hvartheta^{(\pi)} \to 0$ and $\hat{\varrho} \to \varrho$. Therefore, following a similar argument to Theorem \ref{thm:asp-mantel-s-p}, we can show that 
    \begin{align}
        \hH_{1,\phi}^{(\pi)} - \eta_{1,\epsilon} H_{1,(\beta,\gamma)} = o_{\bbP^\pi}(1) \quad \bbP\text{-a.s.}
    \end{align}
    Also recalling from \eqref{eqn:bpi-c-0}, we know that
    \begin{align}
        \hSigma^{(\pi)} - 
        \begin{pmatrix}
            \Sigma_{bb} & 0_{p\times q} \\
            0_{q\times p} & \Sigma_{cc}
        \end{pmatrix}
        =
        o_{\bbP^\pi}(1)
        \quad 
        \bbP\text{-a.s.}
    \end{align}
    Hence the following convergence holds $\bbP\text{-a.s.}$:
    \begin{align*}
        \hV^{(\pi)} - 4\eta_{1,\epsilon} \Sigma_{bb}^{-1} H_{1,\beta} \Sigma_{bb}^{-1} = o_{\bbP^\pi}(1) \quad  \bbP\text{-a.s.}
    \end{align*}

    \item Permuting $\hepsilon_b$: when permuting $\hepsilon_b$, the problem is equivalent to reparameterizing the regression with covariates $ \epsilon_b $ and $ c $. Therefore, following a similar argument as the permuting covariates part, we can prove that the following holds $\bbP\text{-a.s.}$:
    \begin{align}
        \hV^{(\pi)} 
        - 
        4\eta_{1,\epsilon} \diag{\Sigma_{\epsilon_b\epsilon_b}^{-1}, \Sigma_{cc}^{-1}} H_{1, (\epsilon_b, \gamma)} \diag{\Sigma_{\epsilon_b\epsilon_b}^{-1}, \Sigma_{cc}^{-1}}
        = o_\bbP(1)
        \quad 
        \bbP\text{-a.s.}
    \end{align}
    This further gives
    \begin{align}
        F^\top \hV^{(\pi)} F 
        - 
        4\eta_{1,\epsilon} \Sigma_{\epsilon_b\epsilon_b}^{-1} H_{1, \epsilon_b} \Sigma_{\epsilon_b\epsilon_b}^{-1}
        = o_\bbP(1)
        \quad 
        \bbP\text{-a.s.}
    \end{align}

    \item Permuting $\hepsilon$: when permuting residuals, the proof is similar to that of Theorem \ref{thm:asp-mantel-s-p}, for which we obtain
    \begin{align}
        \hV^{(\pi)} - 4\eta_{1, \epsilon} \Sigma^{-1} H_{1,(\beta,\gamma)} \Sigma^{-1} 
        = o_{\bbP^\pi}(1)
        \quad \bbP\text{-a.s.}
    \end{align}

\end{enumerate}
Therefore, for all four permutational regimes, the variance estimator $\hV^{(\pi)}$ converges to the asymptotic variance of $\sqrt{n}F^\top \hw^{(\pi)}$, which in turn concludes that the Wald statistic $W^{(\pi)}$ has an asymptotic distribution of $\chi^2_p$.

\end{proof}

\end{document}

%% file: CustomizedCommands.tex
\usepackage{etoolbox}
\usepackage{tabularx, booktabs, graphicx}

\makeatletter
\newcommand{\ostar}{\mathbin{\mathpalette\make@circled\star}}
\newcommand{\make@circled}[2]{%
  \ooalign{$\m@th#1\smallbigcirc{#1}$\cr\hidewidth$\m@th#1#2$\hidewidth\cr}%
}
\newcommand{\smallbigcirc}[1]{%
  \vcenter{\hbox{\scalebox{0.77778}{$\m@th#1\bigcirc$}}}%
}
\makeatother

\newcommand{\lt}{\left}
\newcommand{\rt}{\right}

\newcommand{\commenting}[1]{}

\renewcommand{\hat}{\widehat}
\renewcommand{\tilde}{\widetilde}

\newcommand{\Var}[1]{{\operatorname{Var}\left\{#1\right\}}}

\newcommand{\Cov}[2]{{\operatorname{Cov}\left\{#1,#2\right\}}}

\newcommand{\Corr}[2]{{\operatorname{Corr}\left\{#1,#2\right\}}}

\newcommand{\E}[1]{{\bbE\left\{#1\right\}}}
\newcommand{\Prob}[1]{{\bbP\left\{#1\right\}}}

\newcommand{\ind}[1]{\boldsymbol{1}\left\{#1\right\}}

\newcommand{\indep}{\perp \!\!\! \perp}

\ifdef{\see}{\renewcommand{\see}[1]{\text{ (#1)}}}{\newcommand{\see}[1]{\text{ (#1)}}}

\newcommand{\diag}[1]{\mathrm{Diag}\left\{#1\right\}}

\def\boxit#1{\vbox{\hrule\hbox{\vrule\kern6pt\vbox{\kern6pt#1\kern6pt}\kern6pt\vrule}\hrule}}

\newcolumntype{P}[1]{>{\centering\arraybackslash}p{#1}}
\newcolumntype{M}[1]{>{\centering\arraybackslash}m{#1}}
\newcolumntype{L}[1]{>{\raggedright\arraybackslash}m{#1}}

\newcommand{\cJ}{{\mathcal{J}}}

\newcommand{\cL}{{\mathcal{L}}}

\newcommand{\cN}{{\mathcal{N}}}

\newcommand{\cK}{{\mathcal{K}}}

\newcommand{\fX}{{\mathfrak{X}}}

\newcommand{\bbP}{{\mathbb{P}}}

\newcommand{\bbS}{{\mathbb{S}}}
\newcommand{\bbR}{{\mathbb{R}}}

\newcommand{\bbE}{{\mathbb{E}}}
\newcommand{\bbN}{{\mathbb{N}}}

\newcommand{\bsb}{{\boldsymbol{b}}}
\newcommand{\bc}{{\boldsymbol{c}}}

\newcommand{\oepsilon}{{\overline{\epsilon}}}

\newcommand{\ozeta}{{\overline{\zeta}}}

\newcommand{\oa}{{\overline{a}}}
\newcommand{\ob}{{\overline{b}}}

\newcommand{\oy}{{\overline{y}}}

\newcommand{\ta}{{\tilde{a}}}
\newcommand{\tb}{{\tilde{b}}}
\newcommand{\td}{{\tilde{d}}}

\newcommand{\ty}{{\tilde{y}}}

\newcommand{\tA}{{\tilde{A}}}
\newcommand{\tB}{{\tilde{B}}}

\newcommand{\tX}{{\tilde{X}}}

\newcommand{\tzeta}{{\tilde{\zeta}}}

\newcommand{\heta}{{\hat{\eta}}}
\newcommand{\hepsilon}{{\hat{\epsilon}}}

\newcommand{\hphi}{{\hat{\phi}}}

\newcommand{\hrho}{{\hat{\rho}}}

\newcommand{\hvartheta}{{\hat{\vartheta}}}
\newcommand{\hvarrho}{{\hat{\varrho}}}

\newcommand{\hv}{{\hat{v}}}
\newcommand{\hw}{{\hat{w}}}

\newcommand{\hH}{{\hat{H}}}

\newcommand{\hV}{{\hat{V}}}

\newcommand{\hSigma}{{\hat{\Sigma}}}

\newcommand{\talpha}{{\tilde{\alpha}}}
\newcommand{\tbeta}{{\tilde{\beta}}}
\newcommand{\tgamma}{{\tilde{\gamma}}}

\newcommand{\tphi}{{\tilde{\phi}}}

\newcommand{\tPhi}{{\tilde{\Phi}}}

